\documentclass{aastex}
\usepackage{spr-astr-addons}             
\usepackage{url}
\usepackage{lscape, morefloats, graphicx, float, longtable}
\usepackage{epstopdf}
\usepackage{graphicx}
\RequirePackage{color}
\usepackage[caption2]{ccaption}                   

\begin{document}

\title{CCD {\it UBV} photometric and {\it Gaia} astrometric study of eight open clusters
- ASCC\, 115, Collinder\, 421, NGC\, 6793, NGC\, 7031, NGC\, 7039, NGC\, 7086, Roslund\, 1 and Stock\, 21}
\slugcomment{Not to appear in Nonlearned J., 45.}
\shorttitle{Photometric and astrometric study of eight clusters}
\shortauthors{T. Yontan, S. Bilir, Z.~F. Bostanc\i, T. Ak, S. Ak, T. G\"uver,  E. Paunzen, 
H. \"Urg\"up, M. \c Celebi, B. A. Akti, S. G\"okmen}

\author{T. Yontan \altaffilmark{1}}
\altaffiltext{1}{Istanbul University, Institute of Graduate 
Studies in Science, Programme of Astronomy and 
Space Sciences, 34116, Beyaz{\i}t, Istanbul, Turkey\\
Corresponding Author: talar.yontan@istanbul.edu.tr\\}

\author{S. Bilir \altaffilmark{2}}
\altaffiltext{2}{Istanbul University, Faculty of Science, Department 
of Astronomy and Space Sciences, 34119, University, Istanbul, Turkey\\}

\author{Z.~F. Bostanc\i~\altaffilmark{2}}
\altaffiltext{2}{Istanbul University, Faculty of Science, Department 
of Astronomy and Space Sciences, 34119, University, Istanbul, Turkey\\}

\author{T. Ak \altaffilmark{2}}
\altaffiltext{2}{Istanbul University, Faculty of Science, Department 
of Astronomy and Space Sciences, 34119, University, Istanbul, Turkey\\}

\author{S. Ak \altaffilmark{2}}
\altaffiltext{2}{Istanbul University, Faculty of Science, Department 
of Astronomy and Space Sciences, 34119, University, Istanbul, Turkey\\}

\author{T. G\"uver \altaffilmark{2, 3}}
\altaffiltext{2}{Istanbul University, Faculty of Science, Department 
of Astronomy and Space Sciences, 34119 University, Istanbul, Turkey\\}
\altaffiltext{3}{Istanbul University Observatory Research and Application 
Center, Beyaz{\i}t, 34119, Istanbul, Turkey\\}

\author{E. Paunzen \altaffilmark{4}}
\altaffiltext{4}{Department of Theoretical Physics and Astrophysics, 
Masaryk University, Kotl\'a\u rsk\'a 2, 611 37 Brno, Czech Republic\\}

\author{H. \"Urg\"up \altaffilmark{1}}
\altaffiltext{1}{Istanbul University, Institute of Graduate 
Studies in Science, Programme of Astronomy and 
Space Sciences, 34116, Beyaz{\i}t, Istanbul, Turkey\\}

\author{M. \c Celebi \altaffilmark{1}}
\altaffiltext{1}{Istanbul University, Institute of Graduate 
Studies in Science, Programme of Astronomy and 
Space Sciences, 34116, Beyaz{\i}t, Istanbul, Turkey\\}

\author{B. A. Akti \altaffilmark{1}}
\altaffiltext{1}{Istanbul University, Institute of Graduate 
Studies in Science, Programme of Astronomy and 
Space Sciences, 34116, Beyaz{\i}t, Istanbul, Turkey\\}

\and 

\author{S. G\"okmen \altaffilmark{5}}
\altaffiltext{5}{Akdeniz University, Faculty of Sciences, 
Department of Space Sciences and Technologies, 07058, 
Antalya, Turkey\\}

\begin{abstract}
In this study, we carried out CCD {\it UBV} photometry of eight open clusters, ASCC 115, Collinder 421, NGC 6793, NGC 7031, NGC 7039, NGC 7086, Roslund 1, Stock 21, and determined  their reddening, metallicity, distance, age, and mass functions. We used new {\it Gaia} Data Release 2 (DR2) astrometric data to separate cluster member stars from the field stars and obtain precise structural and astrophysical parameters. To identify cluster member stars we utilized an unsupervised membership assignment code (UPMASK), which is based on the photometric and astrometric data. The density distributions for the open clusters show good fits with the empirical King model except for Roslund 1 and Stock 21 not having central concentration. The colour excesses and metallicities were derived separately using $U-B\times B-V$ two-colour diagrams. Keeping these parameters as constants, we simultaneously calculated distance moduli and ages of the clusters from $V\times B-V$ and $V \times U-B$ colour-magnitude diagrams using PARSEC theoretical isochrones. Taking into account {\it Gaia} DR2 proper motion components and parallaxes of the member stars, we also calculated mean proper motions and distances for the clusters. Distances derived both from isochrone fitting to colour-magnitude diagrams of the clusters and {\it Gaia} DR2 trigonometric parallaxes are compatible with each other. Slopes of the mass functions of the eight open clusters are in good agreement with \citet{Salpeter55} value of 1.35.  
\end{abstract}


\section{Introduction}
Open star clusters (OCs) are homogeneous stellar systems that formed under the same physical conditions along Galactic plane with rich gas and dust. They contain from tens to a few thousand stars located at very similar distances, having similar age and initial chemical composition, but they are found in different ranges of stellar mass. Therefore, these systems are individually ideal laboratories to study stellar formation and evolution and they can be used to constrain and test the stellar evolution theories. OCs do not only provide information about stellar physics, dynamics and evolution, but also represent the disc structure of the Milky Way. OCs are excellent to examine structural, dynamical and chemical evolution of the Galaxy, because of the fact that the amount of interstellar reddening towards them, their chemical abundances, distances and ages can be determined with great accuracy constituting two-colour diagrams (TCDs) and colour-magnitude diagrams (CMDs) from {\it UBV} photometric data, and comparing these diagrams with stellar models and isochrones.

Homogeneous analysis and precise member selection methods of the cluster stars are important to obtain  precise astrophysical parameters of the cluster, interpret the cluster astrophysical parameters and understand the structure of the Milky Way. Using different data and analysis methods led to discrepant values \citep{Dias14, Netopil15}. Various data of different quality, isochrone-fitting methods and non-sensitive determination of the physical member stars can cause incompatible results and degeneracy between determined parameters of clusters \citep{Anders04, King05}. 

The {\it Gaia} second  data release \citep[hereafter {\it Gaia} DR2,][]{Gaia18} provide unprecedented astronomical information, such as positions, parallaxes and proper motions, for more than 1.3 billion objects. This allows us to accurately separate cluster member stars from field star contamination. In our previous studies \citep[cf.][]{Yontan15, Bostanci15, Ak16}, we used proper motions of stars located through cluster field collected from PPMXL \citep{Roeser10} and calculated their membership probabilities utilizing a method given by \citet{Balaguer98}. 

\citet{Cantat-Gaudin18} determined the membership probabilities of 401488 stars in 1229 open clusters using {\it Gaia} DR2 photometric and astrometric data. They also estimated mean proper motion components and distances of these clusters. Their results show that the youngest clusters are  located near the Galactic plane and they are tracers of the spiral arms of the Galaxy, while older objects are scattered from the Galactic plane, and prone to be located at larger distances from the Galactic center. Recently, \citet{Bossini19} found the age, distance modulus, and extinction of 260 open clusters using only the {\it Gaia} DR2 photometric and astrometric data. They estimated the membership probabilities of the stars in the cluster fields by making use of a statistical method described by \citet{Cantat-Gaudin18}, which is based on the {\it Gaia} DR2 photometry, proper motions and parallaxes. In this study, we used the method descried by \citet{Cantat-Gaudin18} to separate the cluster members from the field stars. 

In this paper, our goal is to determine precise astrophysical parameters of eight open clusters (ASCC 115, Collinder 421, NGC 6793, NGC 7031 NGC 7039, NGC 7086, Roslund 1 and Stock 21), which were not well studied in the past. Combining {\it Gaia} DR2 proper motion components and trigonometric parallaxes with observational CCD {\it UBV} data and using independent methods, we determined mean proper motion, reddening, photometric metallicity, distance, age and mass function of each cluster. We summarized the physical parameters given in previous studies and derived in this study of eight clusters in Table 1.   
\begin{table*}
\setlength{\tabcolsep}{1.8pt}
{\tiny
  \centering
  \caption{Astrophysical parameters of the eight open clusters, which were derived from literature and this study. Columns represent cluster names, Galactic coordinates ($l, b$), colour excesses ($E(B-V)$), distance moduli ($\mu_V$), distances ($d$), iron abundances ([Fe/H]) and ages ($t, \log t$) estimated using CMDs and TCDs of each cluster. Additionally, distances and mean proper motion components ($\langle\mu_{\alpha}\cos \delta\rangle$, $\langle\mu_{\delta}\rangle$) and mean distances ($\langle d_{Gaia}\rangle$) obtained from {\it Gaia} DR2 are also given for each.}
    \begin{tabular}{lcccccccccccc}
\hline
Cluster & $l$  & $b$  & $E(B-V)$ & $\mu_V$ & $d$ & [Fe/H]  & $t$ & $\log t$ & $\langle d_{Gaia}\rangle$ & $\langle\mu_{\alpha}\cos \delta\rangle$ & $\langle\mu_{\delta}\rangle $ & Reference \\
            &($^o$)& ($^o$)& (mag)  & (mag) & (pc) & (dex) & (Myr) & (yr) & (pc) & (mas yr$^{-1}$) & (mas yr$^{-1}$) &  \\
\hline
ASCC 115 & 97.45 & -2.54 & 0.15 & 9.356 & 600 & -- & 389 & 8.59 & -- & -- & -- & (1)\\
           &   &   & 0.17$\pm$0.06 & 9.85$\pm$0.20 & 732$\pm$69 & 0 & 225$\pm$25 & 8.35$\pm$0.05 & 755$\pm$14 & -0.524$\pm$0.062 & -0.568$\pm$0.050 & (2) \\
           &   &   &   &   &   &   &   &   &   &   &   &  \\
Collinder 421 & 79.45 & 2.53 & 0.1 & 10.2 & 950 & -- & 1150 & 9.06 & -- & -- & -- & (3)\\
      &   &   & 0.64$^{+0.11}_{-0.12}$ & 12.08$^{+0.48}_{-0.64}$ & 1050$^{+330}_{-800}$ & -- & 250 & 8.40 & -- & -- & -- & (4) \\
      &   &   & 0.75$\pm$0.05 & 12.80$\pm$0.17 & 1245$\pm$103 & -0.03$\pm$0.10 & 125$\pm$25 & 8.10$\pm$0.08 & 1220$\pm$99 & -3.662$\pm$0.076 & -8.309$\pm$0.082 & (2) \\
      &   &   &   &   &   &   &   &   &   &   &   &  \\
NGC 6793 & 56.17 & 3.30 & 0.17 & 10.73 & 1100 & -- & 436 & 8.64 & -- & -- & -- & (3) \\
      &   &   & 0.33$\pm$0.04 & 9.95$\pm$0.14 & 610$\pm$40 & -0.09$\pm$0.10 & 500$\pm$50 & 8.70$\pm$0.04 & 607$\pm$37 & 3.818$\pm$0.055 & 3.611$\pm$0.071 & (2) \\
      &   &   &   &   &   &   &   &   &   &   &   &  \\
NGC 7031 & 91.33 & 2.31 & 1.03 & 12.60 & 760 & -- & -- & -- & -- & -- & -- & (5) \\
      &   &   & 0.85 & 12.41 & 900 & -- & 137 & 8.14  & --  & --  & --  & (6) \\
      &   &   & -- & 12.25 & 910 & -- & 56 & 7.75 & -- & -- & -- & (7) \\
      &   &   & 0.71-0.84 & 11.45-11.70 & 730-686 & -- & 480 & 8.68 & -- & -- & -- & (8) \\
      &   &   & 1.05$\pm$0.05 & 9.6$\pm$0.2 & 831$\pm$72 & --& 224$\pm$25 & 8.35 & -- & -- & -- & (9)\\
      &   &   & 0.93$\pm$0.08 & 13.30$\pm$0.25 & 1212$\pm$146 & -0.15$\pm$0.13 & 65$\pm$5 & 7.81$\pm$0.03 & 1404$\pm$81 & -1.286$\pm$0.081 & -4.177$\pm$0.076 & (2) \\
      &   &   &   &   &   &   &   &   &   &   &   &  \\
NGC 7039 & 87.88 & -1.70 & 0.13 & 9.60 & 700 & -- & 10 & 7.00 & -- & -- & -- & (10)\\
      &   &   & 0.19 & 11.5 & 1535 & -- & 1000 & 9.00 & -- & -- & -- & (11) \\
      &   &   & 0.076 & 9.14$\pm$0.23 & 675$\pm$70 & --& -- & -- & -- & -- & -- & (12) \\
      &   &   & 0.13 & 10.294 & 951 & -- & 680 & 8.83 & -- & -- & -- & (3) \\
      &   &   & 0.16$\pm$0.05 & 9.85$\pm$0.18 & 743$\pm$64& 0.02$\pm$0.07 & 100$\pm$25 & 8.00$\pm$0.10 & 767$\pm$41 & -0.262$\pm$0.100 & -2.509$\pm$0.097 & (2) \\
      &   &   &   &   &   &   &   &   &   &   &   &  \\
NGC 7086 & 94.41 & 0.22 & 0.69 & 12.4 & 1170 & -- & 600 & 8.78 & -- & -- & -- & (13)  \\
      &   &   & -- & 12.5 & 1205 & -- & 85 & 7.93 & --  & --  & --  & (7) \\
      &   &   & 0.83$\pm$0.02 & 13.4$\pm$0.3 & 1500 & -- & 100 & 8.00 & -- & -- & -- & (14) \\
      &   &   & 0.75$\pm$0.05 & 9.9$\pm$0.2 & 955$\pm$84 & -- & 178$\pm$25 & 8.25$\pm$0.06 & -- & -- & -- & (9) \\
      &   &   & 0.75$\pm$0.07 & 13.37$\pm$0.23 & 1618$\pm$182 & -0.10$\pm$0.20 & 150$\pm$25 & 8.18$\pm$0.07 & 1684$\pm$140 & -1.642$\pm$0.086 & -1.644$\pm$0.076 & (2) \\
      &   &   &   &   &   &   &   &   &   &   &   &  \\
Roslund 1 & 54.60 & -3.40 & 0.05 & 9.285 & 670 & -- & 295 & 8.47 & -- & -- & -- & (3) \\
      &   &   & 0.19$\pm$0.03 & 10.20$\pm$0.12 & 836$\pm$48 & -0.12$\pm$0.09 & 700$\pm$50 & 8.84$\pm$0.03 & 883$\pm$54 & -0.895$\pm$0.043 & -5.766$\pm$0.044 & (2) \\
      &   &   &   &   &   &   &   &   &   &   &   &  \\
Stock 21 & 120.05 & -4.83 & 0.4 & 11.447 & 1100 & -- & 525 & 8.72 & -- & -- & -- & (3) \\
      &   &   & 0.41$\pm$0.04 & 12.70$\pm$0.14 & 1931$\pm$127 & -0.01$\pm$0.08 & 450$\pm$25 & 8.65$\pm$0.02 & 1934$\pm$159 & -2.154$\pm$0.063 &  -1.752$\pm$0.059 & (2) \\
\hline
    \end{tabular}
\\
(1) \citet{Kharchenko09}, (2) This study, (3) \citet{Kharchenko05}, (4) \citet{Maciejewski07}, (5) \citet{Svolopoulos61}, (6) \citet{Hoag61}, (7) \citet{Lindoff68}, (8) \citet{Hassan-Barbon73}, (9) \citet{Kopchev08}, (10) \citet{Zelwanowa72}, (11) \citet{Hassan73}, (12) \citet{Schneider87}, (13) \citet{Hassan67}, (14) \citet{Rosvick06}    
}
\end{table*}

\begin{table*}
  \centering
  \caption{Log of the observations. $N$ refers to the number of exposure.}
    \begin{tabular}{lcccc}
\hline
      & & \multicolumn{3}{c}{Filter/(Exposure Time (s)$\times N$)} \\
    Cluster & Observation Date & $U$ & $B$ & $V$ \\
\hline
    ASCC 115 & 08.10.2013 & 360$\times$4, 60$\times$5, 5$\times$5 & 60$\times$5, 6$\times$5, 0.4$\times$5 & 30$\times$6, 3$\times$5, 0.3$\times$5 \\
Collinder 421& 16.08.2012 & 360$\times$3, 60$\times$4 & 90$\times$2, 7$\times$3 & 60$\times$3, 5$\times$4 \\
    NGC 6793 & 26.08.2014 & 300$\times$4, 30$\times$4 & 90$\times$5, 2.5$\times$4 & 60$\times$10, 2$\times$10 \\
    NGC 7031 & 17.08.2015 & 900$\times$3, 300$\times$5 & 600$\times$3, 60$\times$5 & 600$\times$3, 9$\times$5 \\
    NGC 7039 & 10.08.2013 & 300$\times$3, 8$\times$5 & 60$\times$5, 1$\times$5 & 20$\times$5, 1$\times$5 \\
    NGC 7086 & 19.07.2012 & 360$\times$3, 90$\times$3 & 100$\times$3, 10$\times$3 & 60$\times$3, 5$\times$3 \\
   Roslund 1 & 09.08.2013 & 300$\times$3, 40$\times$5 & 60$\times$5, 4$\times$5 & 30$\times$5, 3$\times$5 \\
    Stock 21 & 08.08.2013 & 300$\times$3, 60$\times$4 & 120$\times$4, 7$\times$4 & 60$\times$4, 2$\times$4 \\
\hline
    \end{tabular}
\end{table*}

\section{Observations}

We observed eight open clusters using 1-m Ritchey-Chr\'etien telescope (T100) of the T\"UB\.ITAK National Observatory (TUG) located in Antalya, Turkey. We present inverse coloured $V$-band images of the clusters in Fig. 1. Each frame in Fig. 1 shows a stacked image of the longest exposures obtained. A Spectral Instruments (SI 1100) CCD camera operating at -90$^{o}$C is attached to this telescope. The camera is equipped with 4k$\times$4k format Fairchild 486 BI CCD. The readout noise and the gain of the CCD camera is 4.11 e$^{-}$ and 0.57 e$^{-}$/ADU, respectively. Pixel scale of the telescope-camera system is 0.$^{''}$31 pixel$^{-1}$, allowing an unvignetted field of view of about $21^{'}\times 21^{'}$. A log of observations is given in Table 2. During the observations, we used different exposure times in order to obtain the widest possible flux range in three different bands: $U$, $B$, and $V$. In order to study the fainter stars, images taken with longer exposure times for each filter were integrated. We used IRAF\footnote{IRAF is distributed by the National Optical Astronomy Observatories}, PyRAF\footnote{PyRAF is a product of the Space Telescope Science Institute, which is operated by AURA for NASA} and astrometry.net\footnote{http://astrometry.net} routines together with our own scripts for CCD calibrations and astrometric corrections of the images. Atmospheric extinction and photometric transformation coefficients for each night were obtained from the observations of stellar fields including a few stars selected from \citet{Landolt09}'s catalogue of {\it UBVRI} photometric standard stars. For each field three images were taken per photometric band and some fields could be observed more than once. The aperture photometry packages of IRAF were used to measure the instrumental magnitudes of the standard stars. Once we obtained the instrumental magnitudes of the standard stars, we could easily apply multiple linear fits to these magnitudes and calculated the photometric extinction and transformation coefficients for each night. The extinction and transformation coefficients for each observing night are listed in Table 3. Brightness of the objects in the cluster fields were measured with Source Extractor and PSF Extractor (PSFEx) routines \citep{Bertin96}. We applied aperture correction to instrumental magnitudes of stars in each cluster field and obtained their apparent magnitudes in the Johnson photometric system with transformation equations given by \citet{Janes13}.

\begin{table*}[htbp]
  \centering
  \caption{Derived transformation and extinction coefficients. $k$ and $k^{'}$ are primary and secondary extinction coefficients, respectively, while $\alpha$ and $C$ are transformation coefficients.}
    \begin{tabular}{cccccc}
 \hline
    Filter/Colour index & Observation Date & $k$ & $k'$ & $\alpha$ & $C$ \\
\hline
    $U$ & 19.07.2012 & 0.472$\pm$0.031 & -0.019$\pm$0.033 & -- & -- \\
    $B$ &            & 0.326$\pm$0.024 & -0.057$\pm$0.025 & 0.992$\pm$0.038 & 0.745$\pm$0.036 \\
    $V$ &            & 0.189$\pm$0.007 & -- & -- & -- \\
    $U-B$            & & -- & -- & 0.861$\pm$0.050 & 3.143$\pm$0.046 \\
    $B-V$            & & -- & -- & 0.077$\pm$0.011 & 0.799$\pm$0.017 \\
\hline
    $U$ & 16.08.2012 & 0.677$\pm$0.030 & -0.023$\pm$0.067 & -- & -- \\
    $B$ &            & 0.525$\pm$0.050 & -0.246$\pm$0.079 & 1.253$\pm$0.105 & 0.550$\pm$0.067 \\
    $V$ &            & 0.242$\pm$0.008 & -- & -- & -- \\
    $U-B$&           & -- & -- & 0.836$\pm$0.093 & 3.041$\pm$0.043 \\
    $B-V$&           & -- & -- & 0.084$\pm$0.011 & 0.779$\pm$0.016 \\
\hline
    $U$ & 08.08.2013 & 0.397$\pm$0.014 & +0.007$\pm$0.018 & -- & -- \\
    $B$ &            & 0.240$\pm$0.014 & -0.041$\pm$0.015 & 0.972$\pm$0.024 & 0.481$\pm$0.021 \\
    $V$ &            & 0.118$\pm$0.005 & -- & -- & -- \\
    $U-B$            & & -- & -- & 0.840$\pm$0.027 & 2.893$\pm$0.020 \\
    $B-V$            & & -- & -- & 0.070$\pm$0.003 & 0.538$\pm$0.005 \\
\hline
    $U$ & 09.08.2013 & 0.347$\pm$0.016 & -0.066$\pm$0.018 & -- & -- \\
    $B$ &            & 0.206$\pm$0.014 & -0.032$\pm$0.014 & 0.959$\pm$0.023 & 0.552$\pm$0.022 \\
    $V$ &            & 0.098$\pm$0.005 & -- & -- & -- \\
    $U-B$            & & -- & -- & 0.951$\pm$0.028 & 2.991$\pm$0.0238 \\
    $B-V$            & & -- & -- & 0.068$\pm$0.006 & 0.591$\pm$0.0092 \\
\hline
    $U$ & 10.08.2013 & 0.439$\pm$0.025 & -0.126$\pm$0.024 & -- & -- \\
    $B$ &            & 0.328$\pm$0.042 & -0.097$\pm$0.038 & 1.045$\pm$0.055 & 0.431$\pm$0.060 \\
    $V$ &            & 0.149$\pm$0.006 & -- & -- & -- \\
    $U-B$            & & -- & -- & 1.038$\pm$0.037 & 2.908$\pm$0.038 \\
    $B-V$            & & -- & -- & 0.069$\pm$0.010 & 0.559$\pm$0.015 \\
\hline
    $U$  & 08.10.2013& 0.354$\pm$0.021 & -0.034$\pm$0.031 & -- & -- \\
    $B$  &           & 0.194$\pm$0.013 & -0.022$\pm$0.016 & 0.942$\pm$0.023 & 0.569$\pm$0.019 \\
    $V$  &           & 0.098$\pm$0.003 & -- & -- & -- \\
    $U-B$&           & -- & -- & 0.917$\pm$0.049 & 2.964$\pm$0.033 \\
    $B-V$&           & -- & -- & 0.071$\pm$0.006 & 0.582$\pm$0.009 \\
\hline
    $U$ & 26.08.2014 & 0.512$\pm$0.061 & +0.436$\pm$0.148 & -- & -- \\
    $B$ &            & 0.493$\pm$0.092 & -0.256$\pm$0.093 & 1.251$\pm$0.127 & 0.687$\pm$0.125 \\
    $V$ &            & 0.203$\pm$0.009 & -- & -- & -- \\
    $U-B$            & & -- & -- & 0.294$\pm$0.198 & 3.303$\pm$0.086 \\
    $B-V$            & & -- & -- & 0.069$\pm$0.020 & 0.969$\pm$0.030 \\
\hline
    $U$ & 17.08.2015 & 0.451$\pm$0.074 & +0.237$\pm$0.091 & -- & -- \\
    $B$ &            & 0.355$\pm$0.087 & -0.109$\pm$0.093 & 1.063$\pm$0.131 & 1.417$\pm$0.123 \\
    $V$ &            & 0.153$\pm$0.013 & -- & -- & -- \\
    $U-B$            & & -- & -- & 0.487$\pm$0.130 & 3.938$\pm$0.106 \\
    $B-V$            & & -- & -- & 0.076$\pm$0.031 & 1.520$\pm$0.046 \\
\hline
    \end{tabular}%
  \label{tab:addlabel}%
\end{table*}%

\begin{figure*}
\centering
\includegraphics[scale=.4, angle=0]{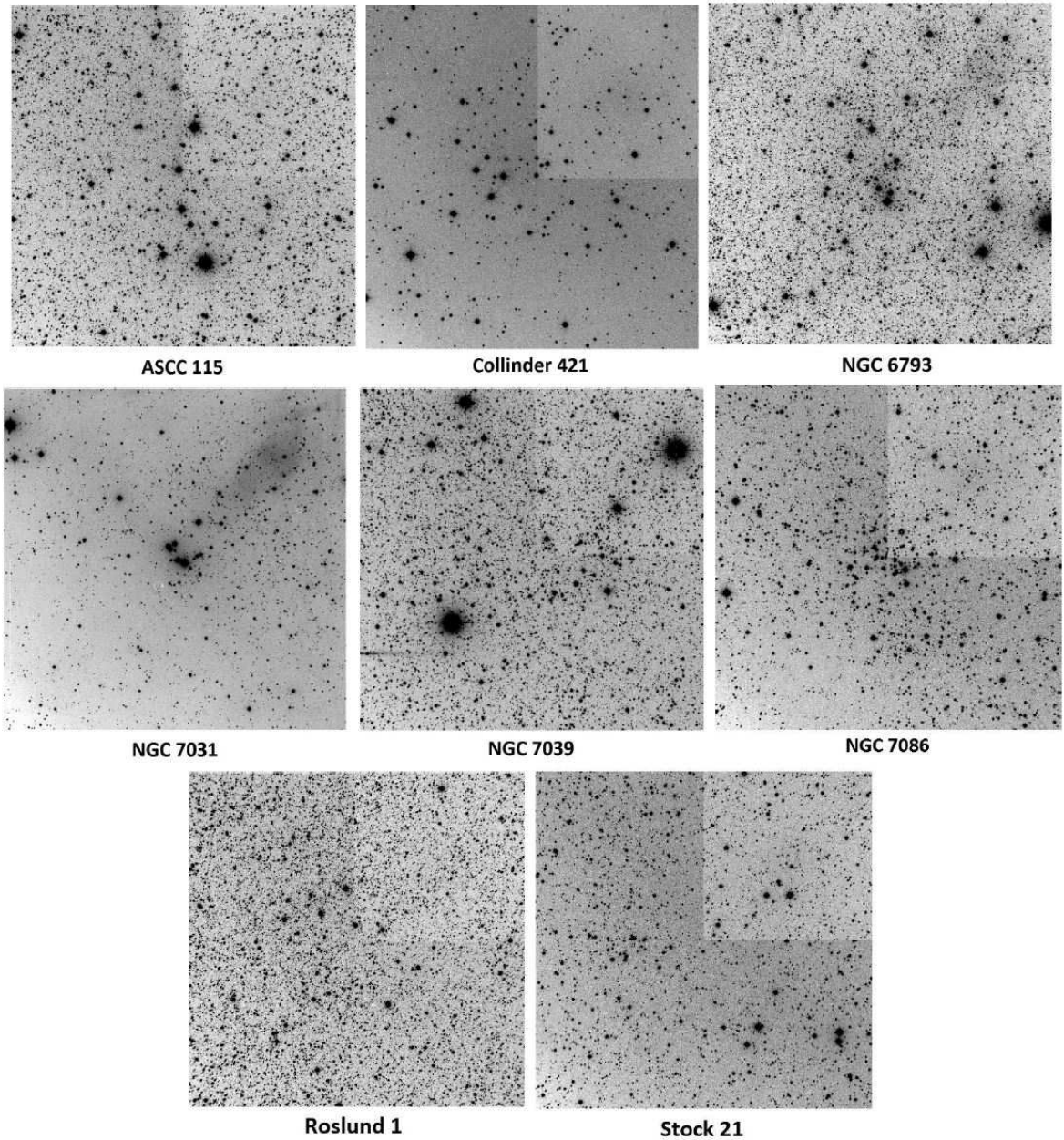}
\caption{Inverse coloured $V$-band images of eight open clusters taken with T100 telescope. The field of view is about 21$\times$21 arcmin (North top and East left).} 
\end {figure*}

\begin{figure*}
\centering
\includegraphics[scale=.8, angle=0]{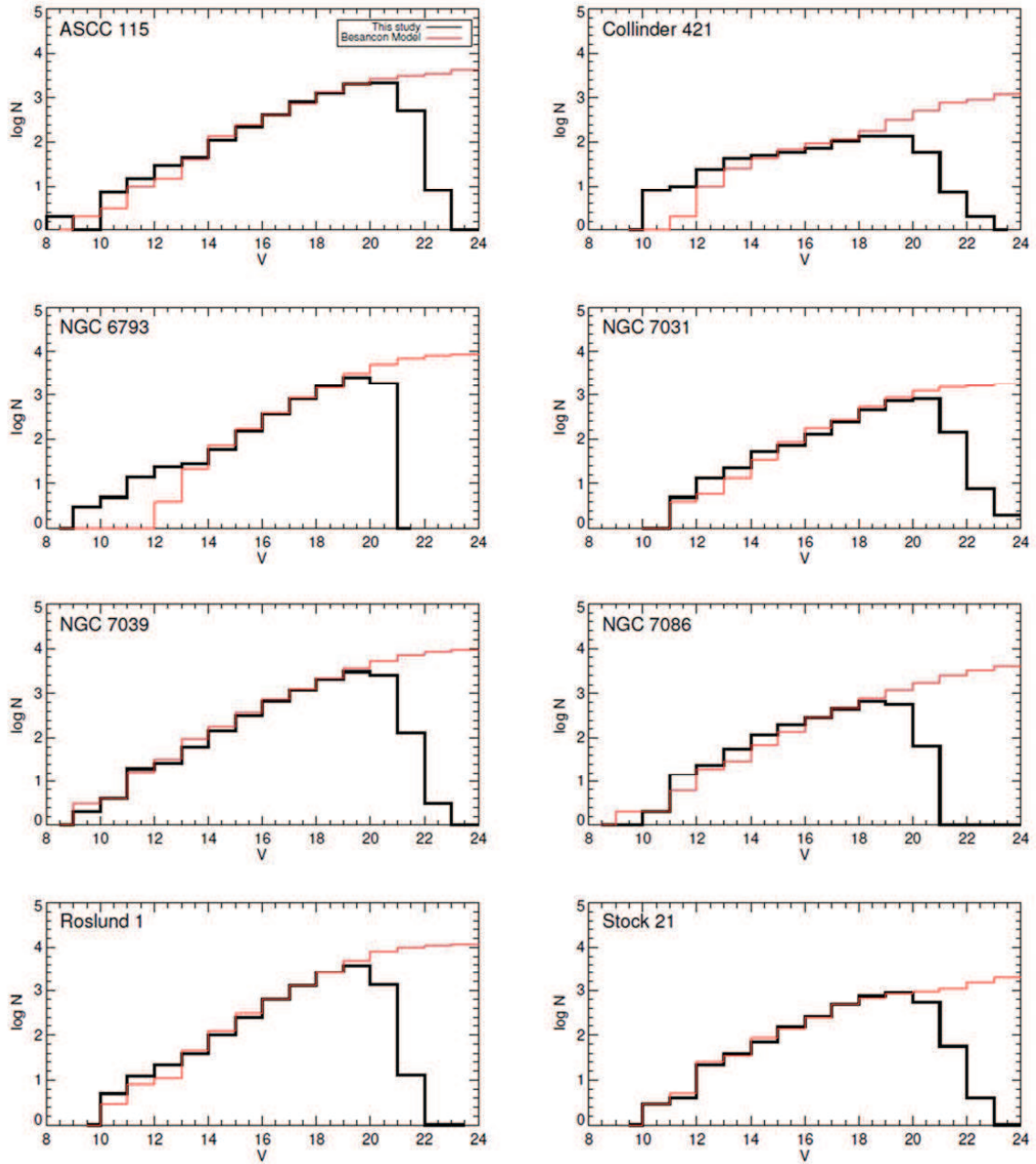}
\caption{Histograms of the $V$-band magnitudes for eight open clusters. Black and red lines show observational and synthetic star count from Besan\c con Galaxy model for each $V$-magnitude interval, respectively.} 
\end {figure*}

\begin{table*}
\setlength{\tabcolsep}{2.2pt}
  \centering
  \caption{Mean internal errors of the photometric measurements for stars in the directions of eight open clusters. $N$ represents the number of stars within the $V$ apparent magnitude range.}
    \begin{tabular}{ccccccccccccccccc}
\hline
            & \multicolumn{4}{c}{ASCC 115} & \multicolumn{4}{c}{Collinder 421} & \multicolumn{4}{c}{NGC 6793} & \multicolumn{4}{c}{NGC 7031} \\
$V$ & $N$ & $\sigma_V$ & $\sigma_{U-B}$ & $\sigma_{B-V}$ & $N$ & $\sigma_V$ & $\sigma_{U-B}$ & $\sigma_{B-V}$ & $N$ & $\sigma_V$ & $\sigma_{U-B}$ & $\sigma_{B-V}$ & $N$ & $\sigma_{V}$ & $\sigma_{U-B}$ & $\sigma_{B-V}$ \\
\hline
     (8,10] &    2 & 0.001 & 0.001 & 0.002 &   1 & 0.000 & 0.001 & 0.001 &   -- & -- & -- & -- & -- & -- & -- & -- \\
    (10,12] &   22 & 0.003 & 0.005 & 0.005 &  18 & 0.001 & 0.002 & 0.002 &   24 & 0.001 & 0.003 & 0.002 &    6 & 0.001 & 0.003 & 0.001 \\
    (12,14] &   75 & 0.001 & 0.002 & 0.002 &  67 & 0.001 & 0.006 & 0.002 &   53 & 0.001 & 0.004 & 0.002 &   35 & 0.002 & 0.008 & 0.003 \\
    (14,16] &  329 & 0.002 & 0.006 & 0.003 & 111 & 0.002 & 0.023 & 0.004 &  215 & 0.002 & 0.017 & 0.004 &  126 & 0.003 & 0.032 & 0.004 \\
    (16,18] & 1237 & 0.006 & 0.019 & 0.009 & 180 & 0.007 & 0.110 & 0.015 & 1280 & 0.005 & 0.064 & 0.013 &  377 & 0.003 & 0.058 & 0.009 \\
    (18,20] & 3258 & 0.020 & 0.074 & 0.036 & 277 & 0.024 & 0.239 & 0.082 & 5563 & 0.020 & 0.128 & 0.066 & 1196 & 0.013 & 0.077 & 0.049 \\
    (20,22] & 2686 & 0.064 & 0.143 & 0.131 &  68 & 0.078 & 0.025 & 0.233 & 2875 & 0.051 & 0.242 & 0.177 &  968 & 0.042 & 0.140 & 0.152 \\
\hline
            & \multicolumn{4}{c}{NGC 7039} & \multicolumn{4}{c}{NGC 7086} & \multicolumn{4}{c}{Roslund 1} & \multicolumn{4}{c}{Stock 21} \\
$V$ & $N$ & $\sigma_V$ & $\sigma_{U-B}$ & $\sigma_{B-V}$ & $N$ & $\sigma_V$ & $\sigma_{U-B}$ & $\sigma_{B-V}$ & $N$ & $\sigma_V$ & $\sigma_{U-B}$ & $\sigma_{B-V}$ & $N$ & $\sigma_{V}$ & $\sigma_{U-B}$ & $\sigma_{B-V}$ \\
\hline
     (8,10] &    3 & 0.001 & 0.001 & 0.001 &   -- & -- & -- & -- & -- & -- & -- & -- & -- & -- & -- & -- \\
    (10,12] &   24 & 0.002 & 0.004 & 0.003 &   16 & 0.001 & 0.002 & 0.002 &   17 & 0.001 & 0.002 & 0.001 &    7 & 0.001 & 0.002 & 0.002 \\
    (12,14] &   90 & 0.002 & 0.003 & 0.002 &   81 & 0.001 & 0.002 & 0.001 &   62 & 0.001 & 0.004 & 0.002 &   61 & 0.001 & 0.003 & 0.002 \\
    (14,16] &  473 & 0.003 & 0.007 & 0.004 &  316 & 0.002 & 0.009 & 0.004 &  358 & 0.002 & 0.012 & 0.004 &  221 & 0.002 & 0.008 & 0.003 \\
    (16,18] & 1829 & 0.008 & 0.025 & 0.011 &  741 & 0.007 & 0.046 & 0.014 & 1933 & 0.008 & 0.037 & 0.012 &  752 & 0.005 & 0.029 & 0.008 \\
    (18,20] & 5012 & 0.031 & 0.100 & 0.048 & 1217 & 0.027 & 0.139 & 0.064 & 6597 & 0.032 & 0.124 & 0.055 & 1619 & 0.019 & 0.117 & 0.032 \\
    (20,22] & 2614 & 0.084 & 0.187 & 0.142 &   67 & 0.072 & 0.191 & 0.157 & 1342 & 0.078 & 0.218 & 0.140 &  599 & 0.060 & 0.221 & 0.108 \\
\hline
    \end{tabular}
\end{table*}

\section{Data Analysis}

\subsection{Photometric data}
We electronically present photometric catalogue of the eight open clusters, which were constructed from all the detected objects in the field of view. Each catalogue contains equatorial coordinates, apparent magnitude $V$, $U-B$, $B-V$ colours, proper motion components ($\mu_{\alpha}\cos\delta, \mu_{\delta}$) and trigonometric parallaxes ($\pi$) from {\it Gaia} DR2, and their errors, respectively. We listed mean internal photometric errors in $V$ band, $U-B$ and $B-V$ colours as a function of $V$ apparent magnitudes in Table 4. We can clearly see that the stars brighter than $V=18$ magnitudes have internal errors smaller than 0.1 mag and the errors increase towards fainter stars. To constrain the photometric quality and perform precise analyses, we first need to determine the photometric completeness limit of the data by constructing $V$ magnitude histograms for the eight clusters (Fig. 2). A cluster region can contain stars much brighter or fainter than this completeness limit. The limit depends on the determination of the magnitude at which the stars start to be fainter since they are below the exposure time threshold during the observations.  

To check both photometric completeness limits and background value of star density profiles we produced synthetic stars for the region of each cluster using Besan\c con Galaxy model\footnote{https://model.obs-besancon.fr/} \citep{Robin96, Robin03}. We took into account Galactic coordinates ($l$, $b$) of eight clusters and assumed the size of the field as 0.111 square degrees, and the magnitude range as $8<V<24$. The frequency distribution constructed according to synthetic $V$ magnitude intervals are shown with observational $V$ magnitude distribution in Fig. 2. The black solid lines represent the observational values, while red lines denote the synthetic stars from Besan\c con Galaxy model. In the Fig. 2 the $V$ magnitude intervals, in which the number of stars detected from observations are higher than the number of model stars, indicate the dominant magnitude range of the clusters. The $V$ magnitude where the number of model stars starts to exceed the number of observed ones correspond to the completeness limit. Comparing magnitude distributions based on both observational and the star count method, the photometric completeness limit was determined for each cluster region. We obtained the photometric $V$-band completeness limits as $V=19$ mag for Collinder 421 and NGC 7086, $V=20$ mag for ASCC 115, NGC 6793, NGC 7031, NGC 7039, Roslund 1 and Stock 21. It can be clearly seen from the distribution that the stars located through the cluster regions are within the determined photometric completeness limits. For further analyses, we took into account stars brighter than the completeness limit of each cluster to offer reliable results in the determination of parameters.

\subsection{Cluster radius and radial stellar surface density}

In order to investigate structural parameters of the clusters we analysed the radial density profiles (RDP). Since ASCC 115, Collinder 421, NGC 6793, NGC 7031, NGC 7039 and NGC 7086 show a distinct central concentration, we could derive RDP only for these open clusters. As can be seen from Figure 1, Roslund 1 and Stock 21 do not show a central concentration, so structural parameters could not be obtained. We note that this fact raises questions regarding their classification as an open cluster. We adopted central coordinates of the clusters as given in SIMBAD database\footnote{http://simbad.u-strasbg.fr/simbad/sim-fid}, and we divided the area into concentric circles, each has a radius 1 arcmin larger than the previous one, and calculated number density of stars that are located in the circles. Then we fitted the RDPs with empirical \citet{King62} formula and determined background stellar density ($f_{bg}$), central stellar density ($f_0$) and core radius ($r_c$) of the clusters, whose results are given in Table 5. In Fig. 3 we present the RPDs with the best fit model shown with red solid lines. Results show that the core radii are generally small and the field star contamination is large. As expected, the contamination increases at fainter magnitudes for each cluster.

We also compared background density value determined from the King model fitting \citep{King62} with those obtained from the Besan\c con Galaxy model \citep{Robin96}. According to Besan\c con Galaxy model, there are $6.607\pm0.018$, $1.245\pm0.043$, $3.444\pm0.026$, $4.517\pm0.022$, $1.583\pm0.038$ and $4.059\pm0.024$ field stars per arcmin$^{2}$ for ASCC 115, Collinder 421, NGC 6793, NGC 7031, NGC 7039 and NGC 7086, respectively. These model background values are represented with dashed blue horizontal lines in Fig. 3. The observational and model background densities are generally in good agreement. The difference between the observed and model background densities are between 0.02 to 0.95 stars arcmin$^{-2}$, while maximum difference occurs in NGC 7086. The discrepancy arises from the Galactic model parameters, since several parameters with fixed values are used in Galaxy models. In addition, Galactic model parameters are a function of Galactic coordinates and absolute magnitude \citep{Karaali04, Bilir06a, Bilir06b, Bilir08, Bilir10}. 

\begin{figure*}
\centering
\includegraphics[scale=.43, angle=0]{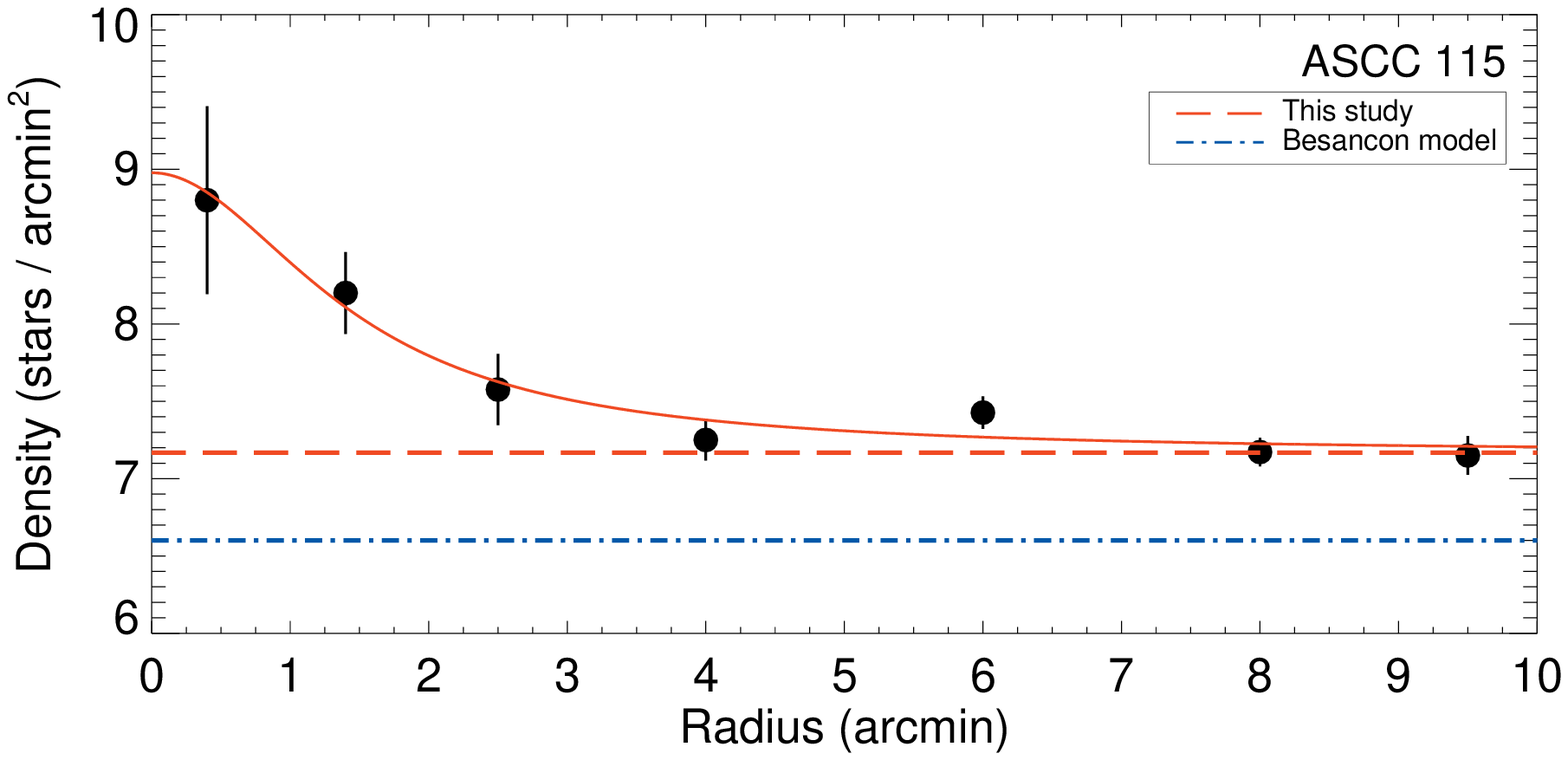}
\includegraphics[scale=.43, angle=0]{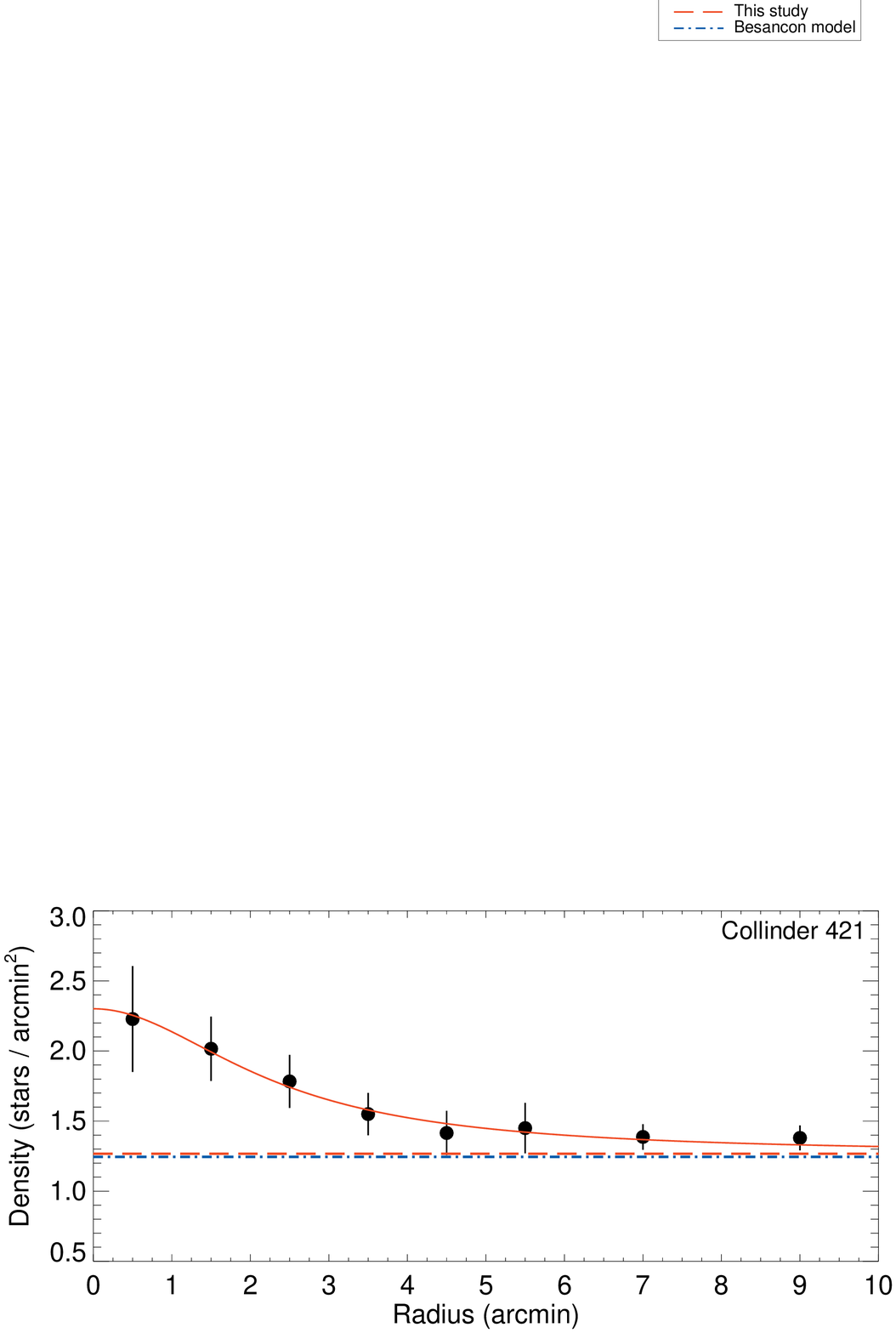}\\
\includegraphics[scale=.43, angle=0]{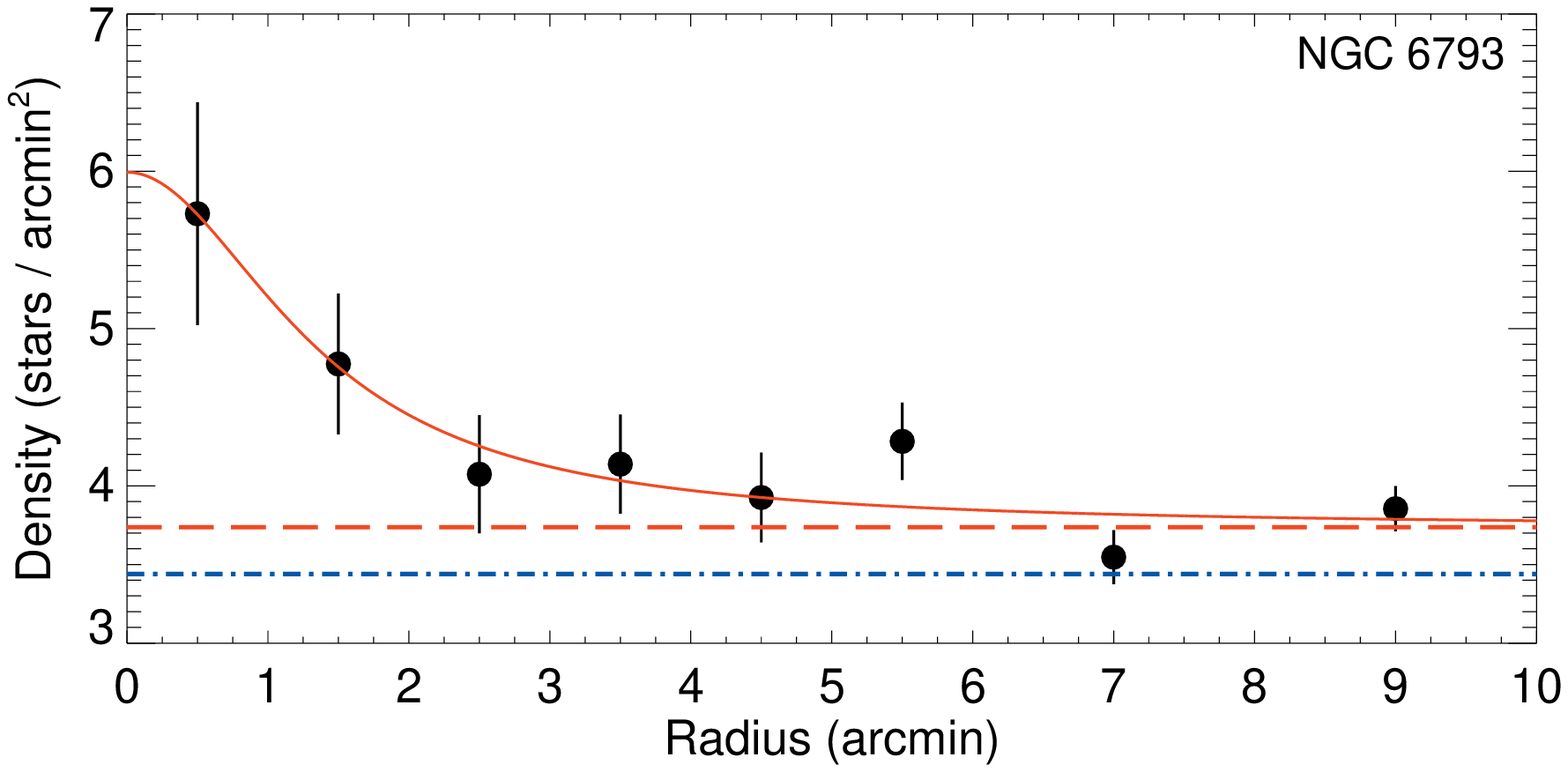}
\includegraphics[scale=.43, angle=0]{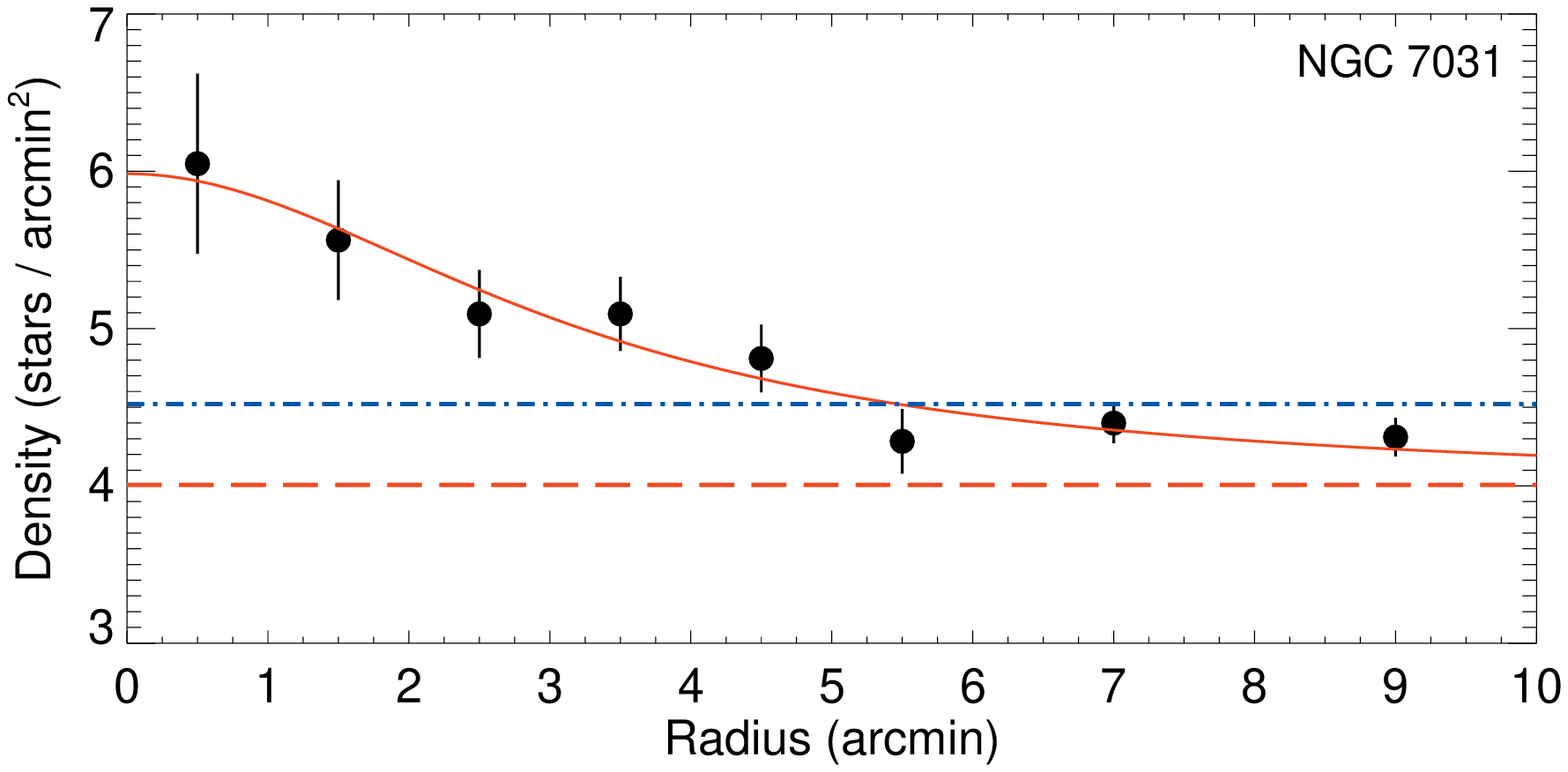}\\
\includegraphics[scale=.43, angle=0]{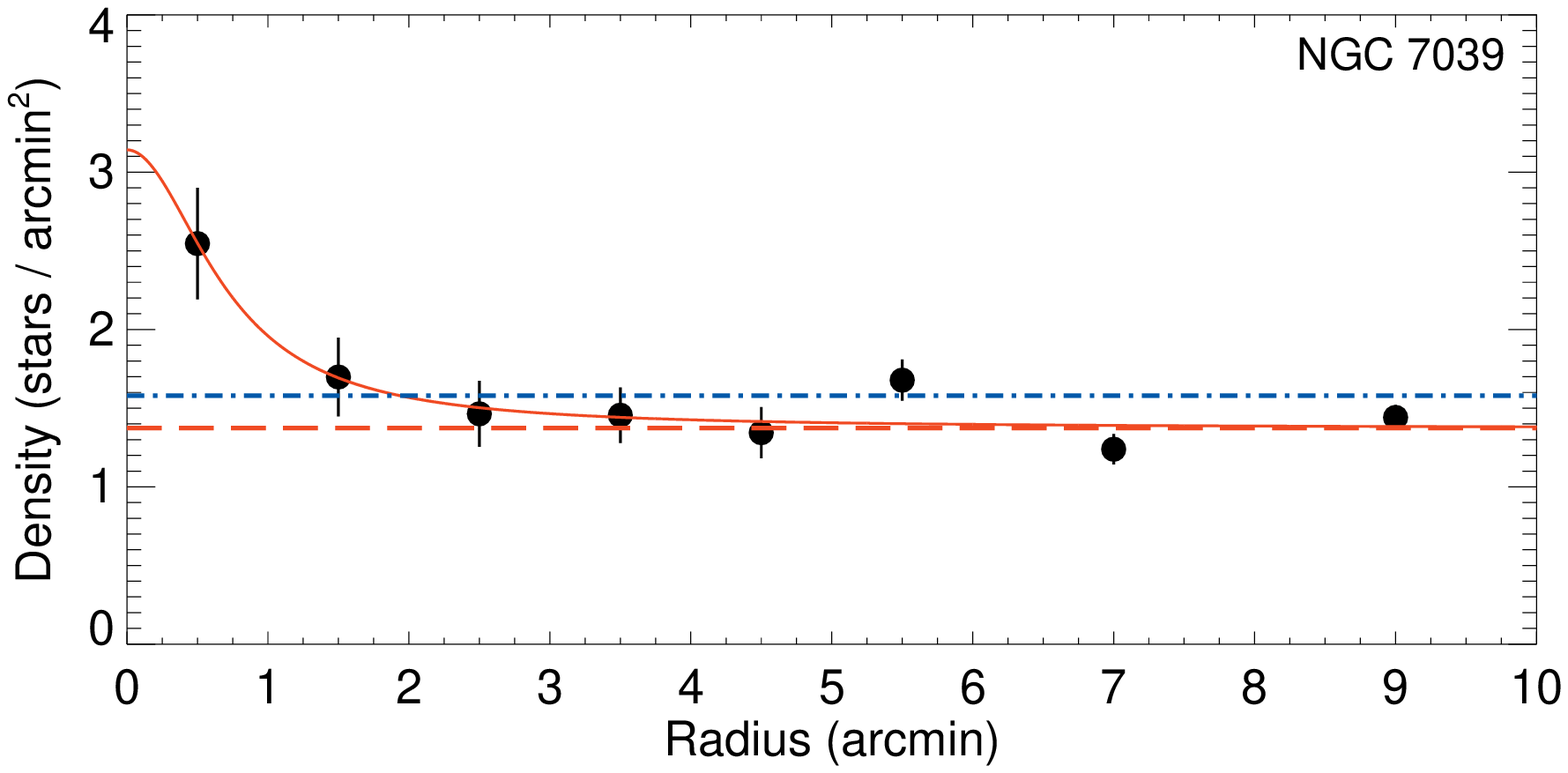}
\includegraphics[scale=.43, angle=0]{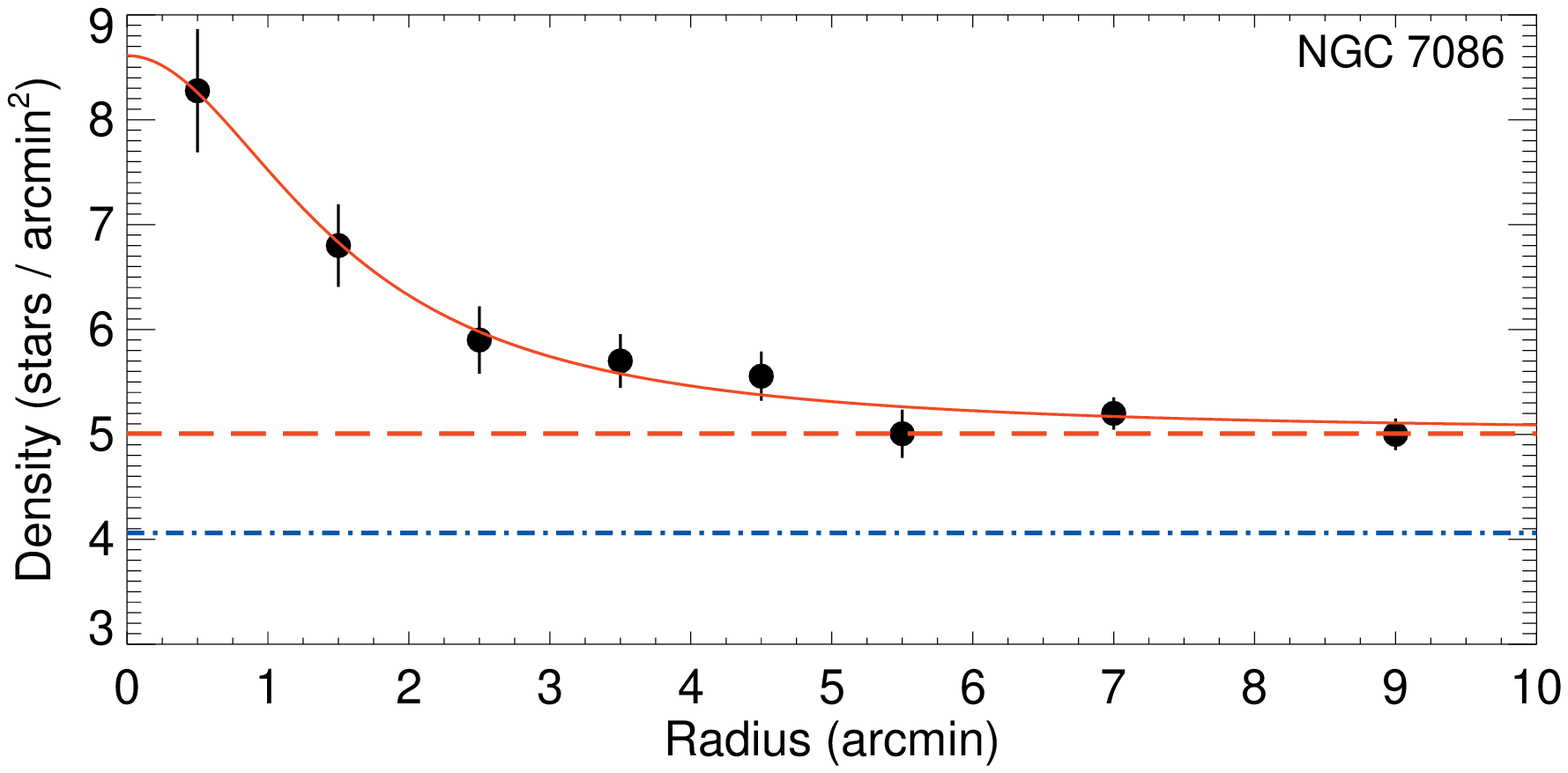}\\
\caption{Stellar density profiles of six open clusters. Errors were determined from $1/\sqrt{N}$, where $N$ indicates the number of stars used in the density
estimation. Red dashed and blue dotted-dash lines represent background stellar density for observational and synthetic data, respectively.} 
\end {figure*}

\begin{table*}
\setlength{\tabcolsep}{5pt}
  \centering
  \caption{The structural parameters of the six open clusters estimated by fitting \citet{King62} model. Here, $f_0$, $r_c$, and $f_{bg}$ parameters represent the central stellar density, the core radius, and the background stellar densities observational and synthetic values are given in fourth and fifth columns, respectively.}
    \begin{tabular}{lcccc}
\hline
    Cluster & $f_0$ & $r_c$ & $f_{bg}$ (Observational) & $f_{bg}$ (Besan\c con) \\
            & (stars arcmin$^{-2}$) & (arcmin) & (stars arcmin$^{-2}$) & (stars arcmin$^{-2}$)  \\
\hline
    ASCC 115 & 1.810$\pm$0.259 & 1.456$\pm$0.459 & 7.168$\pm$0.139 & 6.607$\pm$0.018\\
Collinder 421& 1.035$\pm$0.282 & 2.305$\pm$2.166 & 1.267$\pm$0.338 & 1.245$\pm$0.043\\
    NGC 6793 & 2.256$\pm$0.292 & 1.360$\pm$0.357 & 3.737$\pm$0.170 & 3.444$\pm$0.026\\
    NGC 7031 & 1.978$\pm$0.059 & 3.241$\pm$1.816 & 4.006$\pm$0.443 & 4.517$\pm$0.022\\
    NGC 7039 & 1.769$\pm$0.984 & 0.704$\pm$0.573 & 1.373$\pm$0.194 & 1.583$\pm$0.038\\
    NGC 7086 & 3.602$\pm$0.280 & 1.517$\pm$0.255 & 5.009$\pm$0.171 & 4.059$\pm$0.024\\
\hline
    \end{tabular}%
\end{table*}%

\subsection{CMDs and member stars of the Clusters}
Radial velocities, distances and proper motions can be used alone or combined with each other to determine cluster member stars precisely. Today, cluster members can be identified with large spectroscopic surveys designed for different purposes \citep{Allende-Prieto10, Gilmore12, daSilva16}. Spectroscopic measurements need telescope-time-focused approach and this restricts to understand features of many clusters. In addition, there is no planned spectroscopic survey to measure radial velocities of all stars of the Milky Way. Fortunately, as cluster stars have the same spatial origin, there are various methods to separate cluster members from field stars \citep[e.g.][]{Krone-Martins14,Javakhishvili06,Balaguer98}. These methods generally take into account proper motions of the stars. 

In this study, to determine membership probabilities of the stars, we used {\em UBV} colours and {\it Gaia} DR2 astrometric data, and we utilized Unsupervised Photometric Membership Assignment in Stellar Clusters \citep[UPMASK,][]{Krone-Martins14} method, which gives effective results for clusters located up to distances 2-2.5 kpc from the Sun. UPMASK is a data-driven unsupervised method that depends on minimal physical assumptions about stellar cluster regions. The main idea of this method is classify small groups of stars in various photometric bands using a simple clustering algorithm like a k-means clustering and control these small groups individually to obtain tightly concentrated spatially distributed groups. The basic assumption is that cluster members are clustered and share some similar parameter space properties such as proper motion and parallax, while field stars distributed randomly. According to this assumption, using astrometry gives more reliable results than photometry because of cluster members have the same distance and moving directions. Moreover, method does not assume any probability distributions, such as density profile or isochrones modelling, except spatial uniformity of field stars. 

We applied UPMASK method to the five dimensional astrometric space of equatorial coordinates, proper motions and parallaxes of stars ($\alpha$, $\delta$, $\mu_{\alpha}\cos\delta$, $\mu_{\delta}$, $\pi$) and three-dimensional photometric space ($V$, $U-B$, $B-V$). We set the k-means number between 7 to 12 since region of the eight clusters contain different number of stars. In order to get reliable result, we reach k-means number iterated 1000 times. \citet{Cantat-Gaudin18} successfully applied UPMASK method for 1229 open cluster regions and determined membership probabilities of stars located through each cluster using k-means values between 10 to 25, because of their cluster samples are more concentrated with stars. We calculated the membership probabilities ($P$) of each cluster and showed the histograms of membership probabilities and vector point diagrams (VPDs), as shown in \citet{Gao18}, in Figs. 4 and 5, respectively. We adopted stars with $P\geq50\%$ as the most likely cluster members for further analyses. We showed the stars with $P\geq50\%$ on the $V\times B-V$ colour-magnitude diagrams in Fig. 6 and used them during the analyses. Moreover, we calculated their distances according to their trigonometric parallaxes. Their mean distances and proper motions are listed in Table 1.

\begin{figure*}
\centering
\includegraphics[scale=.45, angle=0]{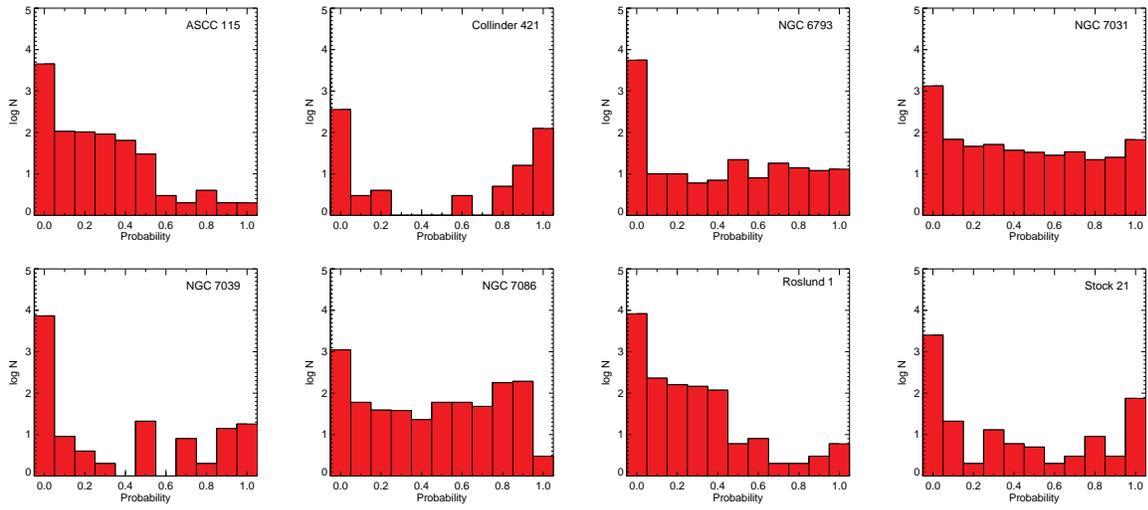}
\caption{Histograms of the membership probabilities for open clusters.} 
\end{figure*}

\begin{figure*}
\centering
\includegraphics[scale=.55, angle=0]{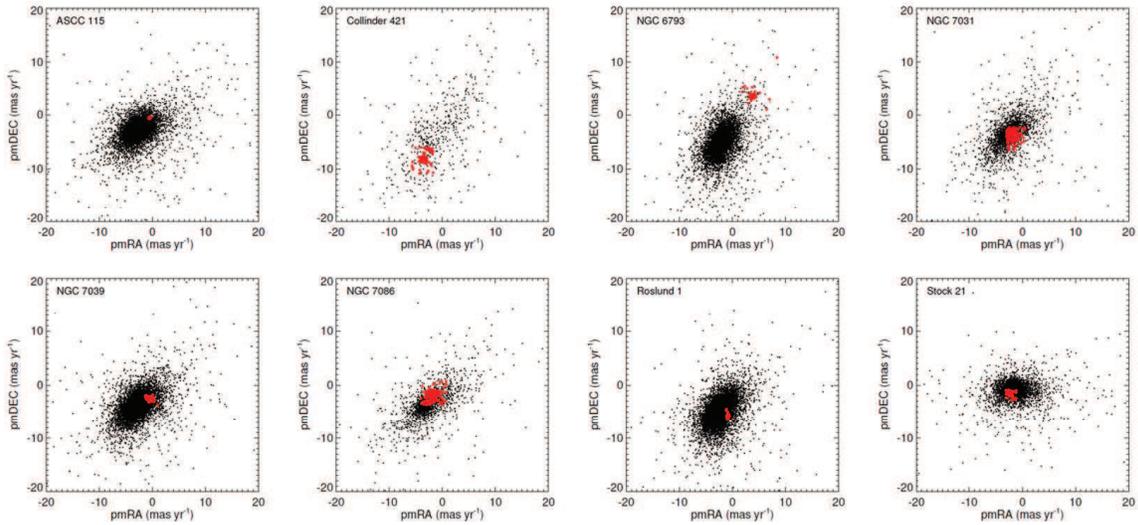}
\caption{Proper motion vector point diagrams (VPDs) of the clusters. The red and black dots indicate the high-probability ($P\geq50\%$) and low-probability $P<50\%$) member stars, respectively.} 
\end{figure*}

We determined 13, 149, 87, 209, 63, 543, 27 and 97 likely members with $P\geq50\%$ for ASCC 115, Collinder 421, NGC 6793, NGC 7031, NGC 7039, NGC 7086, Roslund 1 and Stock 21, respectively. It is noteworthy that 13 member stars for ASCC 115 have been identified with UPMASK method even the RDP of this cluster shows significant central concentration. This is because, we only consider members with $P\geq50\%$ for a uniform approach in the study. If this probability limit value was assumed to be lower (such as 40\%) for the ASCC 115, more than 100 stars could have been used in the study.

\begin{figure*}
\centering
\includegraphics[scale=.20, angle=0]{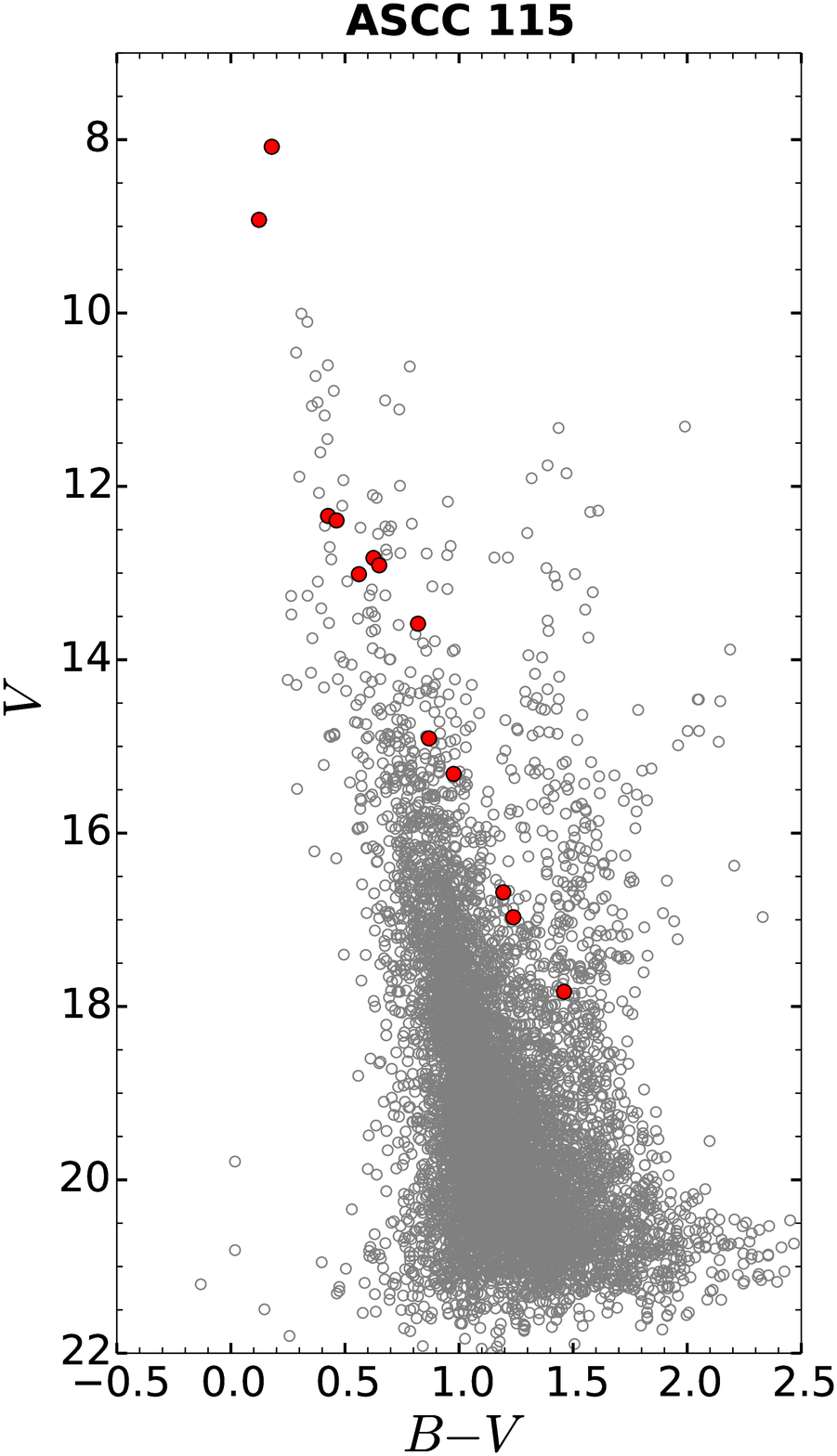}
\includegraphics[scale=.20, angle=0]{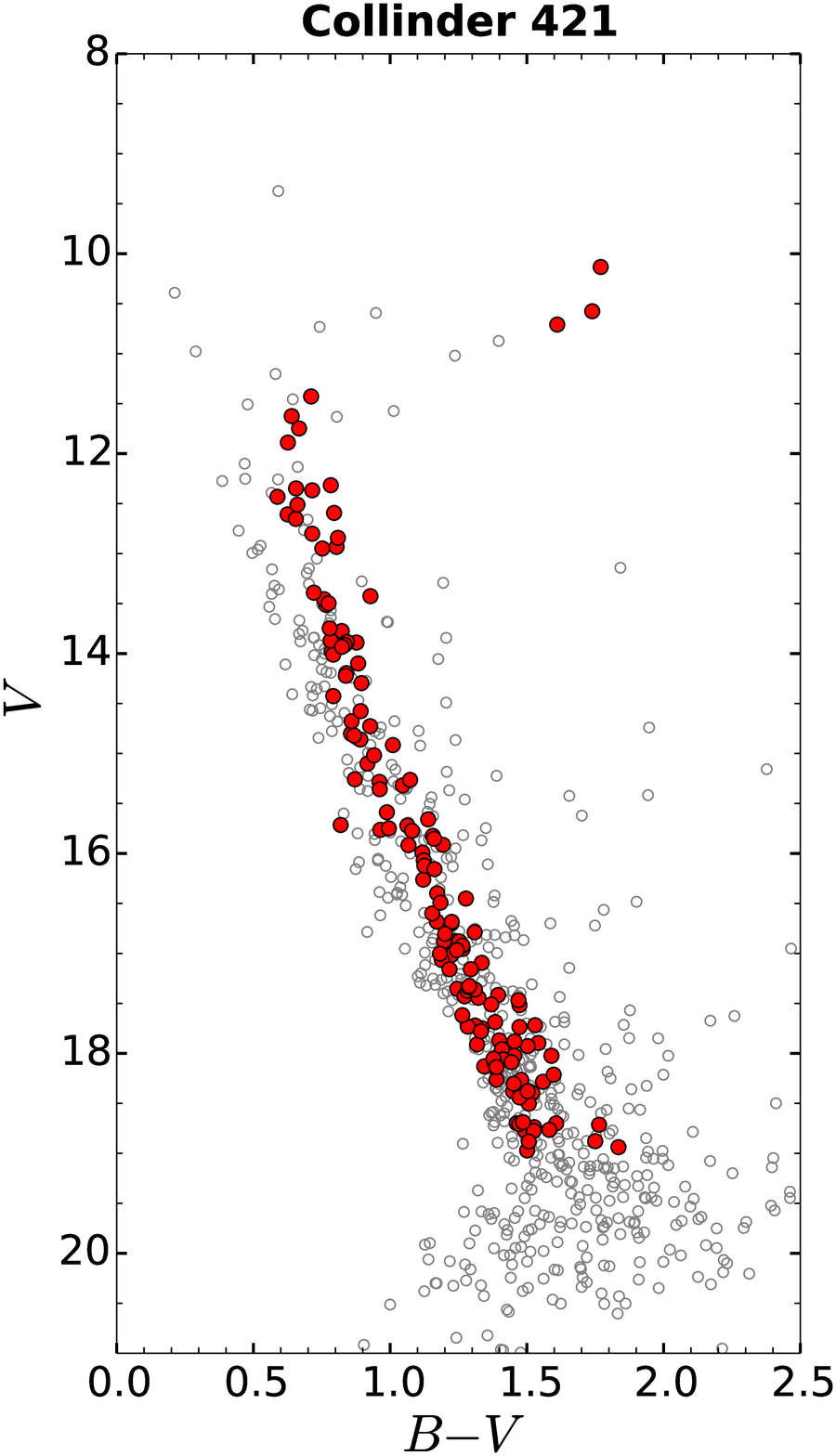}
\includegraphics[scale=.20, angle=0]{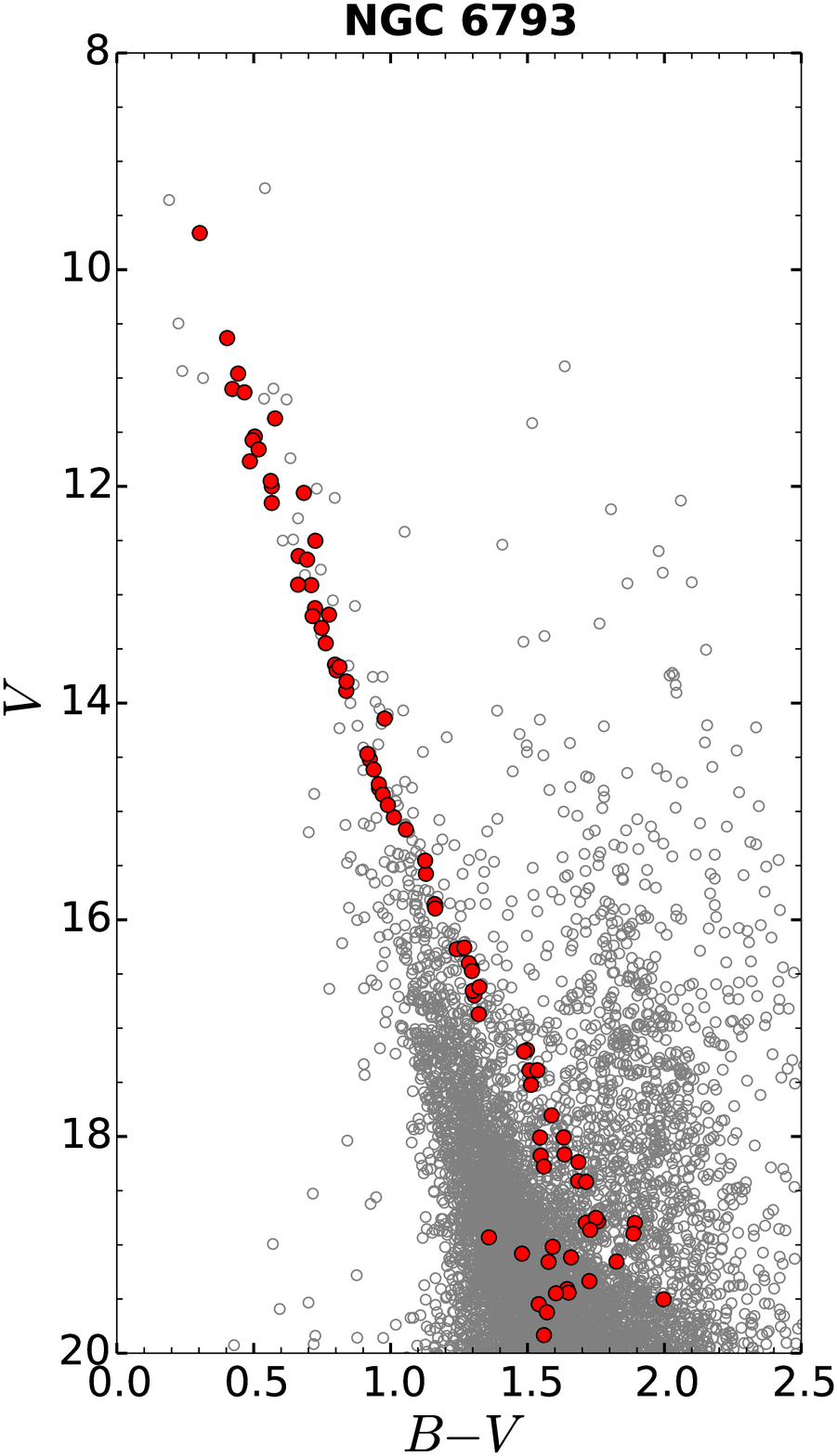}
\includegraphics[scale=.20, angle=0]{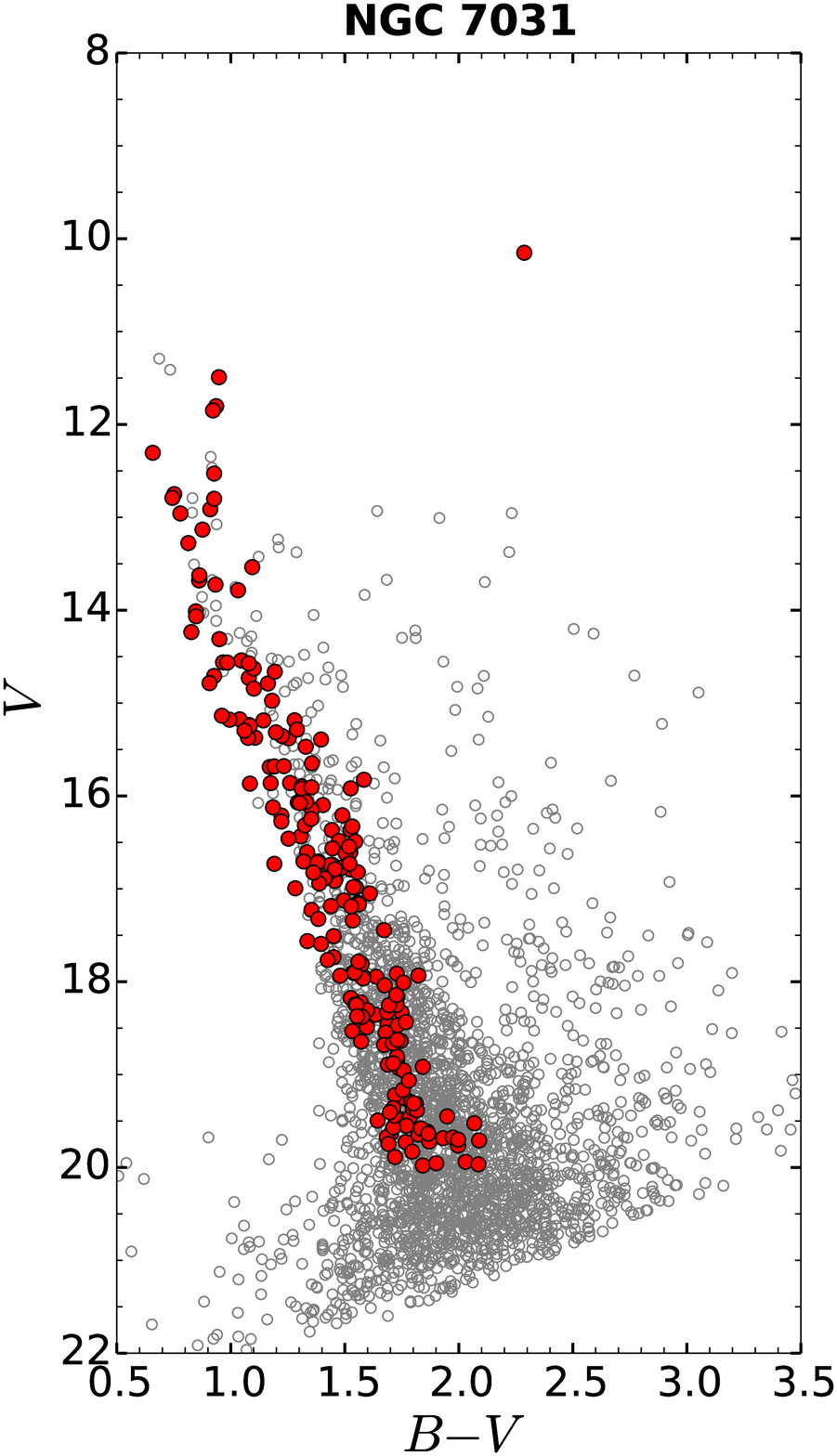}\\
\includegraphics[scale=.20, angle=0]{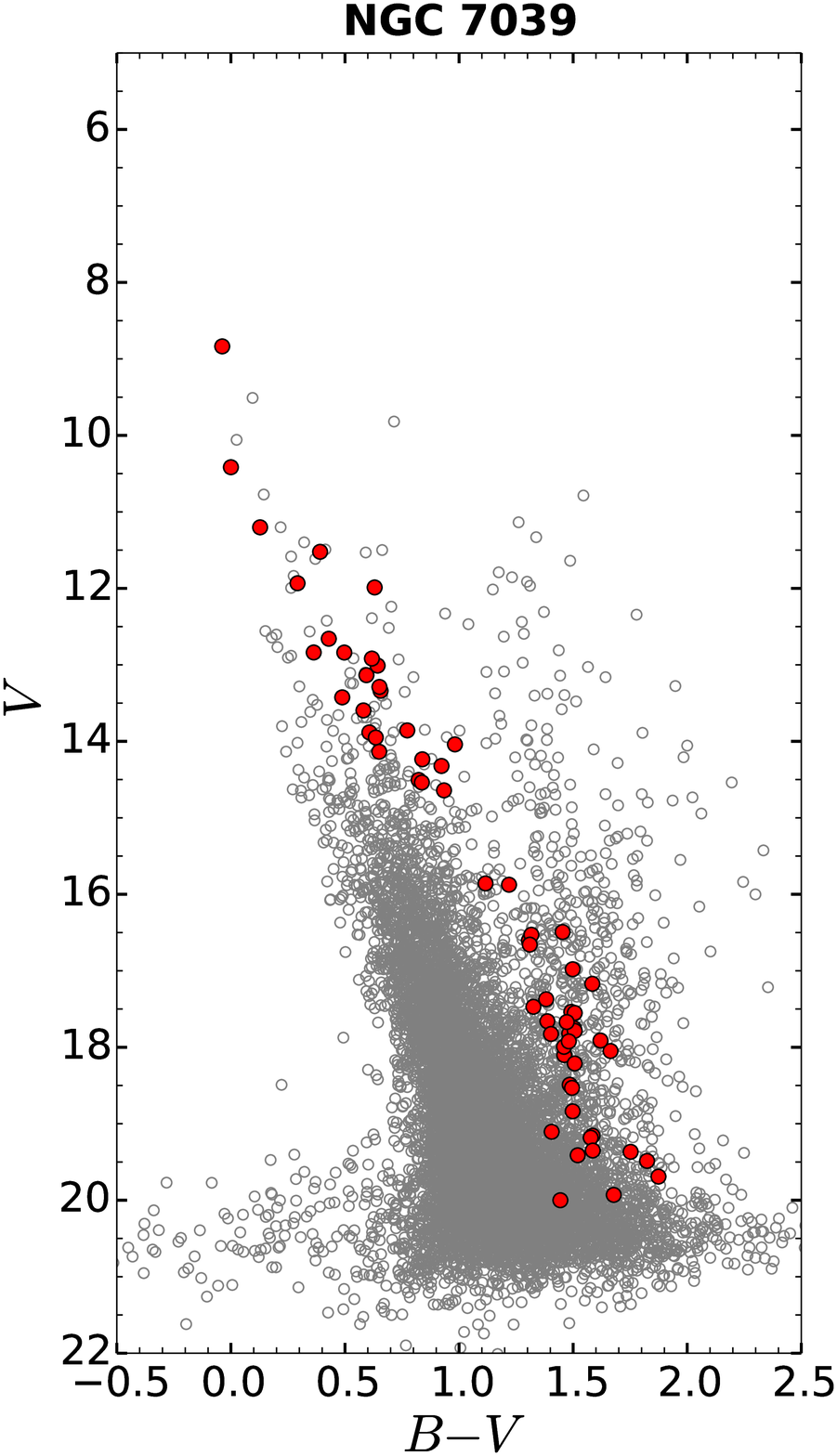}
\includegraphics[scale=.20, angle=0]{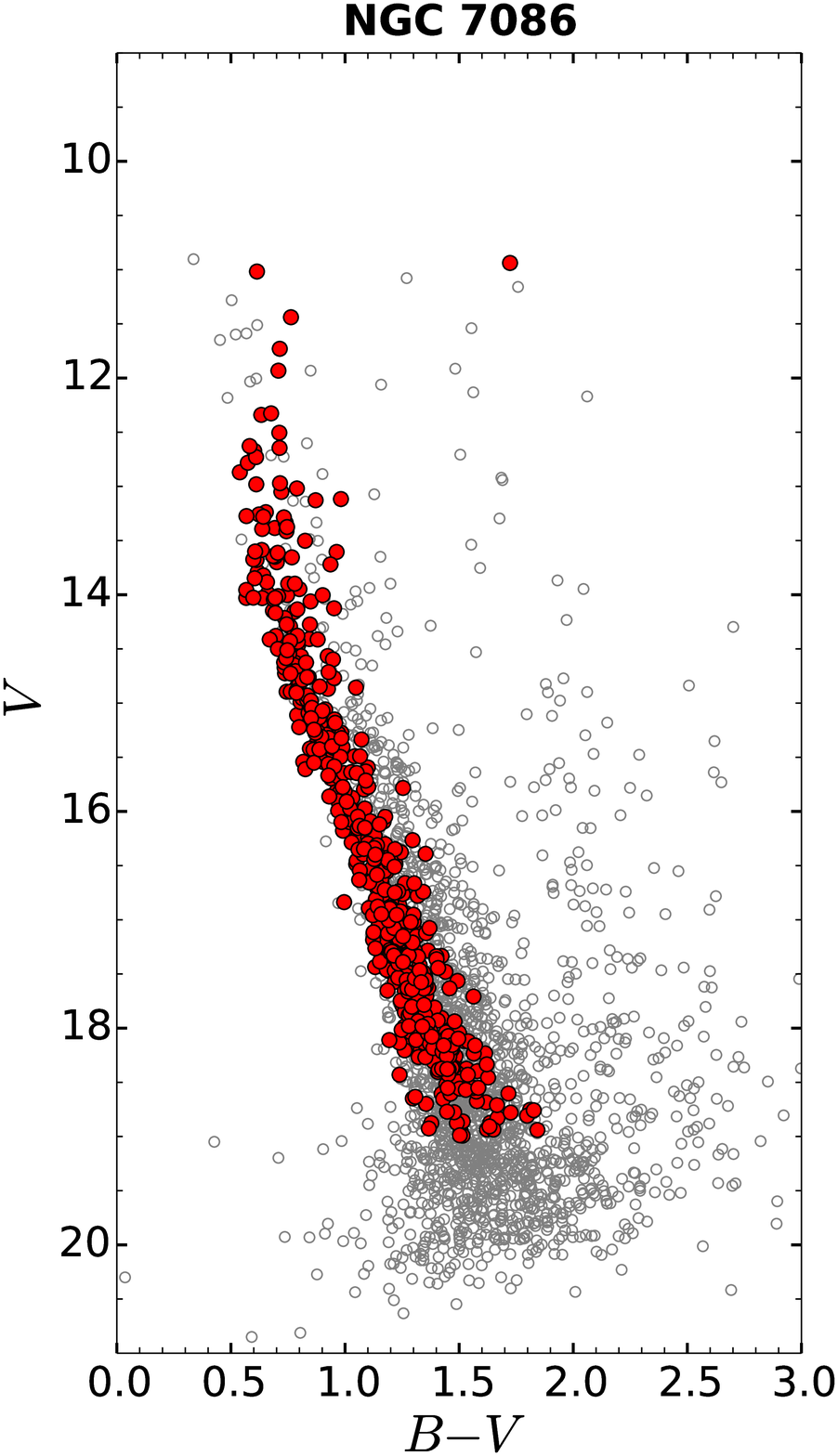}
\includegraphics[scale=.20, angle=0]{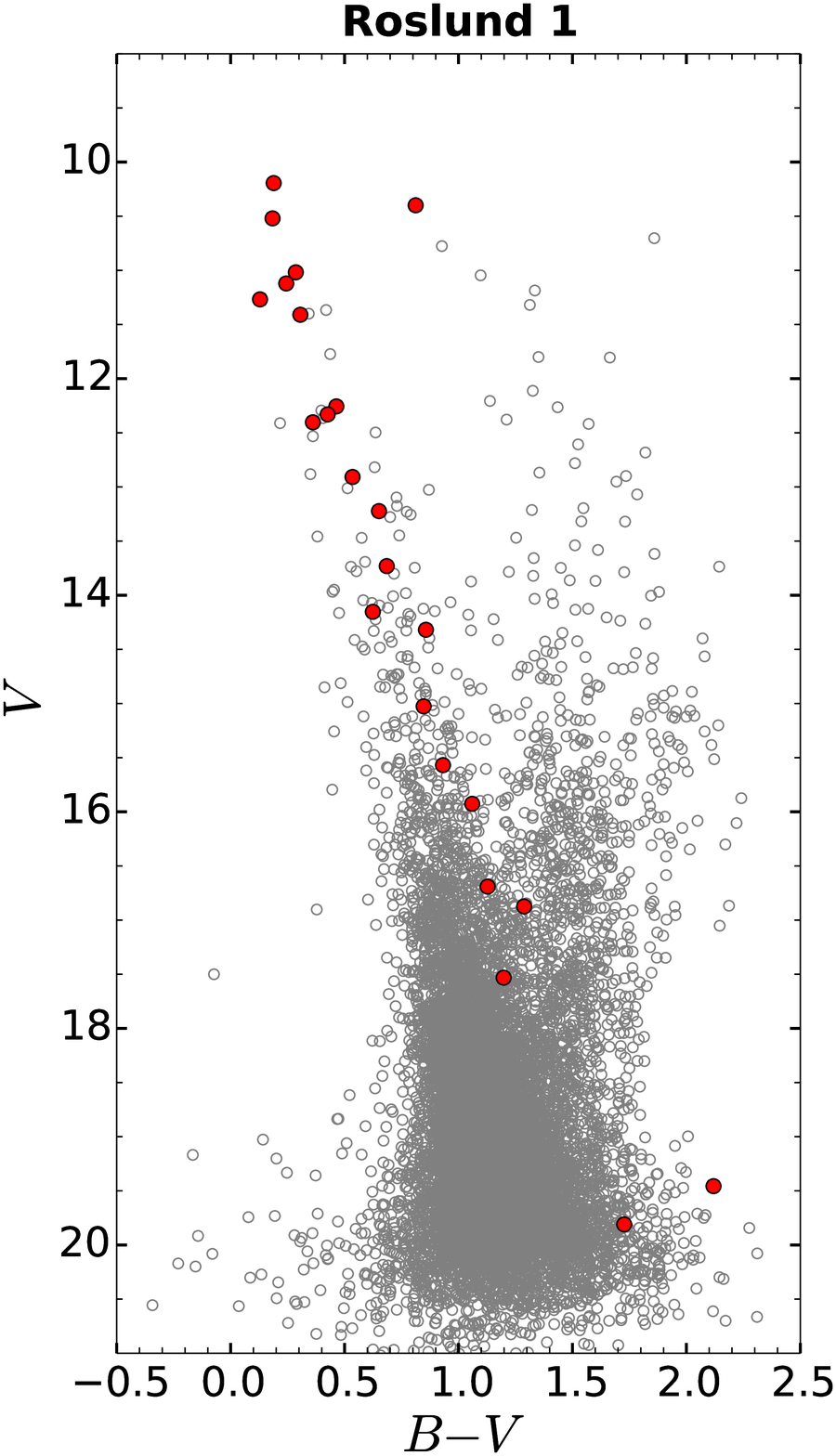}
\includegraphics[scale=.20, angle=0]{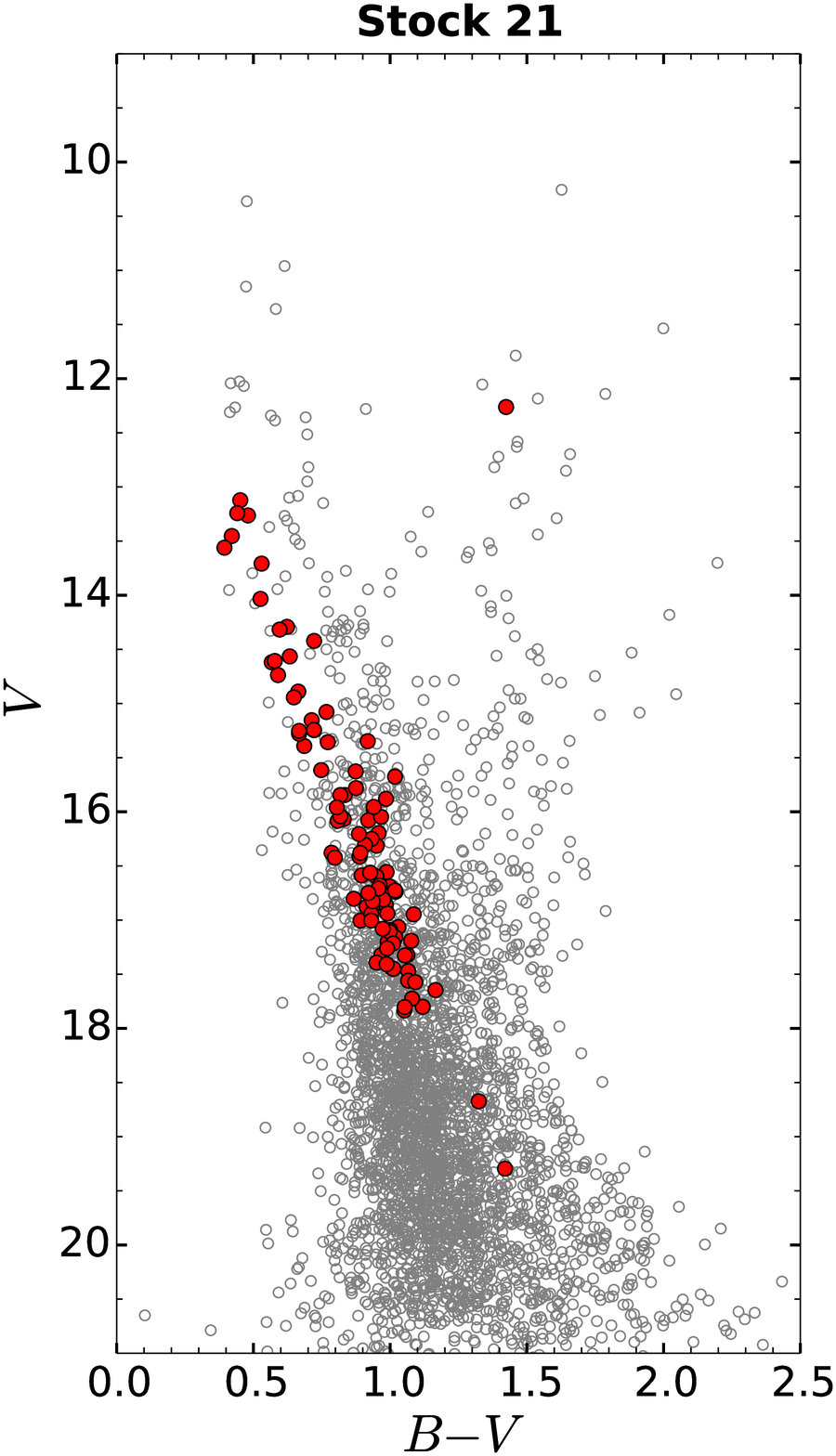}\\

\caption{$V\times B-V$ CMDs of the eight open clusters. Red dots indicate the most probable cluster stars ($P\geq50 \%$) while grey circles show the stars with $P<50 \%$ or field stars.} 
\end {figure*}

\section{Astrophysical parameters of the clusters}
In this section, we summarize the methods used to obtain astrophysical parameters of the eight open clusters \citep[for details see][]{Bilir16, Bostanci18}. We separately determined reddening, distance moduli and ages for each cluster via fitting models to the main-sequence stars that were already selected as members of clusters.  

\subsection{The reddening}
Since TCDs and CMDs are effected from interstellar reddening, we first derived $E(U-B$) and $E(B-V)$ colour excesses. In order to do this, TCDs of main-sequence stars with $P\geq50\%$ of eight clusters were compared with the solar metallicity ZAMS of \citet{Sung13}. The de-reddeneded ZAMS curve of \citet{Sung13} was compared with the positions of main-sequence stars of the clusters and shifted according to the equation of $E(U-B)=0.72\times E(B-V)+0.05\times E(B-V)^2$ \citep{Garcia88} with steps 0.01 mag within the range $0<E(B-V)\leq1$, and the most likely colour excesses of clusters were determined by minimum $\chi^2$ analyses. The final reddened TCDs with the best fits are shown in Fig. 7. We list the results in Table 1. The errors show $\pm 1\sigma$ statistical uncertainties.

\begin{figure}
\centering
\includegraphics[scale=.33, angle=0]{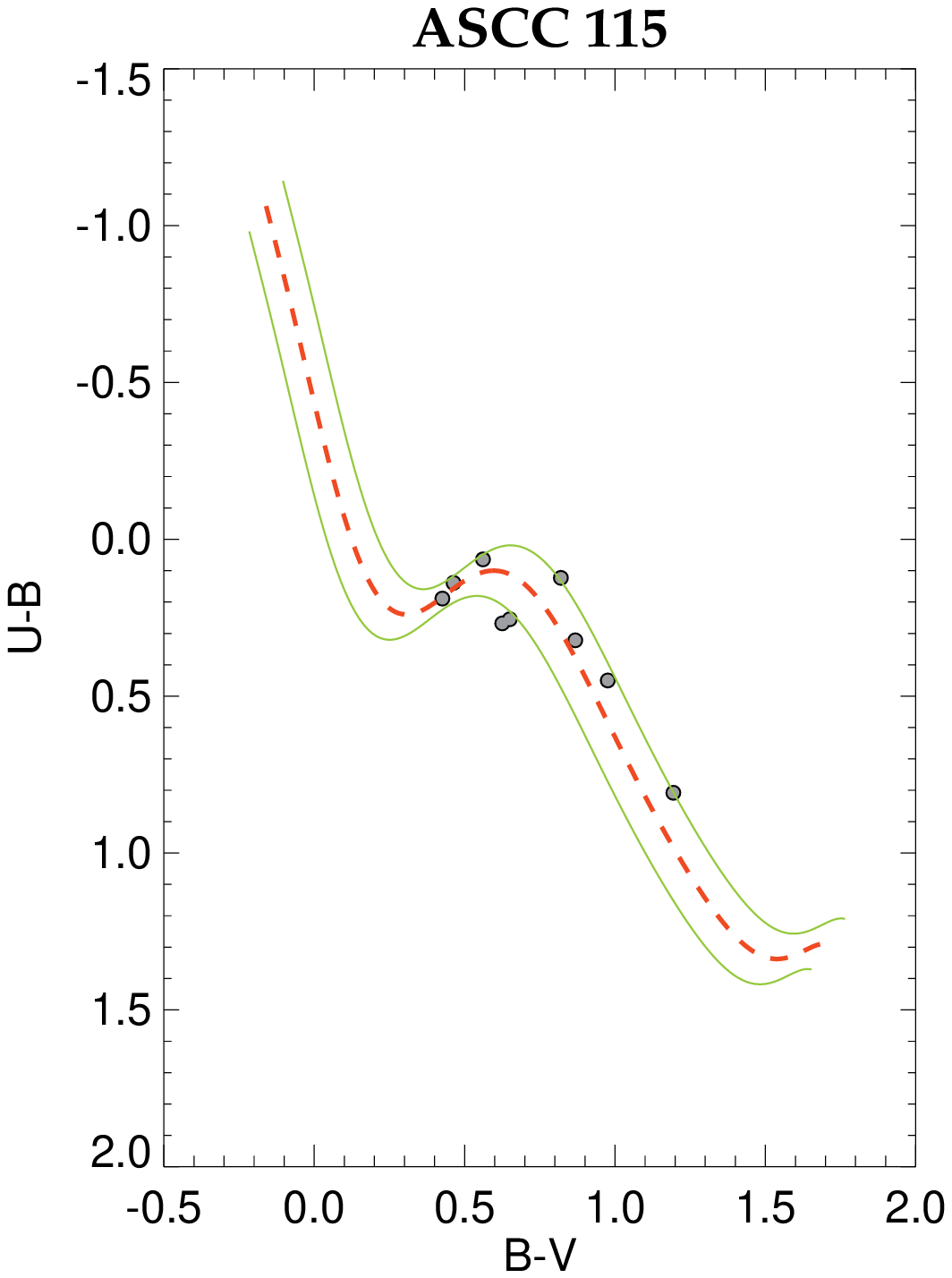}
\includegraphics[scale=.33, angle=0]{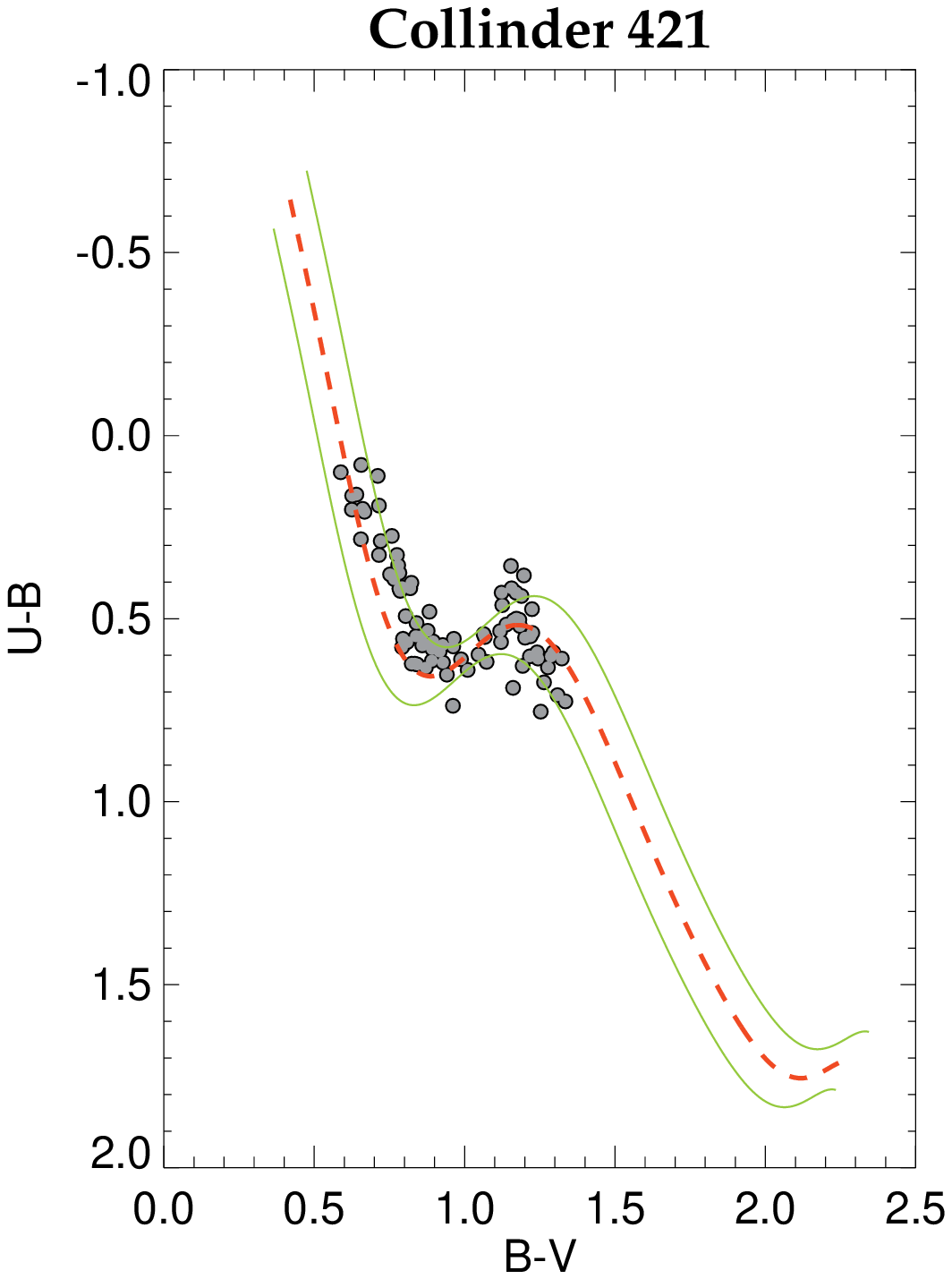}
\includegraphics[scale=.33, angle=0]{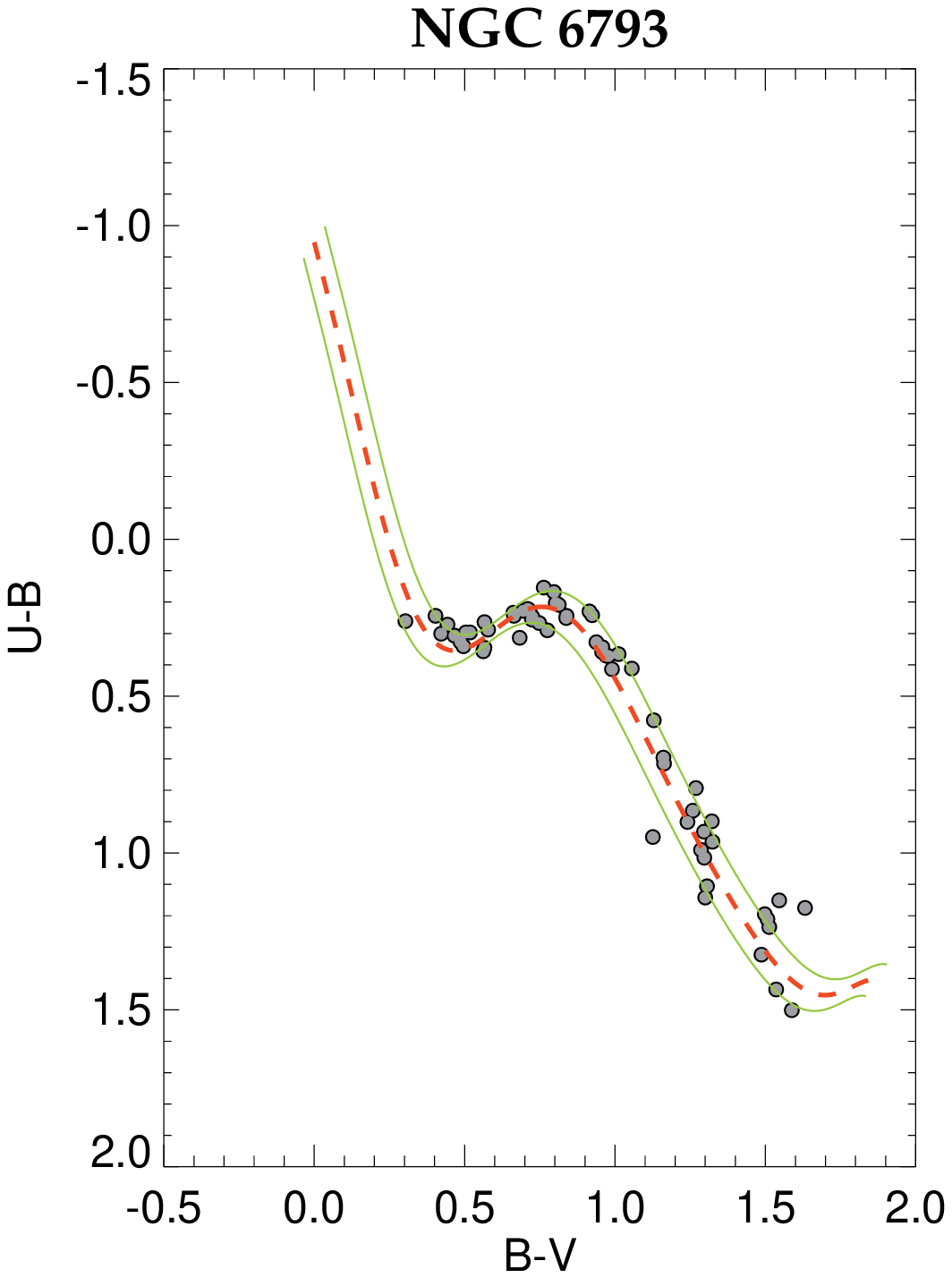}
\includegraphics[scale=.33, angle=0]{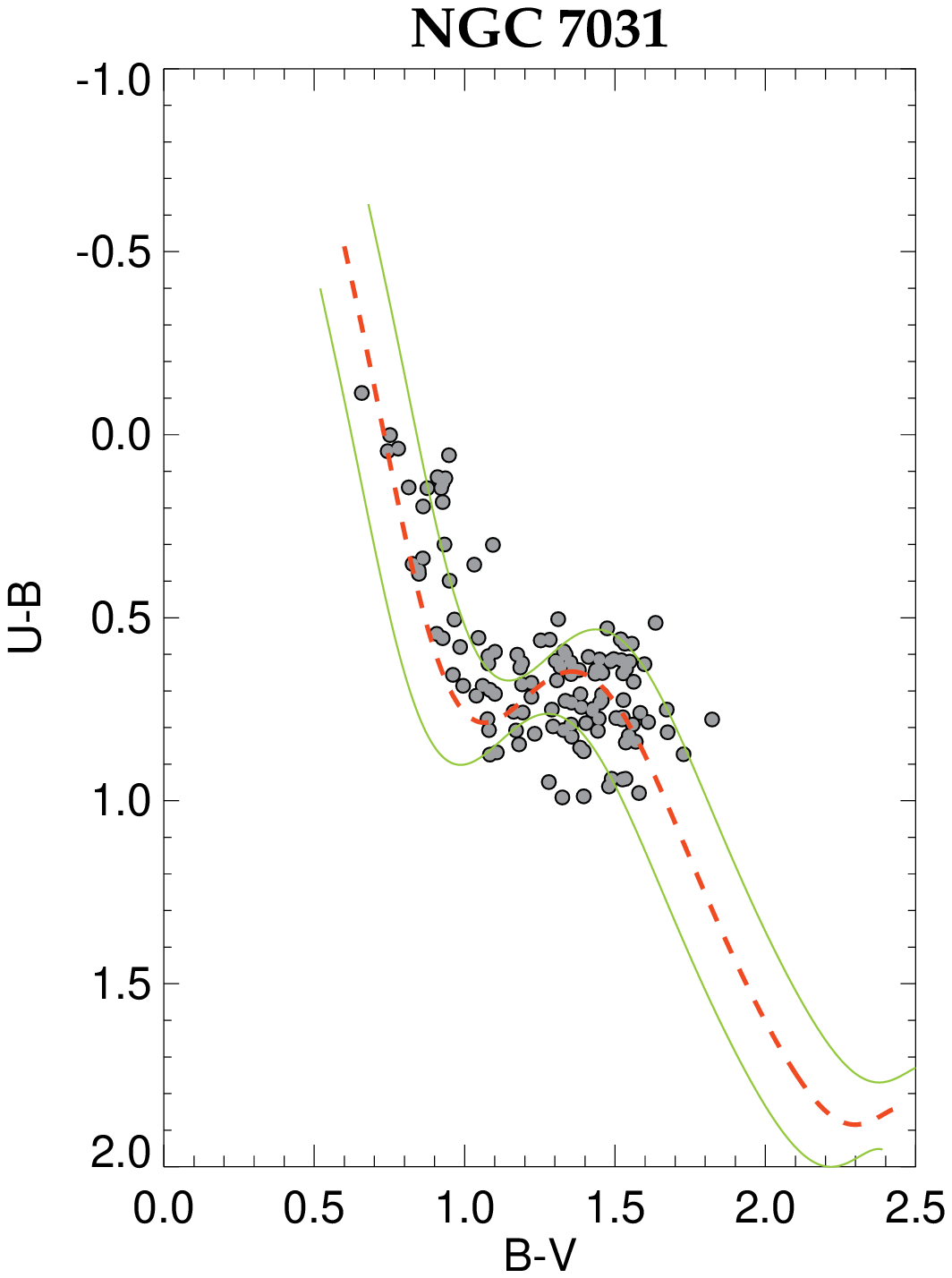}\\
\includegraphics[scale=.33, angle=0]{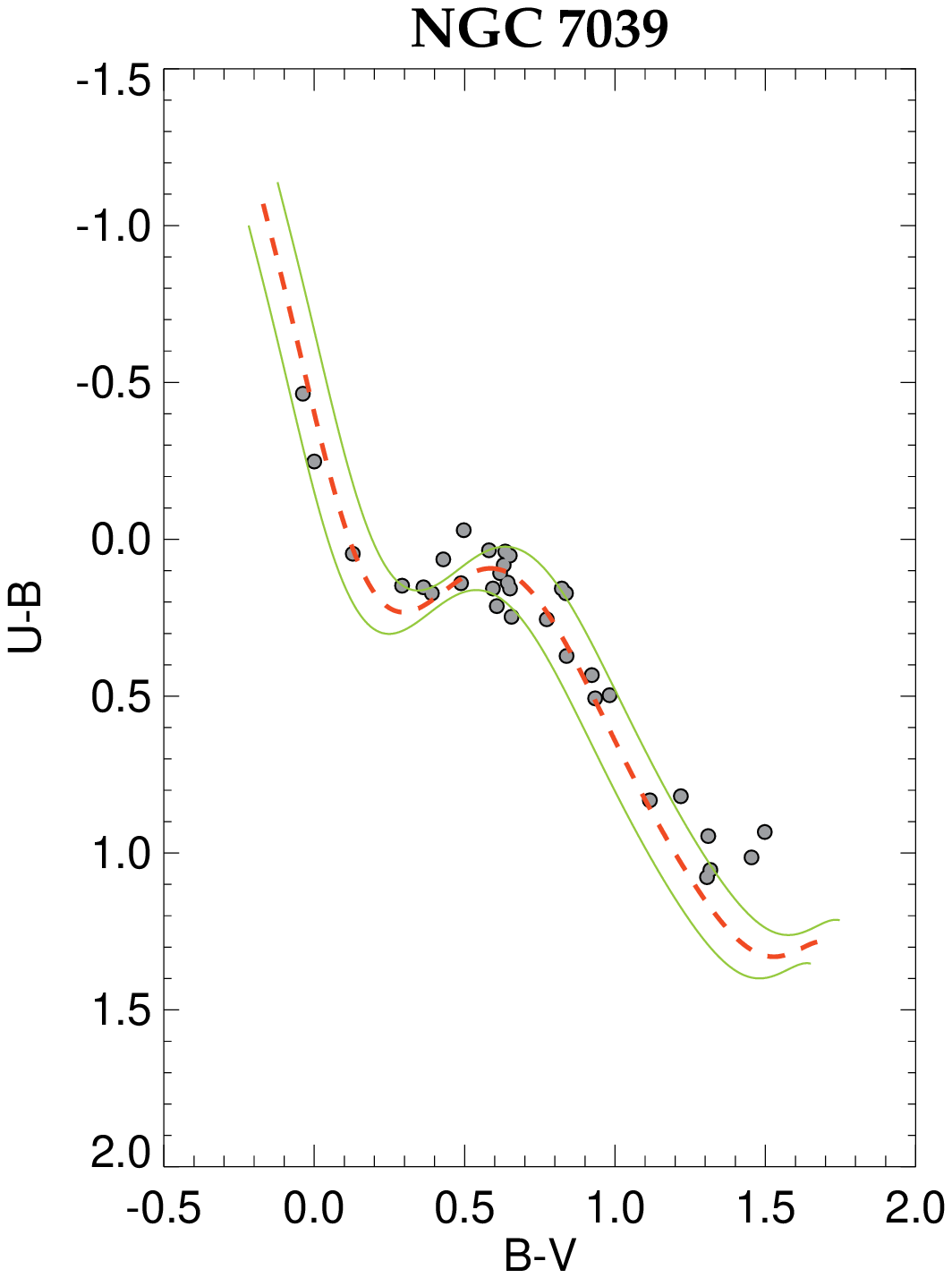}
\includegraphics[scale=.33, angle=0]{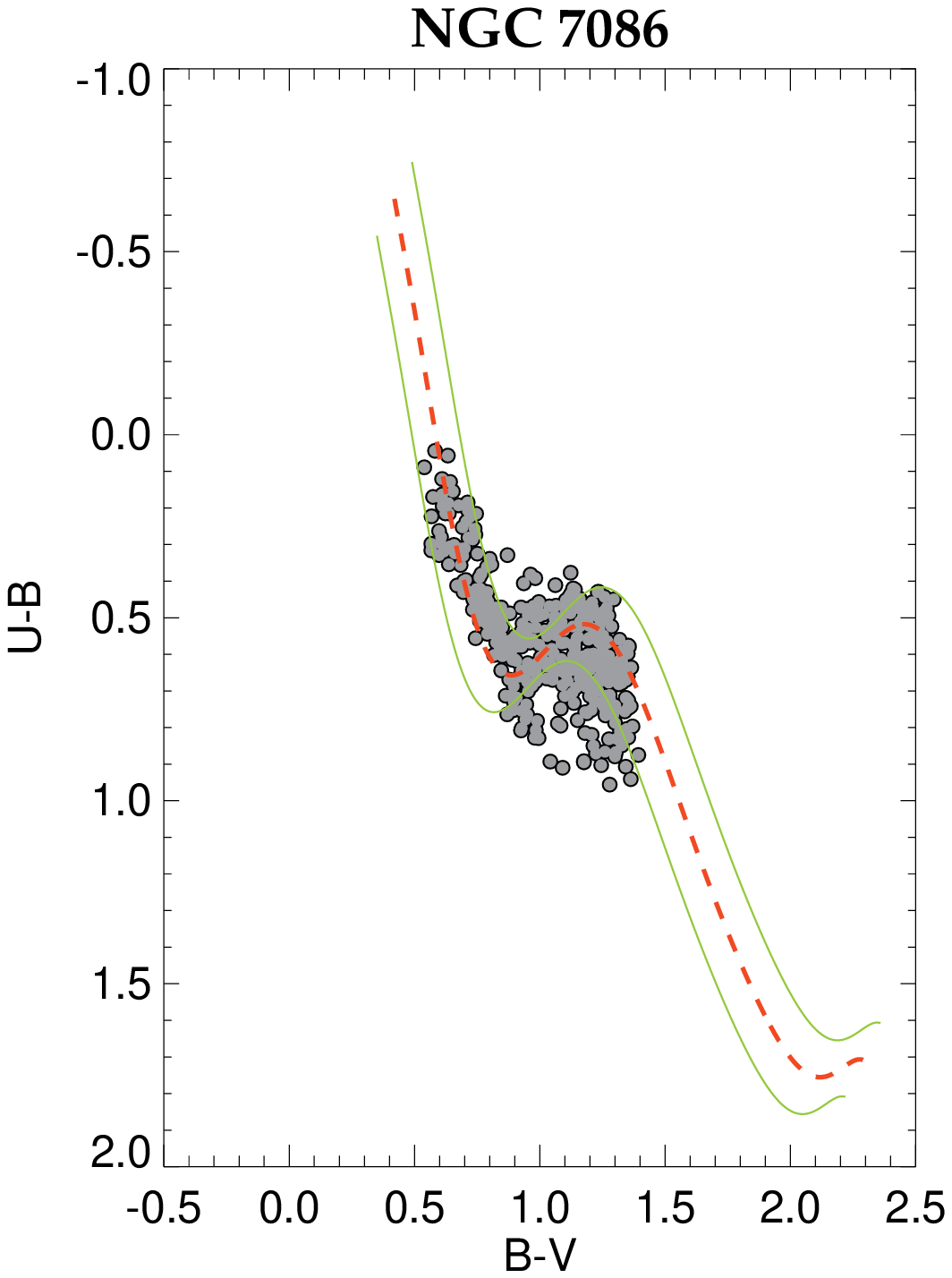}
\includegraphics[scale=.33, angle=0]{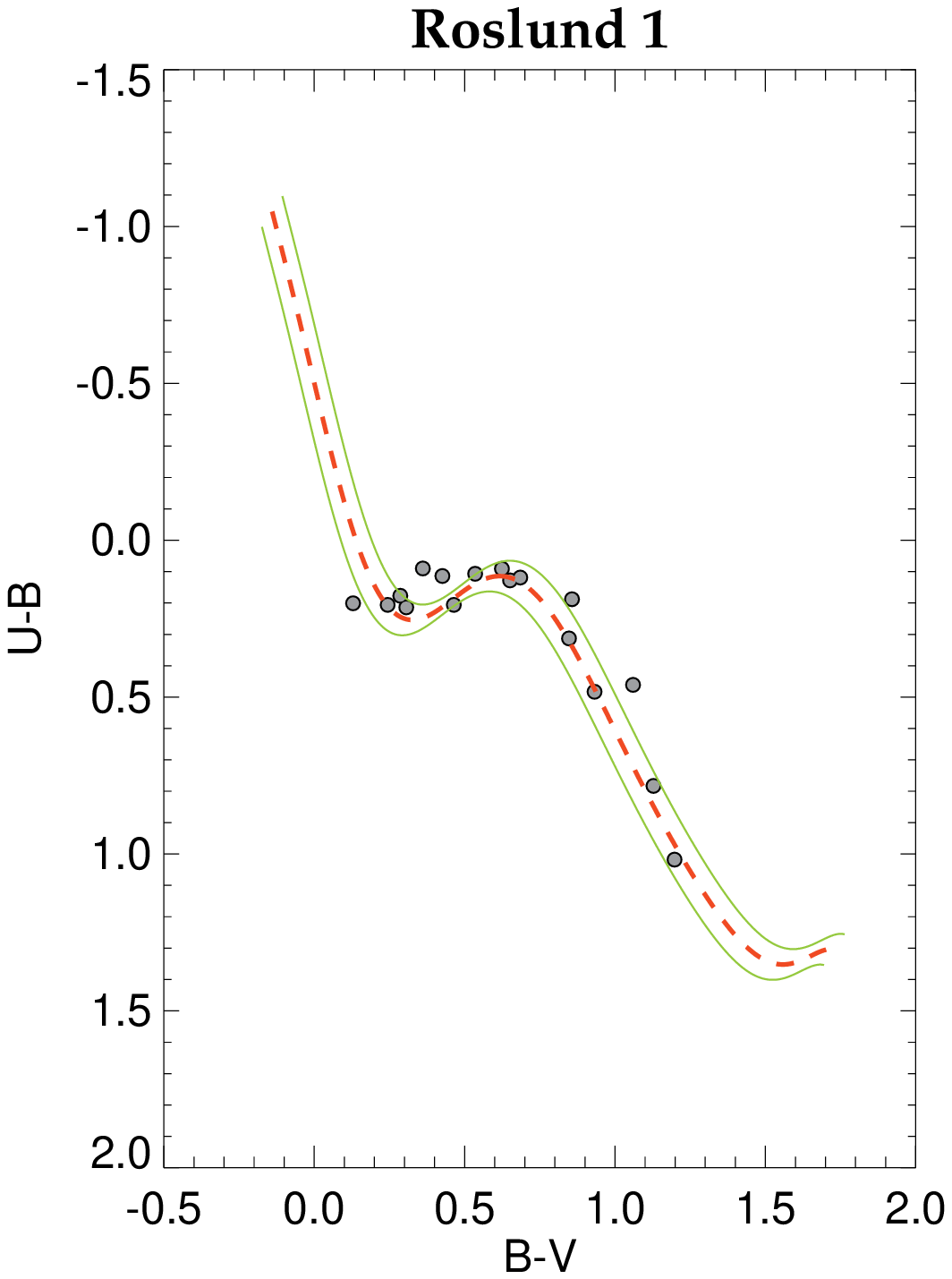}
\includegraphics[scale=.33, angle=0]{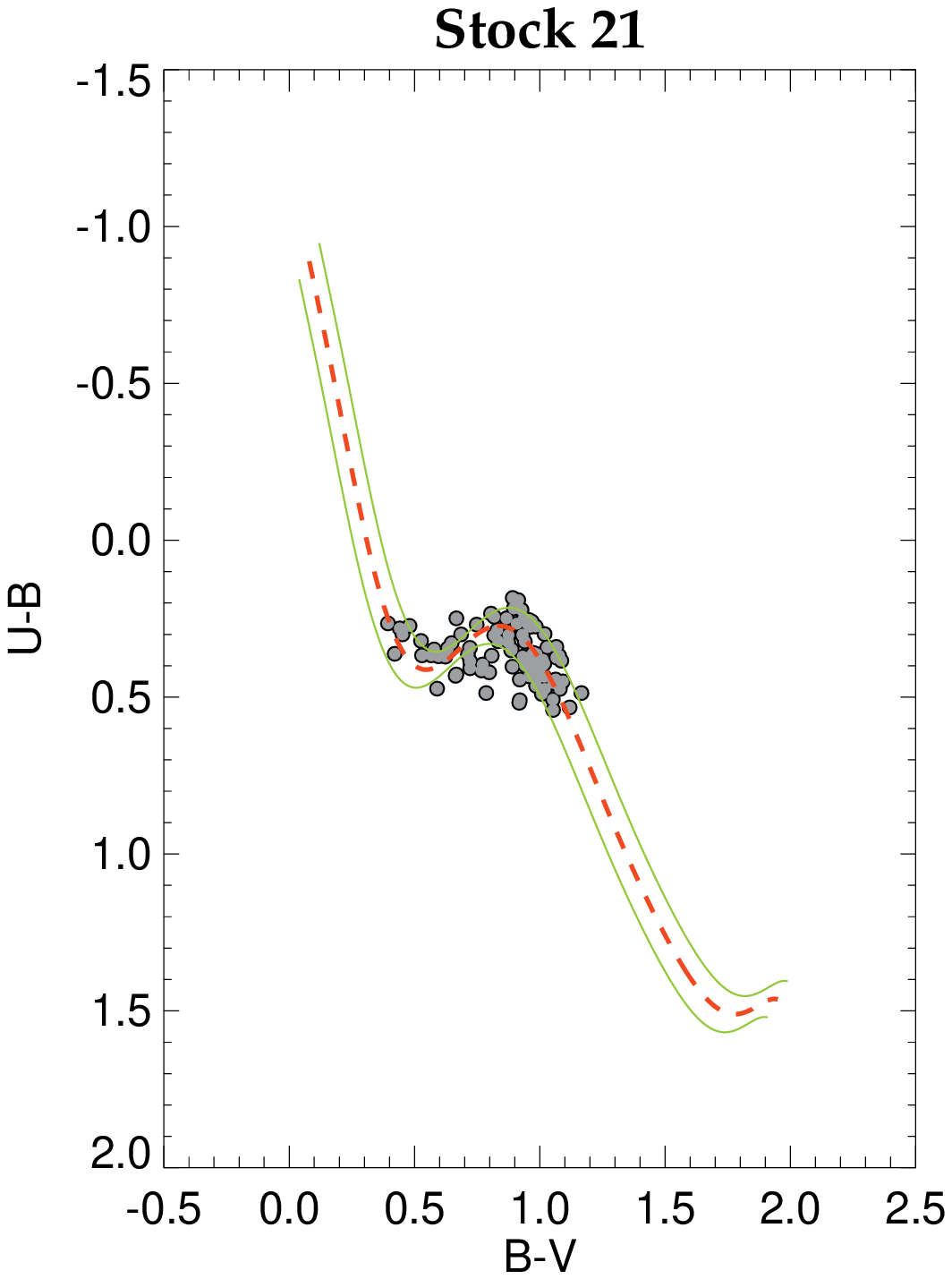}\\
\caption{$U-B \times B-V$ TCDs of the member main-sequence stars for each cluster. The reddened ZAMS of \citet{Sung13} and $\pm1\sigma$ standard deviations are represented with red dashed lines and green solid lines, respectively.} 
\end {figure}

\subsection{Metallicities of the clusters} 
We utilized the method by \citet{Karaali11} to obtain the photometric metallicity of the clusters. Since the method considers only the F-G type main-sequence stars, we selected the main-sequence stars with $P\geq50\%$ in the colour index interval $0.3\leq(B-V)_0\leq0.6$ mag \citep{Cox00} for the analysis. First, we calculated ultraviolet (UV) excesses ($\delta =(U-B)_{0, H}-(U-B)_{0,S}$, where $H$ and $S$ denote Hyades and star having the same $(B-V)_0$, respectively) and normalised their $\delta$ differences to the UV-excess at $(B-V)_0=0.6$ mag (i.e. $\delta_{0.6}$). Then, we fitted distributions of normalised $\delta_{0.6}$ UV-excesses with a Gaussian and we have taken the Gaussian peak as the normalised $\delta_{0.6}$ UV-excess value for each cluster. Note that we were not able to derive precise photometric metallicity of the cluster ASCC 115,  since the photometric selection yields only one main-sequence star in $0.3<(B-V)_0\leq 0.6$ mag colour index interval. Thus, we assumed that this cluster has the solar metallicity. We present the TCDs and histograms of the normalised $\delta_{0.6}$ UV-excesses in Fig. 8. We take the 1$\sigma$ width of the best fit Gaussian model as the statistical uncertainty of our measurements. We then found the [Fe/H] metallicity from the equation \citep{Karaali11} as below;

\begin{equation}
{\rm [Fe/H]}=0.105-3.557\times\delta_{0.6}-14.316\times\delta_{0.6}^2.
\end{equation} 
We then transformed the [Fe/H] metallicities to the mass fraction $Z$ of all elements heavier than helium using the equations of Jo Bovy\footnote{https://github.com/jobovy/isodist/blob/master/isodist/\\Isochrone.py} who analytically derived it from PARSEC isochrones:

\begin{equation}
Z_X={10^{{\rm [Fe/H]}+\log \left(\frac{Z_{\odot}}{1-0.248-2.78\times Z_{\odot}}\right)}},
\end{equation}      

\begin{equation}
Z=\frac{(Z_X-0.2485\times Z_X)}{(2.78\times Z_X+1)}.
\end{equation}
Here, $Z_X$ is the intermediate operation function and $Z_{\odot}$ is the solar mass fraction taken as 0.0152 \citep{Bressan12}. 

\subsection{Distance moduli and the ages of the clusters}
Keeping the reddening and metallicity values found above as constants, we shifted the theoretical isochrones from the PARSEC V1.2 synthetic stellar library \citep{Bressan12, Tang14, Chen14} on the observed $V\times U-B$ and $V\times B-V$ CMDs until a theoretical isochrone fits with the observational data, in order to obtain distance modulus and age of each open cluster studied here. In the distance calculation for each open cluster, we assumed the standard selective absorption coefficient as $R_V=3.1$ \citep{Schultz75}. We considered the errors in distance moduli of the clusters for the determination of errors in the distances \citep{Carraro17}. We show CMDs with the best fit theoretical isochrones in Fig. 9 and list the distance moduli, distances and ages of the clusters in Table 1. 

In particular, the presence of a small number of cluster member stars ($P\geq50\%$) on the blue side of the cluster isochrone towards the faint apparent magnitudes is remarkable. The number of stars in NGC 6793, NGC 7031 and NGC 7039 in this region are 12 ($53\leq P(\%)\leq56$), 16 ($53\leq P(\%)\leq87$) and 30 ($P=50\%$), respectively. Median values of the membership probabilities of these stars for each cluster are less than $P=60\%$. We expect this scattering in this region due to insensitivity of the  photometric and astrometric data towards to fainter apparent magnitudes ($V=18$ mag).
 
\begin{figure*}
\centering
\includegraphics[scale=.33, angle=0]{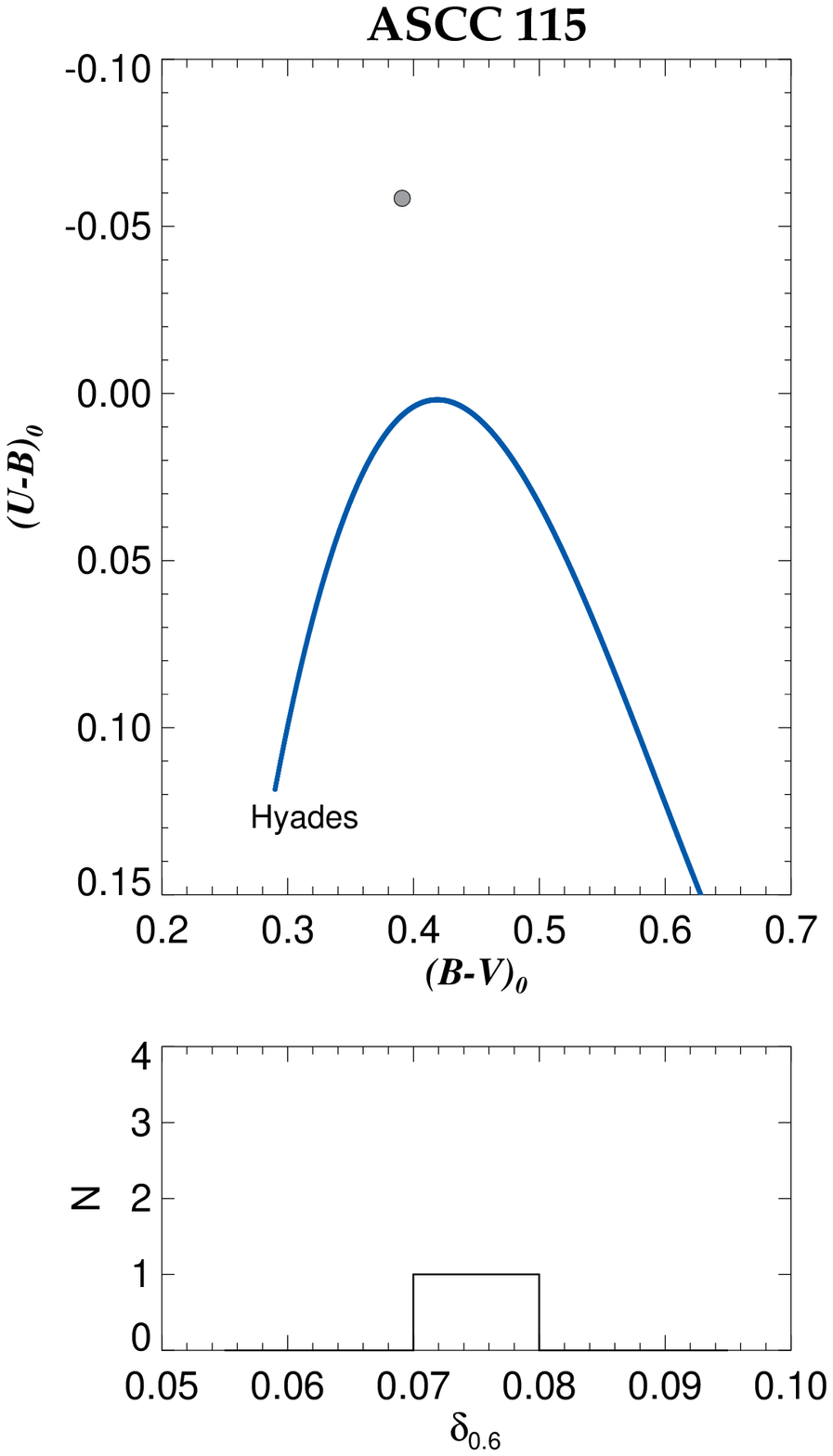}
\includegraphics[scale=.33, angle=0]{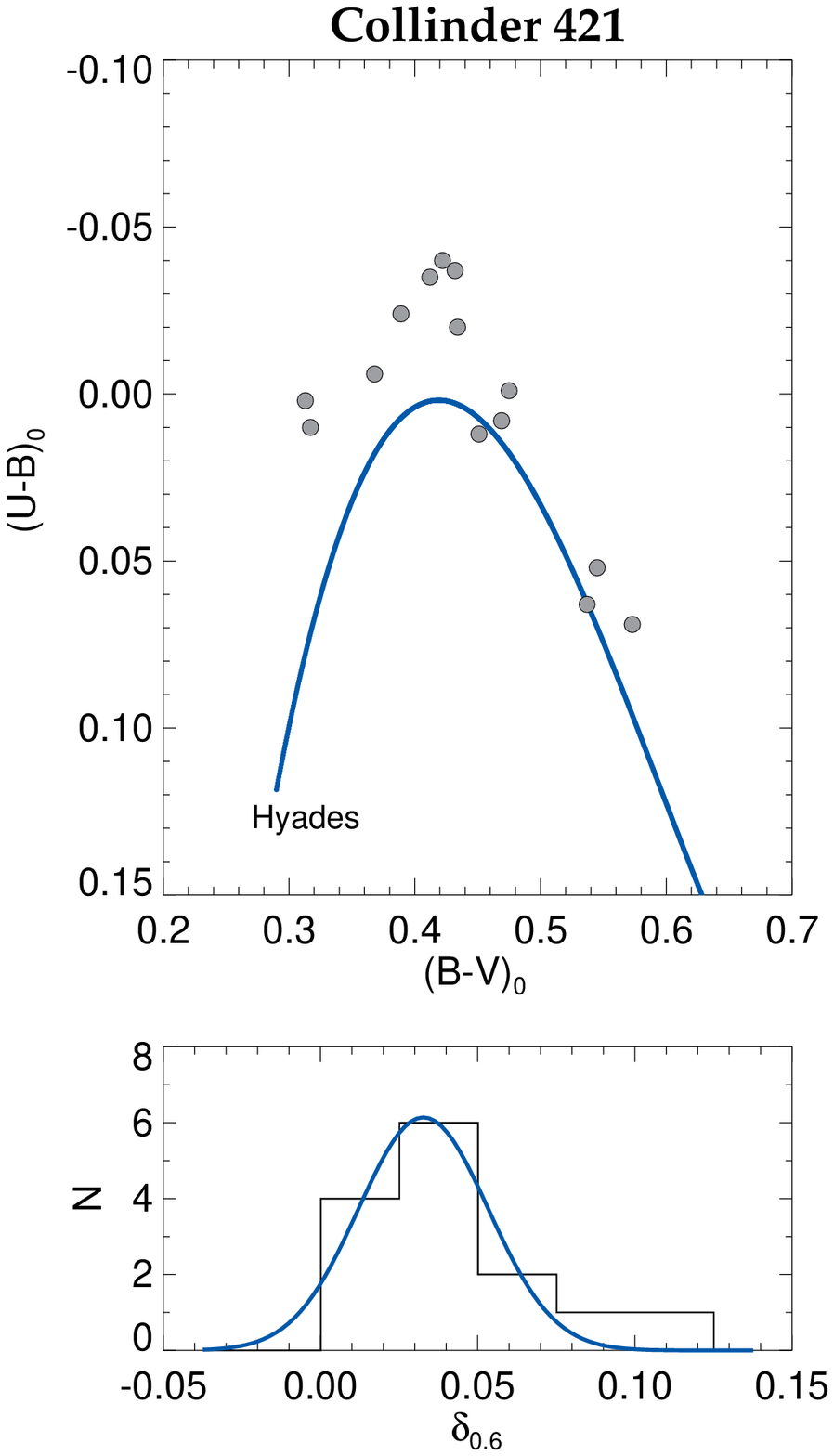}
\includegraphics[scale=.33, angle=0]{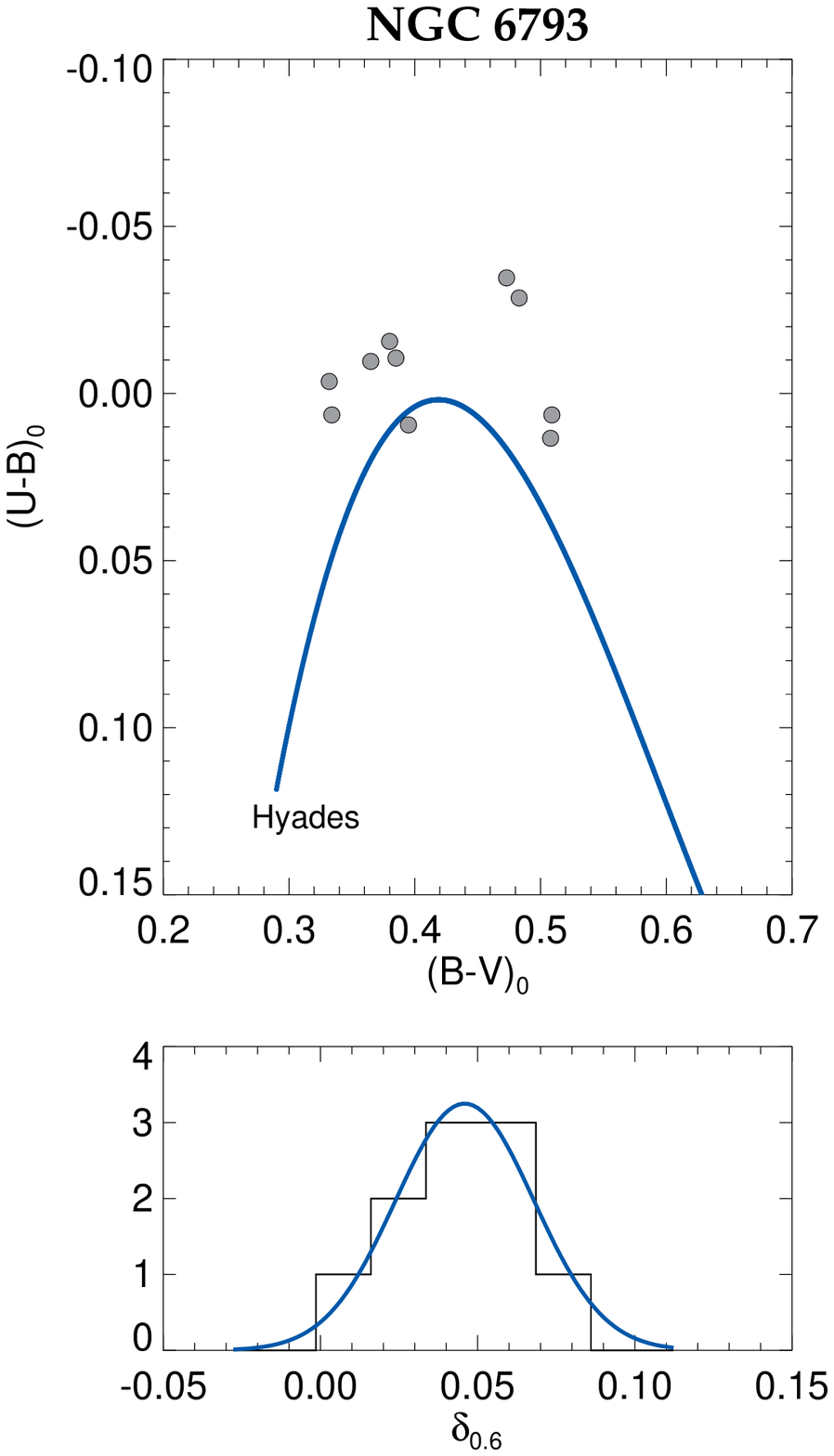}
\includegraphics[scale=.33, angle=0]{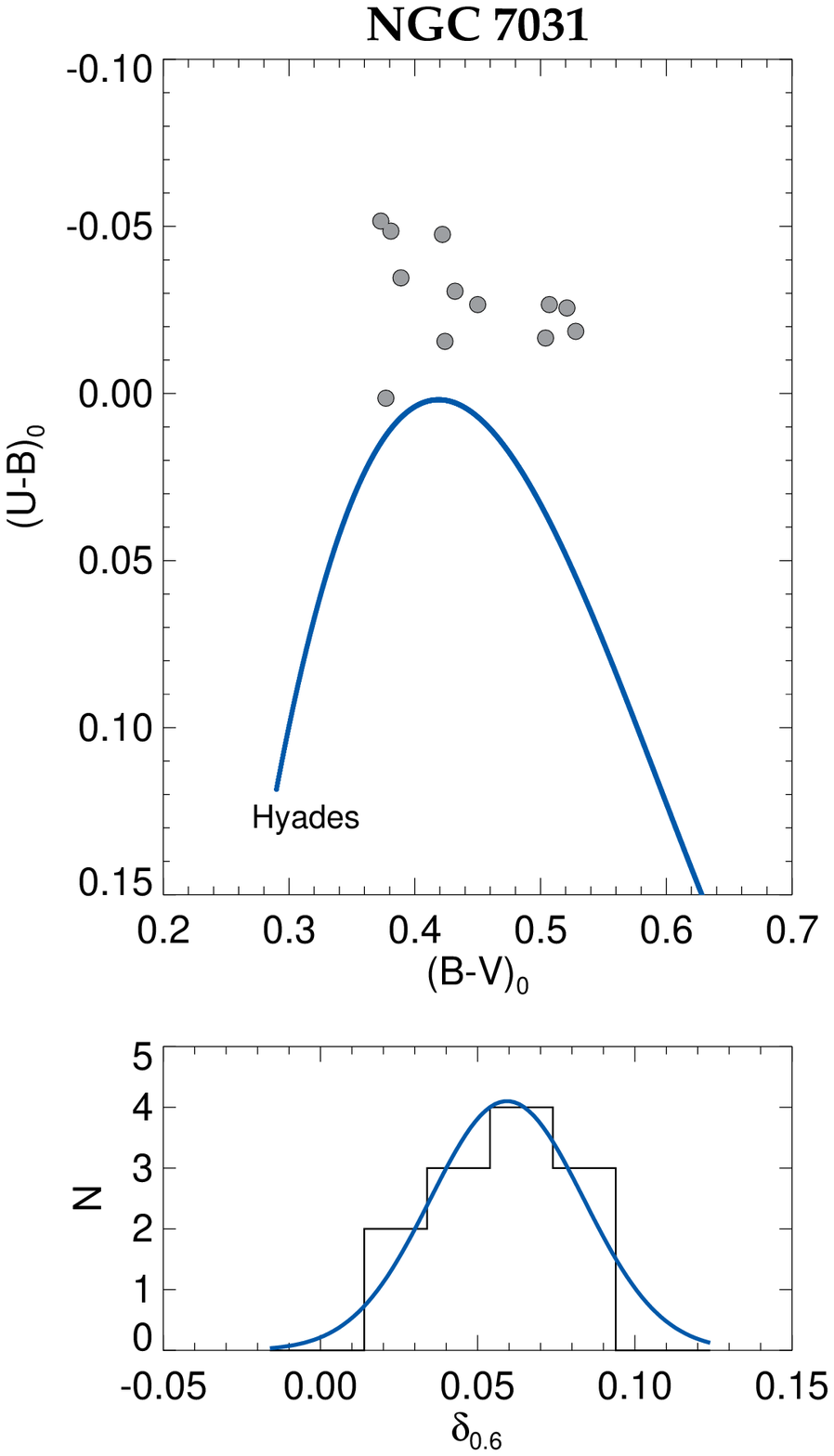}\\
\includegraphics[scale=.33, angle=0]{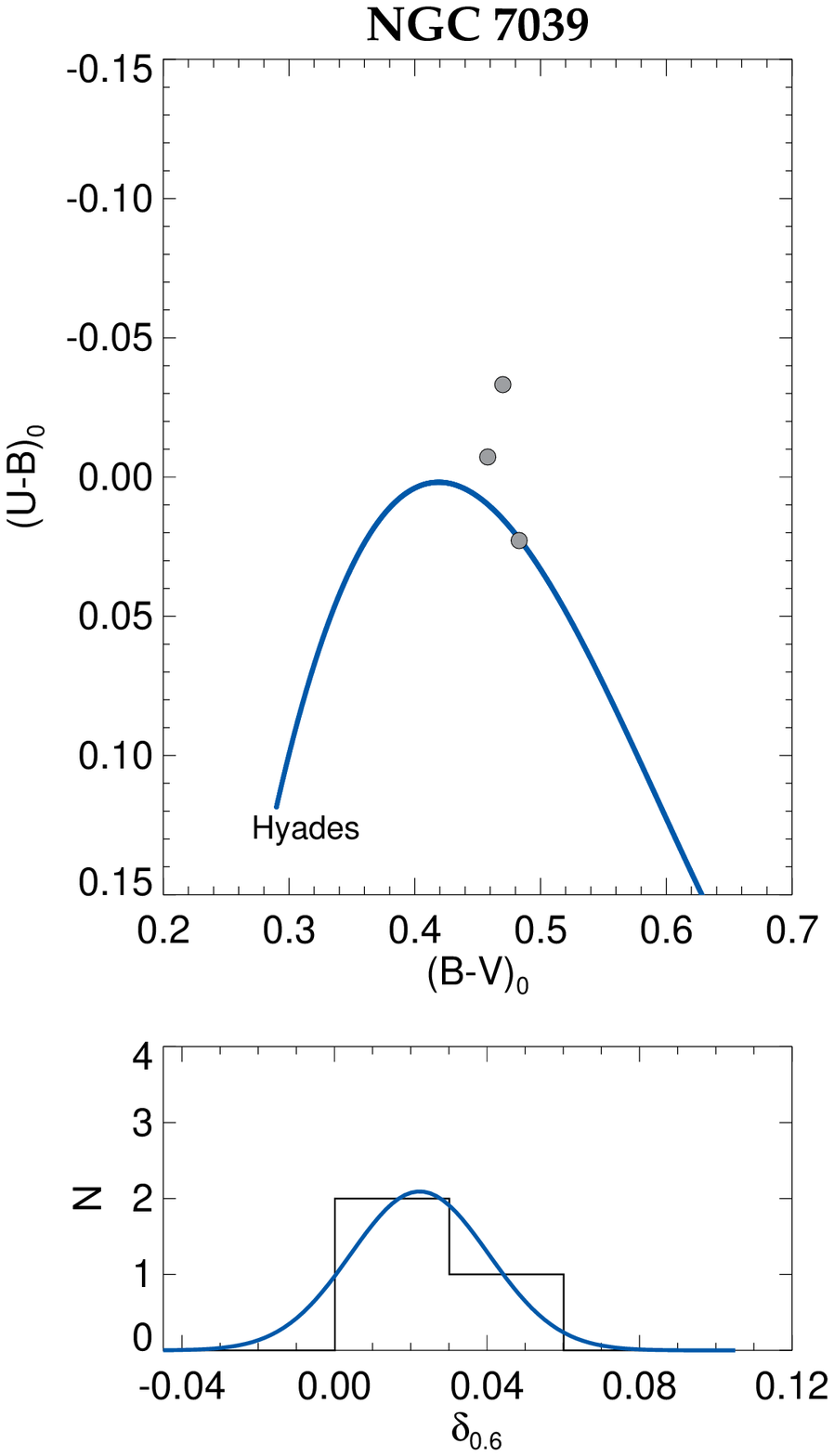}
\includegraphics[scale=.33, angle=0]{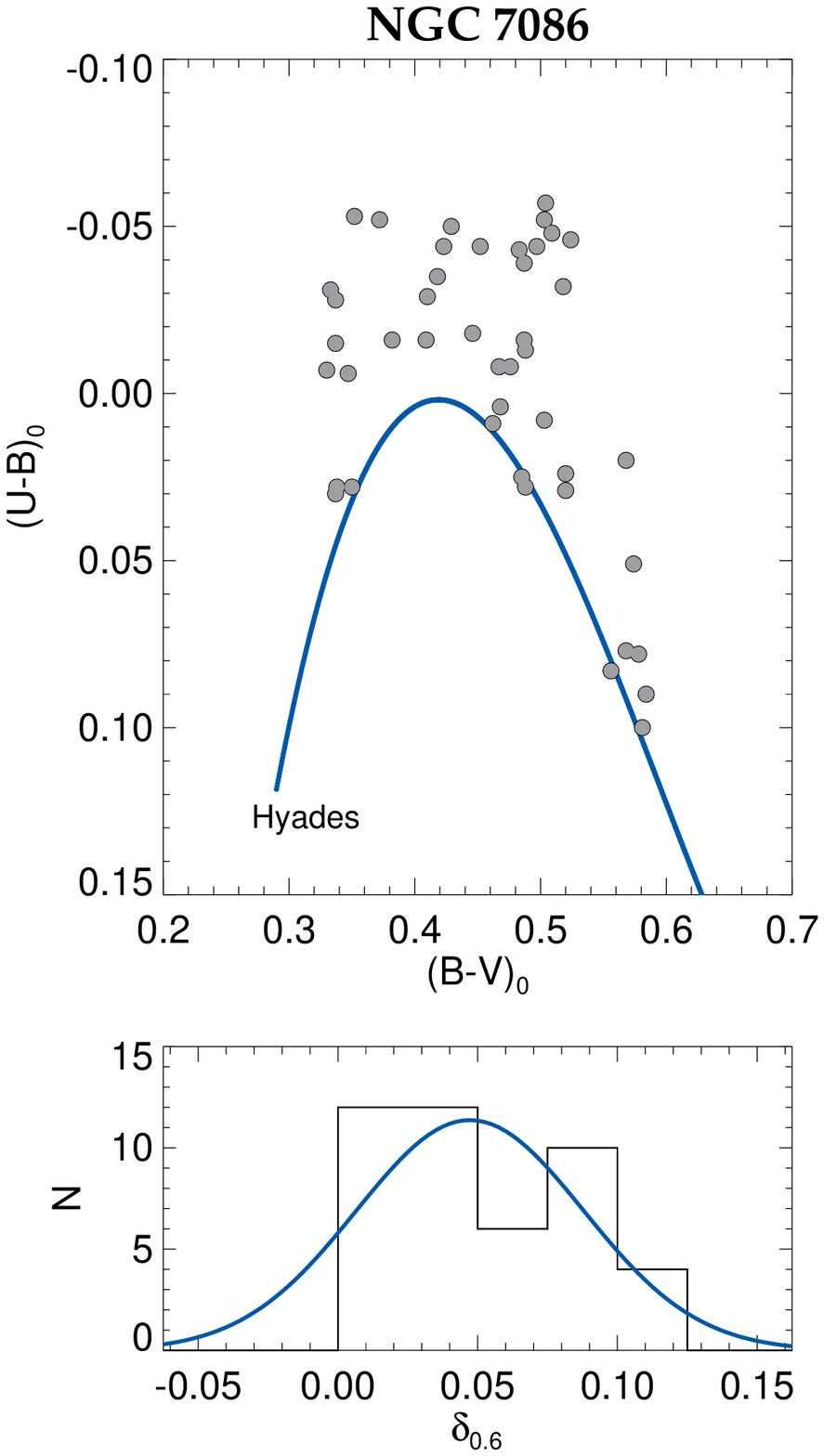}
\includegraphics[scale=.33, angle=0]{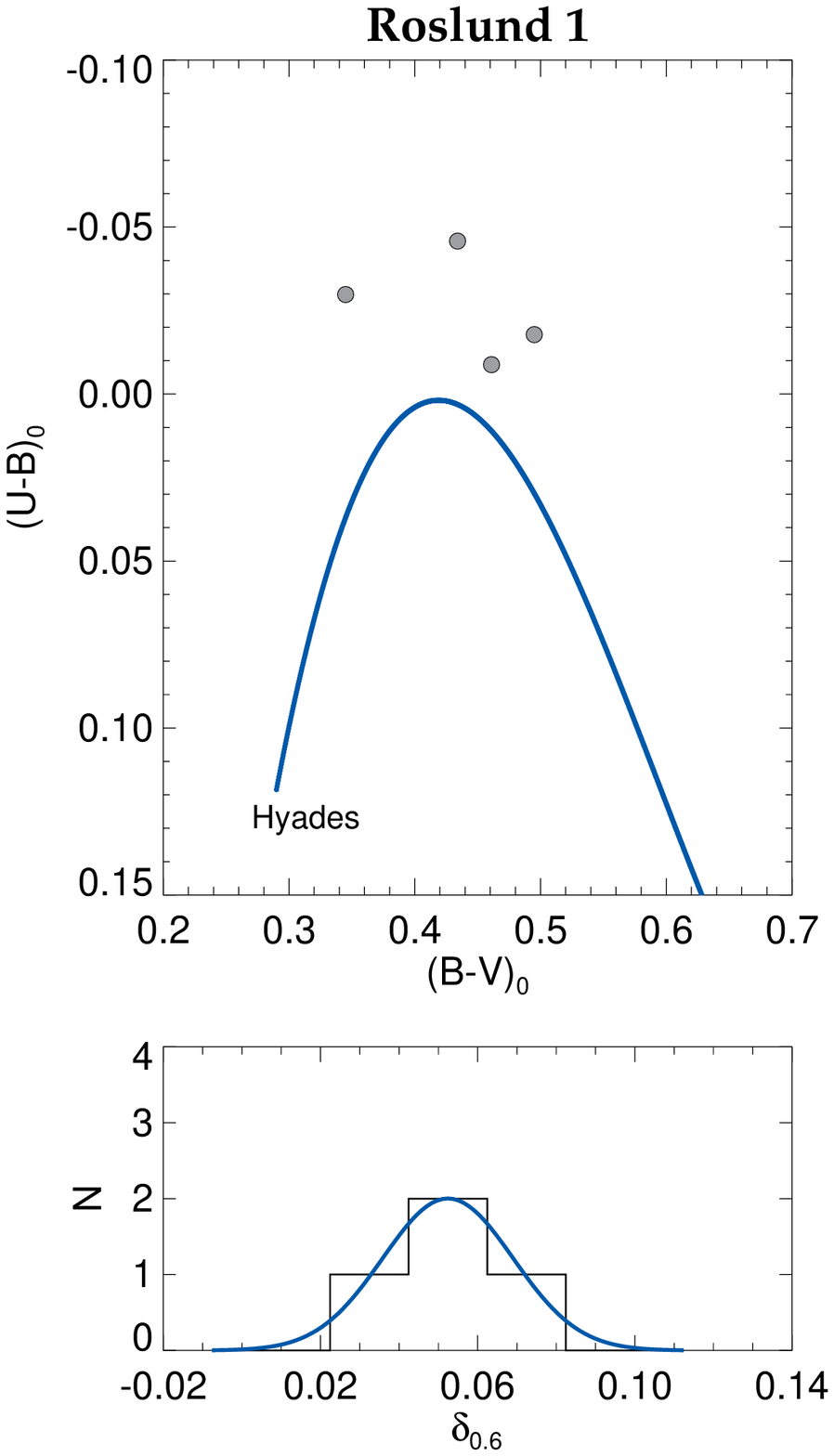}
\includegraphics[scale=.33, angle=0]{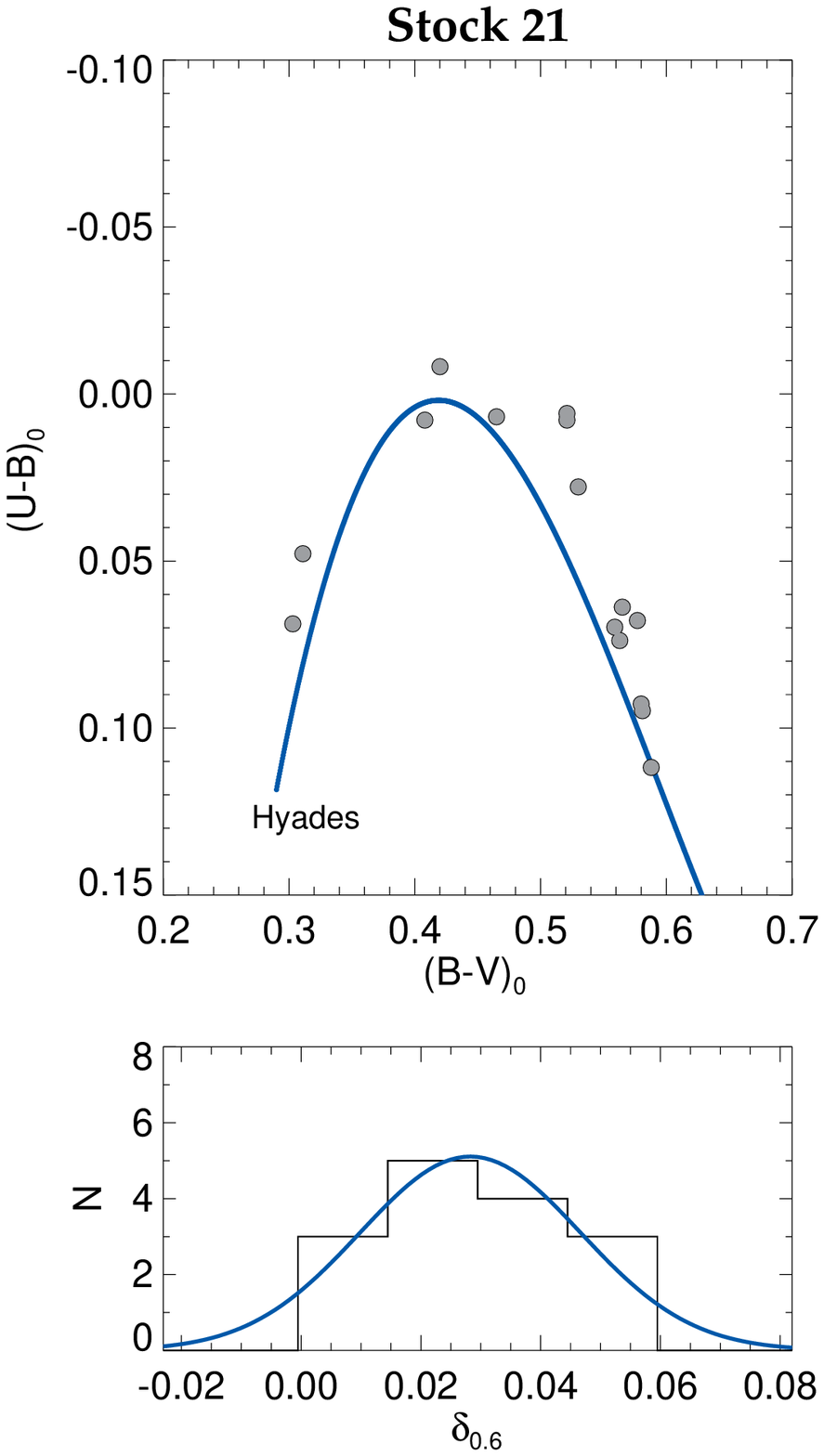}\\
\caption{$(U-B)_0\times(B-V)_0$ TCDs (upper panel) and the histograms (lower panel) for the normalised UV-excesses of the main-sequence stars used for the metallicity estimation of open clusters. The solid lines in the upper and lower panels represent the main-sequence of Hyades cluster and the Gaussian fit to the histogram, respectively.} 
\end{figure*}

\begin{landscape}
\begin{figure}
\centering
\includegraphics[scale=.75, angle=0]{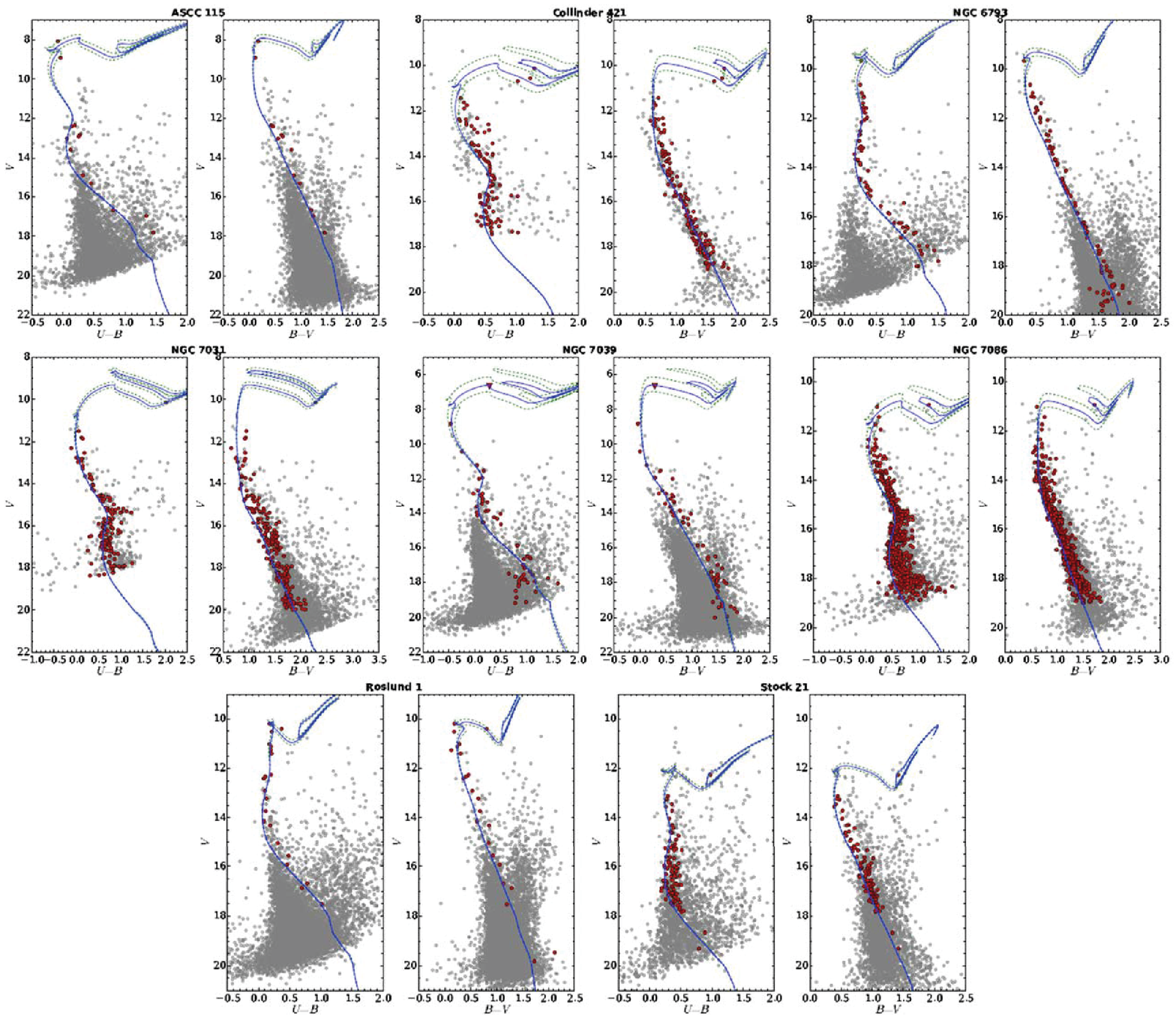}
\caption{$V\times U-B$ and $V\times B-V$ CMDs for the eight open clusters. Red filled circles denote the most probable members of each cluster. Blue lines and green dashed lines are the best fit theoretical isochrones determined in this study and the isochrones with estimated age plus/minus its error, respectively.} 
\end{figure}
\end{landscape}

\subsection{Mass functions of the clusters}
We first identified a high degree polynomial function between absolute magnitudes and masses of the main-sequence stars from the PARSEC isochrones fitted best for each cluster. Then, we converted $V$ apparent magnitudes of the most likely main-sequence stars of the clusters to absolute magnitudes using the distance moduli of the clusters, and calculated their theoretical masses from the derived absolute magnitude-mass relation. We obtained mass function (MF) slope from the relation of $\log\xi(M)= -(1+X)\times \log M + \rm constant$, where $dN$ denotes the number of stars in a mass interval $dM$ with central mass $M$, and $X$ is the slope of the MF. Fig. 10 shows the data with linear fits giving the value of the MF slope for each open cluster. We present the results of fitting procedures in Table 6. MF slopes for eight clusters are in very good agreement with the value of 1.35 of \citet{Salpeter55}.

\begin{table}
\setlength{\tabcolsep}{2.2pt}
  \centering
  \caption{The slopes ($X$) of the mass functions of eight open clusters. $N$ is number of main-sequence stars in the clusters.}
    \begin{tabular}{lccc}
\hline
    Cluster & $N$ & $X$ & Mass Range \\
\hline
    ASCC 115 &  13 & -1.36$\pm$0.75 & 0.6 $<M/M_{\odot}<$ 3.6 \\
Collinder 421& 146 & -1.14$\pm$0.17 & 1.0 $<M/M_{\odot}<$ 4.4 \\
    NGC 6793 &  87 & -1.31$\pm$0.31 & 0.5 $<M/M_{\odot}<$ 2.7 \\
    NGC 7031 & 208 & -1.42$\pm$0.16 & 0.7 $<M/M_{\odot}<$ 5.6 \\
    NGC 7039 &  62 & -1.32$\pm$0.37 & 0.4 $<M/M_{\odot}<$ 2.4 \\
    NGC 7086 & 538 & -1.33$\pm$0.12 & 0.9 $<M/M_{\odot}<$ 4.0 \\
   Roslund 1 &  25 & -1.10$\pm$0.28 & 0.8 $<M/M_{\odot}<$ 2.4 \\
    Stock 21 &  96 & -1.32$\pm$0.16 & 0.8 $<M/M_{\odot}<$ 2.5 \\
\hline
    \end{tabular}
\end{table}

\begin{figure*}
\centering
\includegraphics[scale=.9, angle=0]{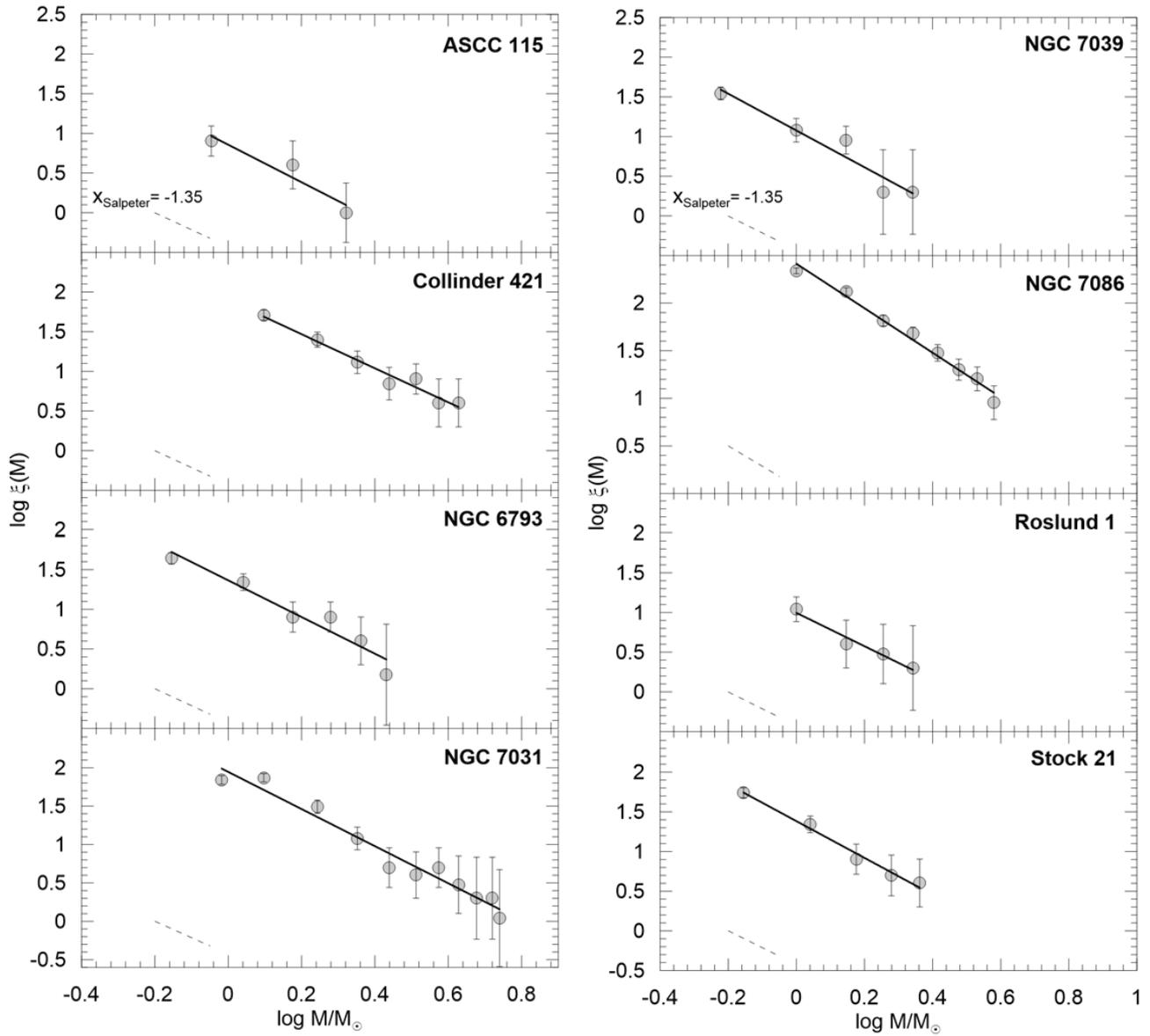}
\caption{Mass functions of eight clusters determined from the selected member stars. The solid line shows the mass function of the cluster.} 
\end{figure*}

\section{Discussion and conclusions}
\subsection{Comparison with the literature}
 
In this study, we obtained basic astrophysical parameters (reddening, photometric metallicity, distance modulus, distance and age) from CCD {\it UBV} observations of eight open clusters. Dense star cluster regions, less accurate membership selections, high extinction through the cluster field, or different analysis methods can affect parameter determination and cause degeneration between parameters. As a result, the values given by different authors for the same cluster can have large discrepancies, which affects the studies on understanding the Galactic structure using open clusters.

We used independent methods to obtain reddening, metallicity, distance modulus, distance and age of each cluster to suffer less from degeneracy and to present homogeneous values in this study. Analyses were performed for each cluster using the same methods. During the analyses, we have taken into account the stars that were selected as probable cluster members and brighter than $V$-magnitude completeness limits of our observations.

{\bf ASCC 115}: There are no comprehensive CCD {\it UBV} photometric study for ASCC 115 in the literature. \citet{Kharchenko09} used $B$ and $V$ magnitudes taken from ASCC-2.5 \citep{Kharchenko01} and calculated the reddening as $E(B-V)=0.15$ mag, distance modulus as $\mu=9.36$ mag, distance $d=600$ pc and age $\log t {\rm(yr)}=8.59$. Our results are generally compatible with \citet{Kharchenko09}, although we determined a younger age for the ASCC 115 (Table 1) in this study. As there is only one F-G type main-sequence star in the $0.3\leq(B-V)_0\leq0.6$ mag range, we assumed its metallicity as the Solar value ($Z=0.0152$). We constructed the histogram of distances calculated from {\it Gaia} DR2 parallaxes for stars adopted as cluster members ($P\geq50\%)$, and took the mode of this histogram as {\it Gaia} distance of the cluster, $d_{Gaia}=755\pm14$ pc. This value is in very good agreement with the distance derived via isochrone fitting ($d=732\pm69$ pc) in this study. Hence, we conclude that our results for the parameters of ASCC 115 open cluster are reliable.

{\bf Collinder 421}: Collinder 421 is a poorly studied cluster and CCD $U$ band observations have not been presented in the literature so far. There are large differences in reddening, distance modulus and age values between different studies of the cluster (see Table 1). As a result of our independent analyses, the age is younger and reddening is higher than those in previous studies. Also, using selected member stars ($P\geq50\%$) and {\it Gaia} DR2 parallaxes, we calculated {\it Gaia} distance of the cluster as $d_{Gaia}=1220\pm99$ pc. This value is in very good agreement with the distance $d=1245\pm103$ pc determined via isochrone fitting in this study. From the agreement between these two distances we conclude that obtained cluster age, reddening and metallicity are accurately found in this study. 

Moreover, \citet{Maciejewski07} performed radial density profile analyses of Collinder 421 and obtained central stellar density as $f_0=2.67\pm0.46$ stars arcmin$^{-2}$, core radius $r_c=1.1\pm0.3$ arcmin and background stellar density $f_{bg}=0.93\pm0.07$ stars arcmin$^{-2}$. The structural parameters obtained in this study are almost compatible with their results (see Table 5). Discrepancy among the parameters are due to the number of stars and photometric $V$ band limit used in our study. The astrophysical parameters determined in this study are based on well-defined cluster member stars, {\it UBV} photometric data, $V$-band limit value and independent analysis methods.

{\bf NGC 6793}: \citet{Kharchenko05} used Johnson {\it BV} photometric data, and determined reddening, distance, distance modulus and age of the cluster as $E(B-V)=0.17$ mag, $d=1100$ pc, $\mu=10.73$ mag and $\log t~{\rm yr}=8.64$, respectively. We determined age of the cluster as $\log t~{\rm(yr)}=8.70\pm0.04$, which is very close to the value published by \citet{Kharchenko05}. When we compared other parameters, it is seen that reddening is two times higher and distance is shorter than those given by the authors. The {\it Gaia} distance is calculated as $d_{Gaia}=607\pm37$ pc, which is nearly the same with the distance $d=610\pm40$ obtained from the isochrone fitting method in our study. This shows that not only our distance, but also the reddening, metallicity and age values are determined reliably in this study. The methods used by \citet{Kharchenko05} and this study are different, this may be one of the reasons behind the differences.

{\bf NGC 7031}: There are many photographic studies done by various authors between 1960's and 1970's for this cluster (see Table 1). According to these studies, the reddening $E(B-V)$, distance ($d$), and age ($t$) of the cluster differ from 0.71 to 1.03 mag, 686 to 910 pc, and 56 to 480 Myr, respectively. \citet{Kopchev08} used CCD {\it BV} photometric observations and determined reddening, distance, distance modulus and age of NGC 7031 as $E(B-V)=1.05\pm0.05$ mag, $d= 831\pm72$ pc, $\mu=9.6\pm0.2$ mag and $t=224\pm25$ Myr, respectively.

Published studies reveal that NGC 7031 has a relatively high reddening and the cluster distance is less than 1 kpc. However, the inferred cluster ages show a very large scatter. Reddening determined in this study is compatible with those given in the literature. In addition, cluster has a younger age $t=65\pm5$ Myr, which is close to the value given by \citet{Lindoff68}, and a larger distance than the values given in the literature. The {\it Gaia} distance of the cluster is calculated as $d_{Gaia}=1404\pm81$ pc, which is in agreement with our results ($d=1212\pm146$ pc) within the errors.

{\bf NGC 7039}: Photoelectric and photographic {\it UBV} studies of the cluster were published by \citet{Zelwanowa72} and \citet{Hassan73}. In addition to these, the first CCD {\it BV} photometric results were given by \citet{Kharchenko05}. Authors determined almost the same reddening for the cluster, while they presented a distance ranging between 700 and 1535 pc, and age $7<\log t~{\rm (yr)}<9$  (see Table 1). 

We could not detect the turn-off point or giant stars due to the lack of stars brighter than $V=8$ mag in our catalogue. Although there are brighter stars in the cluster field, observing limit of the T100 telescope correspond to about $V=8$ mag. According to previous studies \citep{Hassan73, Kharchenko05} NGC 7039, seems to be a young-intermediate age open star cluster. Therefore, there may be brighter stars on the turn-off point, sub-giant or giant star region. 

In the cluster field, there is one bright star HD 201935 ($V=6.671$, $B-V=0.354$, $U-B=0.285$ mag) that is out of our brightness limit. It is emphasized in SIMBAD database that HD 201935 is a member of NGC 7039. The trigonometric parallaxes for this star measured from {\it Hipparcos} and {\it Gaia} satellites are $\pi=1.90\pm0.38$ and $\pi=0.5715\pm0.0297$ mas, respectively. It can be seen that the relative parallax error ($\sigma_{\pi}/\pi=0.052$) for {\it Gaia} measurement is very small. However, the {\it Hipparcos} trigonometric parallax of HD 201935 given by \citet{vanLeeuwen07} is more compatible with the cluster distance determined from PARSEC isochrone fittings in this study.

The reddening and distance obtained in this study are in good agreement with those given in previous studies, except with the distance presented in \citet{Hassan73} and the age we estimated for the NGC 7039 is younger than the previous studies. Moreover, the {\it Gaia} distance calculated as $d_{Gaia}=767\pm41$ pc is in good agreement with our results ($d=743\pm64$ pc).

{\bf NGC 7086}: From the results of photographic and photometric measurements done by \citet{Hassan67} and \citet{Lindoff68}, the mean values of reddening, distance modulus and distance are $E(B-V)=0.70$ mag, $\mu=12.50$ mag and $d=1190$ pc. However, the age of the cluster was given as $t=600$ Myr and $t=85$ Myr by \citet{Hassan67} and \citet{Lindoff68}, respectively. \citet{Rosvick06} presented the first CCD {\it B, V,} and {\it R} observations and determined cluster's parameters from {\it BV} photometry and spectra. From spectral analysis of five member stars, they estimated reddening as $E(B-V)=0.83\pm0.02$ mag, the distance modulus as $\mu=13.4\pm0.3$ mag via fitting the ZAMS of \citet{Vandenberg89}'s to the cluster main-sequence, and finally the age of the cluster as 100 Myr by comparing \citet{Girardi00}'s metal-poor ($Z=0.008$) isochrone with cluster's CMD. Moreover, authors pointed out they found the distance modulus and age roughly because of lack of metallicity for NGC 7086. Also, \citet{Kopchev08} utilized CCD {\it BV} photometry and they obtained the astrophysical parameters of the cluster.  
 
Although there are two CCD {\it BV} studies available for NGC 7086, we could only reach the data of \citet{Rosvick06} by private communication and compared $V$ magnitudes and $B-V$ colours in two studies. As a result of cross-matching two photometric data with the limiting magnitude $V=18$ mag, we detected 284 identical stars in both catalogs. The mean differences and standard deviations between $B-V$ colour and $V$ magnitudes are calculated as $\langle \Delta (B-V)\rangle=0.014\pm0.046$ mag and $\langle \Delta V \rangle=0.051\pm0.43$ mag, respectively. We present colour and magnitude differences between our measurements and those of \citet{Rosvick06} in Fig. 11. We conclude that our photometric data are in agreement with theirs.

\begin{figure}
\centering
\includegraphics[trim=1cm 0cm 2cm 1cm, clip=true, scale=0.55, angle=0]{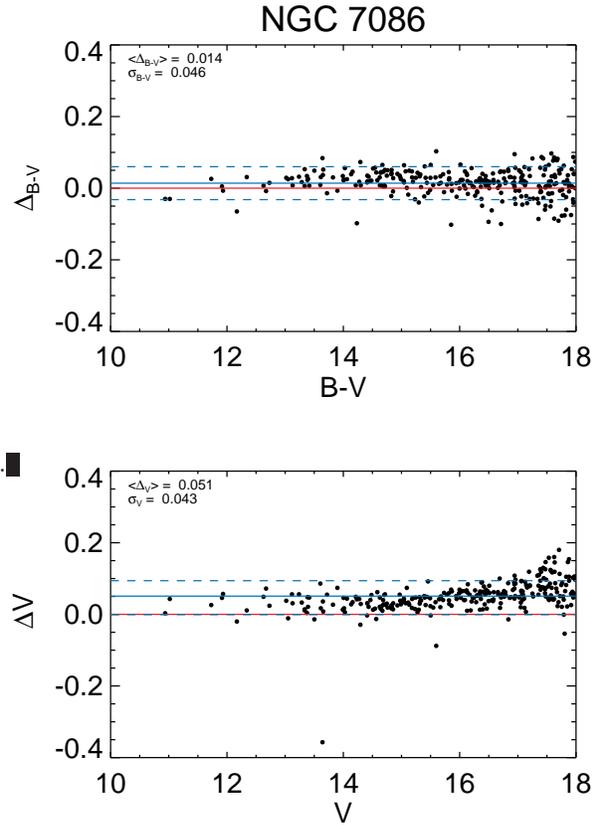}
\caption{Comparison of $B-V$ colours (upper panel) and $V$ apparent magnitudes (lower panel) of NGC 7086 with \citet{Rosvick06}.} 
\end{figure}

The reddening yielded in this study is compatible with the values given in the literature except for \citet{Rosvick06} who suggest that there is more reddening through the cluster field. The distance modulus and distance are in agreement with \citet{Rosvick06}, while the other studies indicate that the cluster is more closer to the Sun than that we suggested. The age of the cluster is generally in good agreement with the literature except the result of \citet{Hassan67}. Moreover, the distance calculated from the {\it Gaia} trigonometric parallaxes is $d_{Gaia}=1684\pm140$ pc, which is matched very well with the value calculated from isochrone fitting, $d=1618\pm182$ pc. We deduce from this consistency that our results are more reliable and do not suffer from degeneracy.  

{\bf Roslund 1}: Analyzing CCD {\it BV} photometric data, \citet{Kharchenko05} calculated the reddening, distance, distance modulus, and age of the cluster, as $E(B-V)=0.05$ mag, $d=670$ pc, $\mu=9.28$ mag and age $\log t~{\rm (yr)}=8.47$ ($t=295$ Myr), respectively. 
 
Astrophysical parameters estimated in this study for Roslund 1 are quite different from those of \citet{Kharchenko05}. We estimated a higher interstellar extinction through the cluster region. In addition, older age isochrones are fitted well with CMDs of the cluster (Fig. 7). The distance calculated from {\it Gaia} DR2 parallaxes is $d_{Gaia}=883\pm54$ pc, which is in good agreement with estimated distance ($d=836\pm48$ pc) within the errors. Thus, we conclude that our reddening value is reliable for the cluster region.

\begin{table*}[t]
\setlength{\tabcolsep}{1.5pt}
  \centering
\scriptsize{
  \caption{Distance and proper motion values calculated from different methods for eight open clusters.}
    \begin{tabular}{l|ccccc|cccc|c}
\hline
                  & \multicolumn{5}{c}{This Study} &  \multicolumn{4}{c}{\citet{Cantat-Gaudin18}} & \citet{Bailer-Jones18} \\
\hline
     Cluster      & $N$ & $d_{Isochrone}$ & $d_{Gaia}$ & $\mu_{\alpha}\cos \delta$ & $\mu_{\delta}$ & $N$ & $d$ & $\mu_{\alpha}\cos \delta$ & $\mu_{\delta}$ & $d$ \\
                  &     &  (pc)   & (pc) & (mas~yr$^{-1}$)  & (mas~yr$^{-1}$) &  & (pc) & (mas~yr$^{-1}$) & (mas~yr$^{-1}$) & (pc) \\
\hline
    ASCC115       &  13 & 732$\pm$69   &  755$\pm$14  & $-$0.524$\pm$0.062 & $-$0.568$\pm$0.050 &  39 &  746$^{+52}_{-60}$   & $-$0.549$\pm$0.068 & $-$0.543$\pm$0.088 &  736$\pm$15 \\
    Collinder 421 & 149 & 1245$\pm$103 & 1220$\pm$99  & $-$3.662$\pm$0.076 & $-$8.309$\pm$0.082 & --  &   --                 &                 -- &                 -- & 1175$\pm$81 \\
    NGC 6793      &  87 & 610$\pm$40   &  607$\pm$37  &   +3.818$\pm$0.055 &   +3.611$\pm$0.071 & 190 &  589$^{+33}_{-37}$   &   +3.795$\pm$0.186 & +3.544$\pm$0.177   &  594$\pm$38 \\
    NGC 7031      & 208 & 1212$\pm$146 & 1404$\pm$81  & $-$1.286$\pm$0.081 & $-$4.177$\pm$0.076 & 171 & 1365$^{+164}_{-216}$ & $-$1.242$\pm$0.122 & $-$4.205$\pm$0.130 & 1342$\pm$68 \\
    NGC 7039      &  63 & 743$\pm$64   &  767$\pm$41  & $-$0.262$\pm$0.100 & $-$2.509$\pm$0.097 & 119 &  756$^{+53}_{-62}$   & $-$0.320$\pm$0.287 & $-$2.542$\pm$0.215 &  736$\pm$19 \\
    NGC 7086      & 543 & 1618$\pm$182 & 1684$\pm$140 & $-$1.642$\pm$0.086 & $-$1.644$\pm$0.076 & 622 & 1616$^{+225}_{-312}$ & $-$1.656$\pm$0.148 & $-$1.629$\pm$0.143 & 1606$\pm$133\\
    Roslund 1     &  23 & 836$\pm$48   &  883$\pm$54  & $-$0.895$\pm$0.043 & $-$5.766$\pm$0.044 & --  &                   -- &                 -- &                 -- &  875$\pm$28 \\
    Stock 21      &  97 & 1931$\pm$27  & 1934$\pm$159 & $-$2.154$\pm$0.063 & $-$1.752$\pm$0.060 &  62 & 1854$^{+289}_{-422}$ & $-$2.161$\pm$0.067 & $-$1.787$\pm$0.092 & 1828$\pm$146\\
\hline
    \end{tabular}%
}
  \label{tab:addlabel}%
\end{table*}%

{\bf Stock 21}: The astrophysical parameters which are calculated from CCD {\it BV} photometric data, are only presented by \citet{Kharchenko05}. In their study the reddening, distance modulus, distance and age are given as $E(B-V)=0.4$ mag, $\mu=11.45$ mag$, d=1100$ pc and $\log t~{\rm (yr)}=8.72$ ($t=525$ Myr), respectively. The reddening obtained in this study compatible with \citet{Kharchenko05}, while they determined a shorter distance and older age for the cluster. The {\it Gaia} distance of the cluster was calculated as $d_{Gaia}=1934\pm159$ pc, a value which is almost the same as the distance estimated from isochrone fitting, $d=1931\pm127$ pc. \citet{Kharchenko05} used {\it BV} photometric data that were compiled from the literature for the bright stars and utilized simultaneous methods to determine parameters of the Stock 21. For these reasons there can be degeneracy particularly between the reddening and distance and so the age values in their study. 

In this study, cluster member stars were determined from sensitive photometric and astrometric data for the open clusters investigated. However, the low number of cluster member stars and the lack of structural parameters make it doubtful that Roslund 1 and Stock 21 are open clusters. The determination of the radial velocity distribution and model atmospheric parameters of the most probable cluster member stars of Roslund 1 and Stock 21 need to be carried out by using high-resolution spectroscopic observations in order to clarify whether these are open clusters.

\subsection{Astrometric comparison with the literature}

In this study, distances of eight clusters were estimated using PARSEC isochrones \citep{Bressan12}, which were fitted to CMDs of the clusters according to the positions of cluster stars with the membership probabilities $P\geq50\%$ (see Table 7). Considering {\it Gaia} DR2 \citep{Gaia18} proper motions and parallaxes of member stars, we calculated median values of proper motion components and distances (hereafter {\it Gaia} distance) of each cluster. For determining {\it Gaia} distances of the clusters, first we calculated distances of each member star from trigonometric parallaxes via linear method, i.e. $d (pc)=1000/\pi$ (mas). Then, we fitted Gaussian curves to distance histograms (Fig. 12) and calculated {\it Gaia} distances (Table 7) from these fits. Standard deviations of the distance distributions were taken as uncertainties. 

The correction for the zero-point offset to the trigonometric parallaxes of the {\it Gaia} stars \citep{Lindegren18} was not preferred in the study. \citet{Lindegren18} recommended -0.029 mas value for the zero point of trigonometric parallaxes using a sensitive quasar sample from the {\it Gaia} DR2. In our study, when a correction of -0.029 mas was applied to trigonometric parallax values of the cluster member stars, it is seen that the differences in the distances of the clusters vary from 11 to 115 pc. Since these values are within the distance errors ($\pm1\sigma$) calculated for the clusters, we considered that the zero-point offset correction for trigonometric parallaxes is not statistically significant and therefore not needed for the open clusters studied here.

\citet{Bailer-Jones18} emphasized that precise distances cannot be estimated by applying linear method to trigonometric parallaxes for most of the stars in the {\it Gaia} DR2. Instead, they suggested that distances should be obtained from geometric approaches, which take into account probability distribution of stars' distances. Consequently, they inferred distances and distance errors of 1.33 billion stars in {\it Gaia} DR2 using geometric approach. In order to compare distances from linear and geometric approaches, we also constructed distance distributions of clusters' most probable stars calculated in the two approaches and obtained distances by fitting them with Gaussian curves (Fig. 12). Table 7 shows cluster distances obtained from three different methods. 

In this study, membership probabilities of the stars were estimated via UPMASK method \citep{Krone-Martins14} with different k-means values by taking into account {\it UBV} magnitudes, equatorial coordinates, proper motion components and trigonometric parallaxes. \citet{Cantat-Gaudin18} also estimated membership probabilities of the stars in the cluster directions applying UPMASK method and determined mean distances and proper motion components. Note that they used {\it Gaia} magnitudes instead of {\it UBV} magnitudes. As a result, number of cluster member stars are not the same in the two studies. Table 7 reveals that mean distances obtained from isochrone fitting method and trigonometric parallaxes in this study are in agreement with those estimated by \citet{Cantat-Gaudin18}. A similar agreement can be also seen for the proper motion components. As Collinder 421 and Roslund 1 were not studied by \citet{Cantat-Gaudin18}, we could not compare our astrometric results with their estimations. In addition, we compared mean distances of eight clusters in this study with those calculated by geometric method of \citet{Bailer-Jones18}. An inspection of Table 7 shows that the distances obtained from isochrone fitting and trigonometric parallaxes are in agreement with distances from geometric method, as well. From this comparison, we conclude that the geometric method of \citet{Bailer-Jones18} gives very similar distances with those obtained from isochrone fitting and trigonometric parallaxes for the clusters within about 2 kpc from the Sun. We believe that astrophysical parameters of the eight clusters in this study are not suffered from parameter degeneration since these parameters were determined by independent methods and the distances in this study were confirmed by different distance determination methods.

\begin{figure*}
\centering
\includegraphics[scale=.33, angle=0]{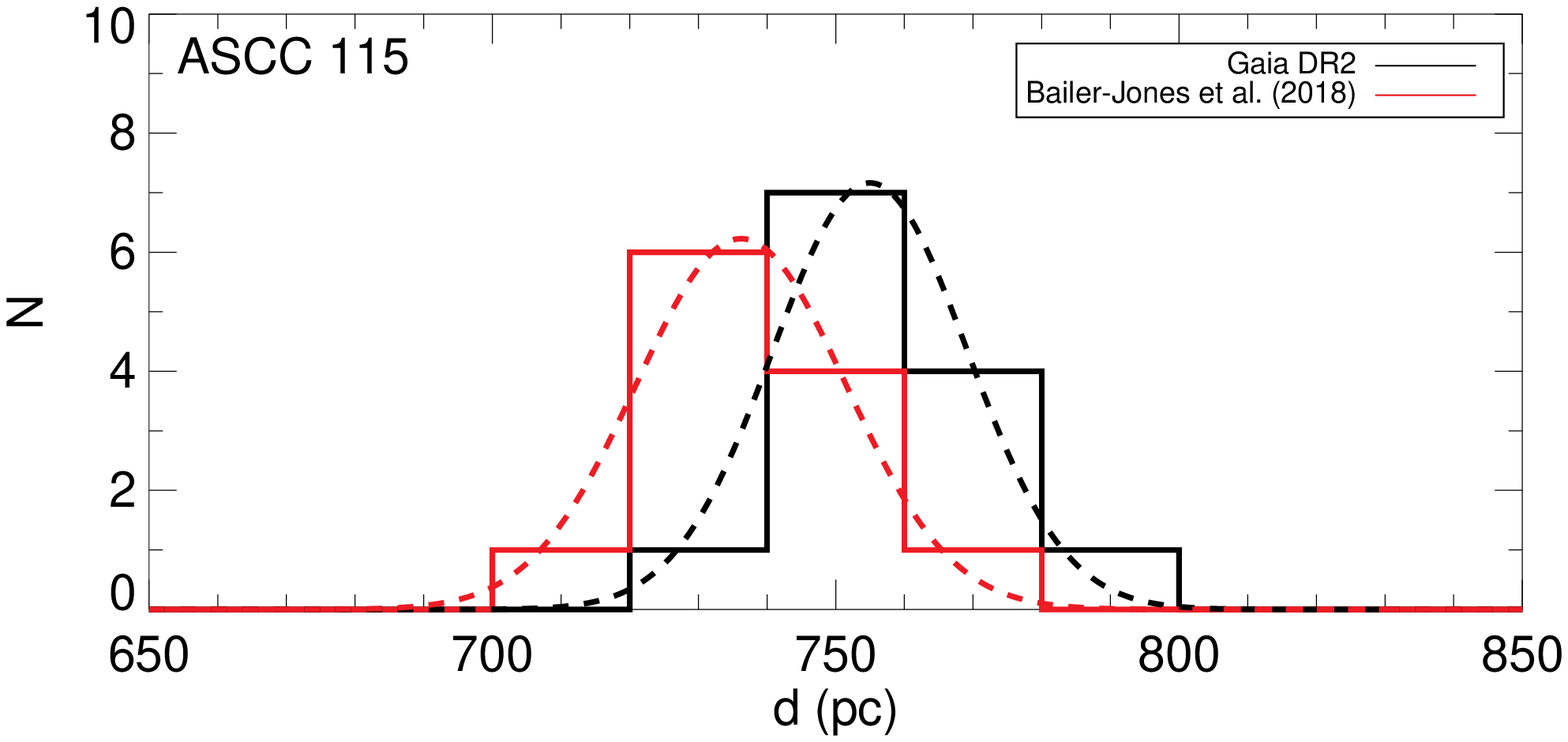}
\includegraphics[scale=.33, angle=0]{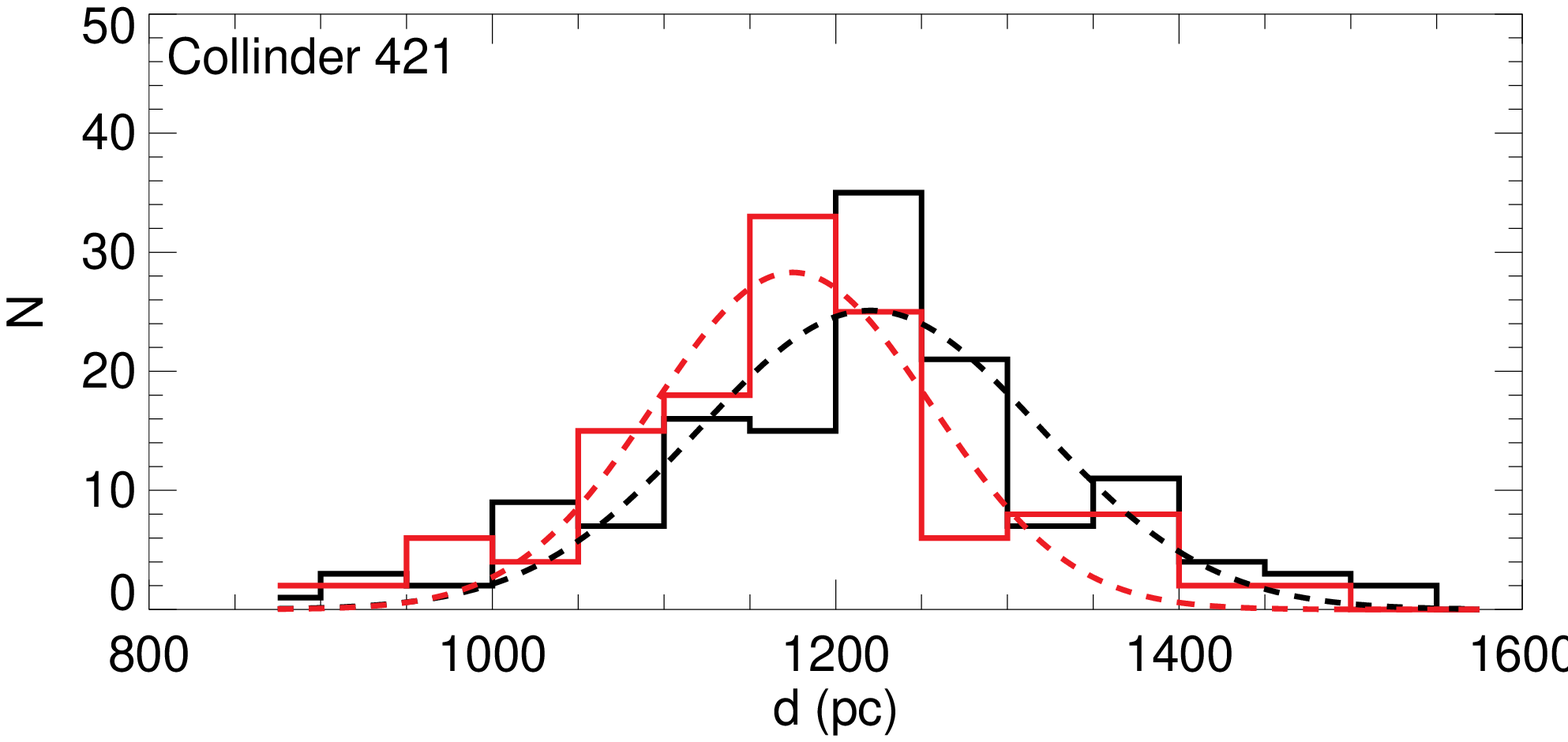}\\
\includegraphics[scale=.33, angle=0]{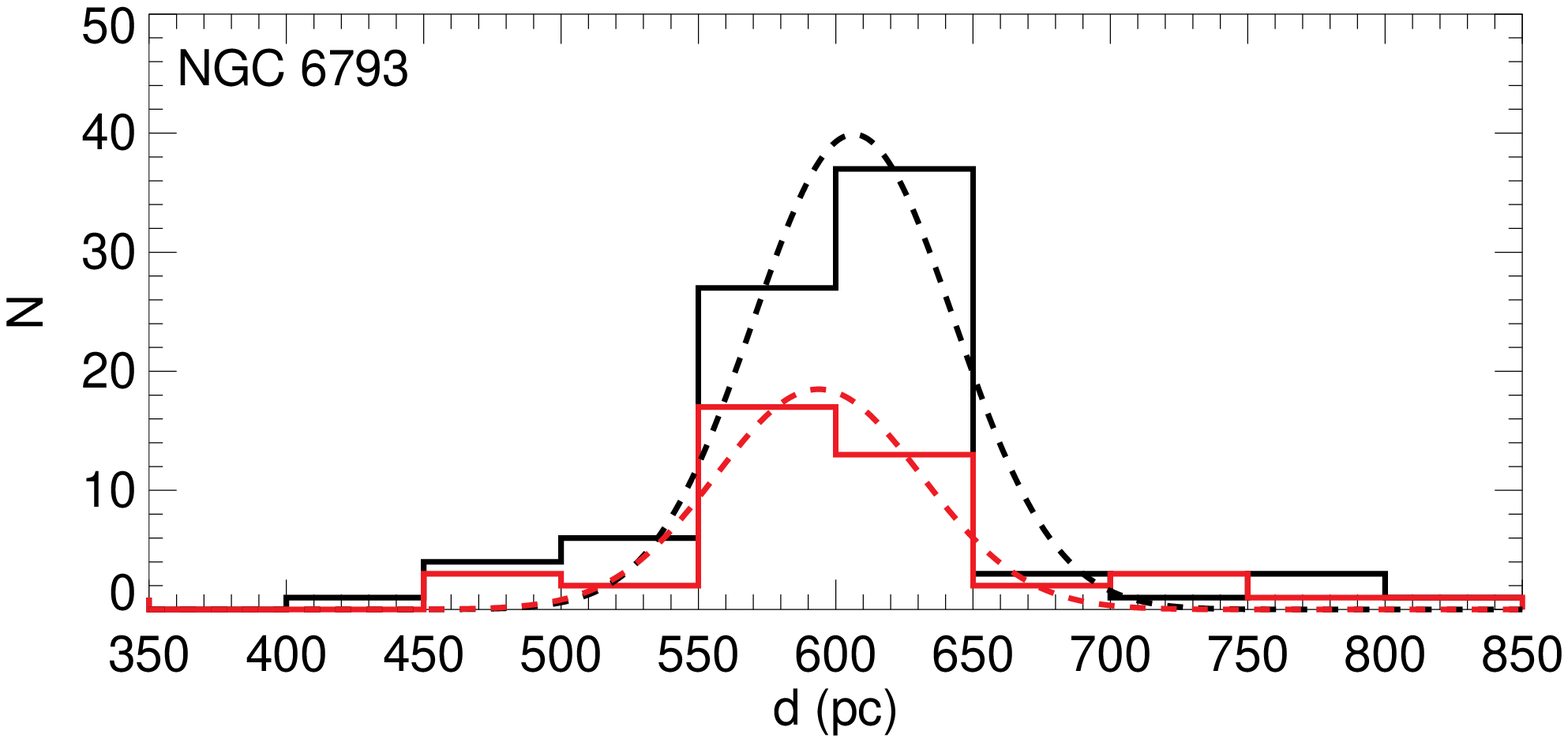}
\includegraphics[scale=.33, angle=0]{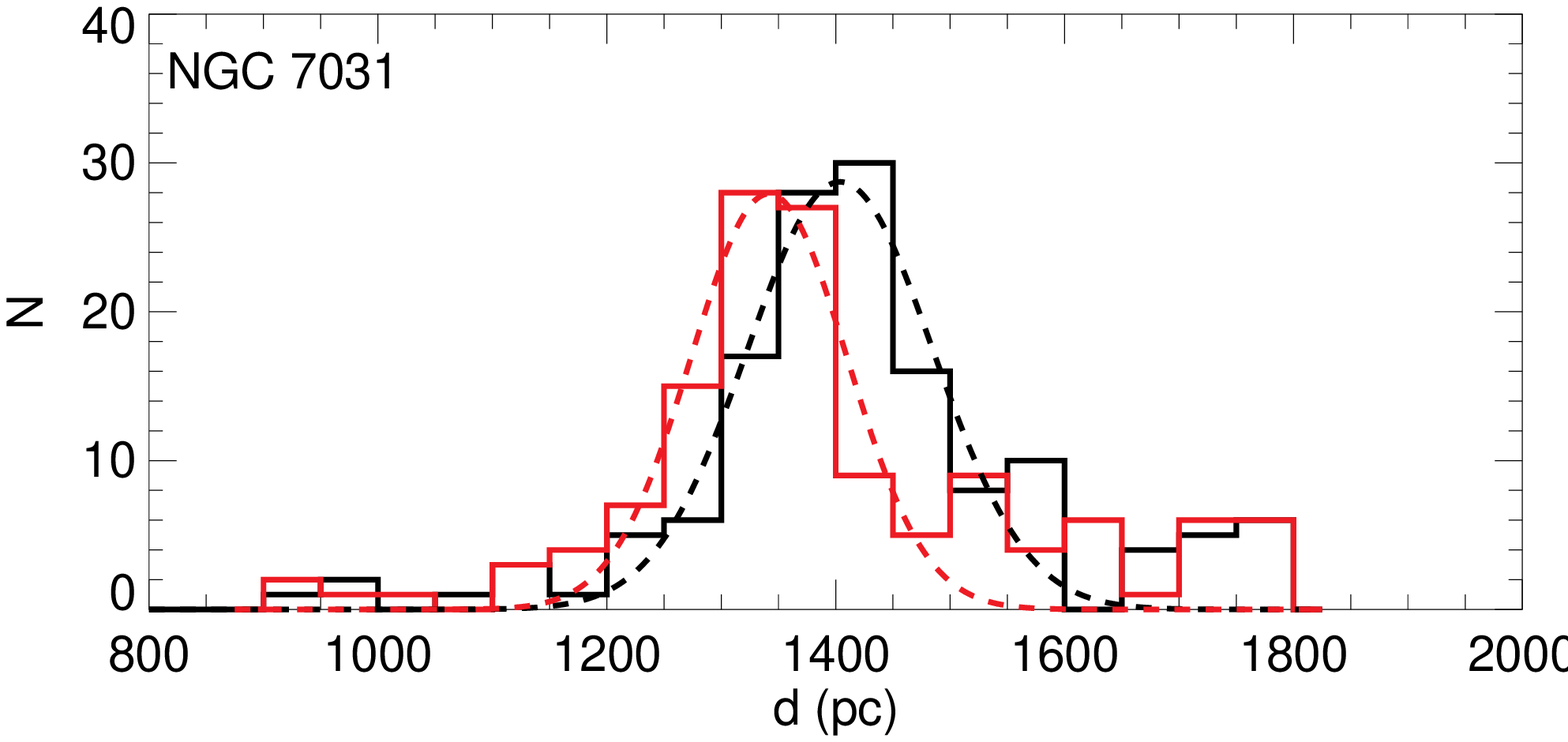}\\
\includegraphics[scale=.33, angle=0]{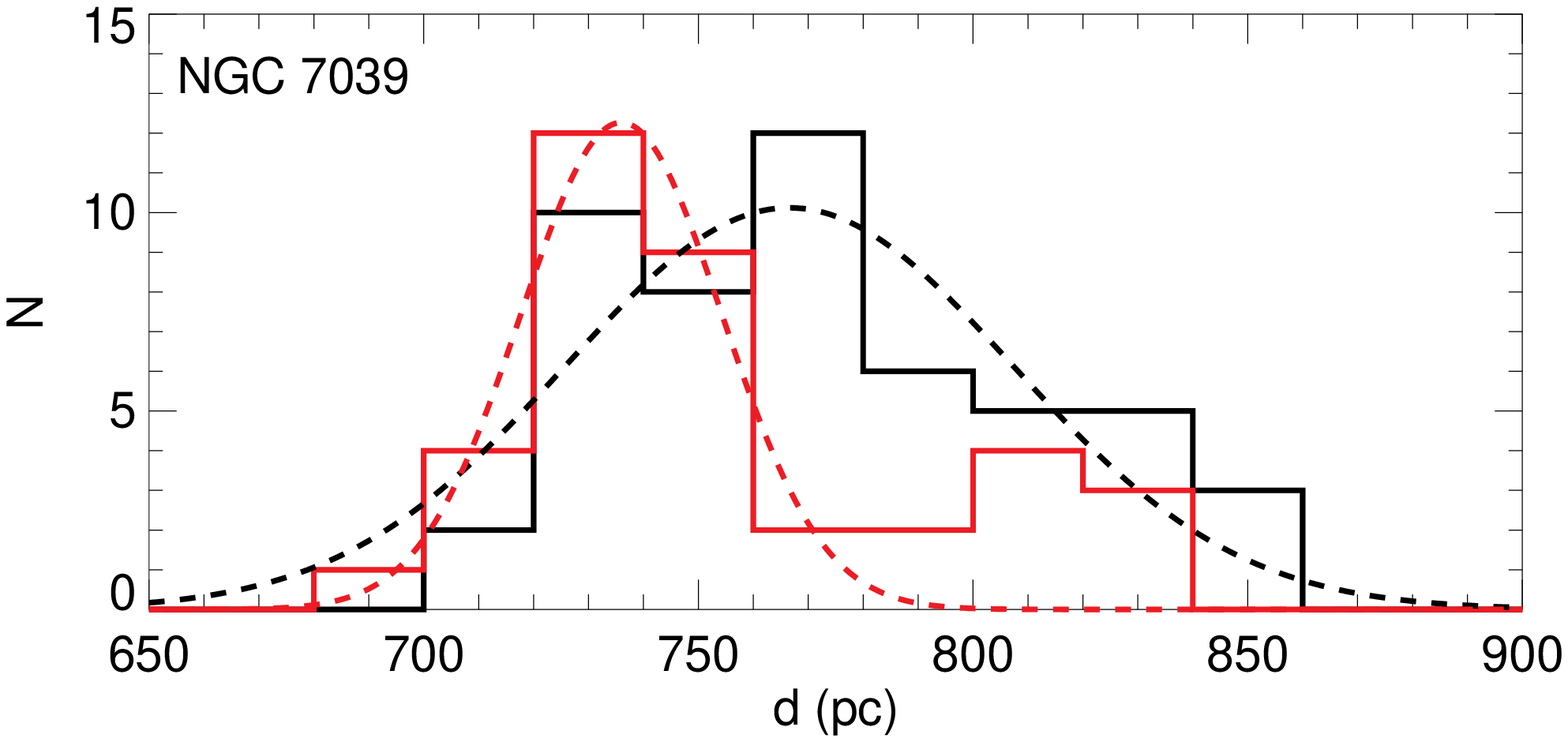}
\includegraphics[scale=.33, angle=0]{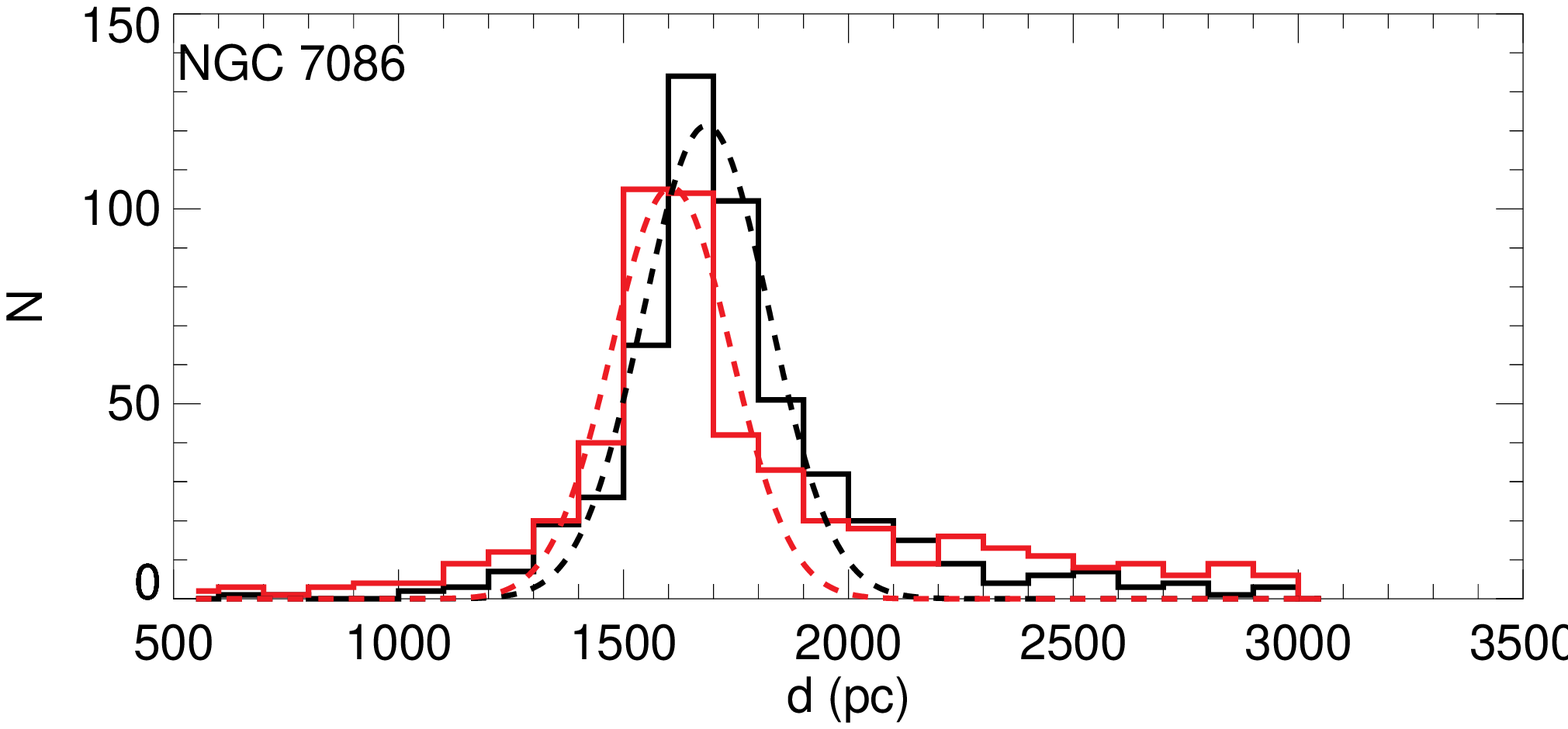}\\
\includegraphics[scale=.33, angle=0]{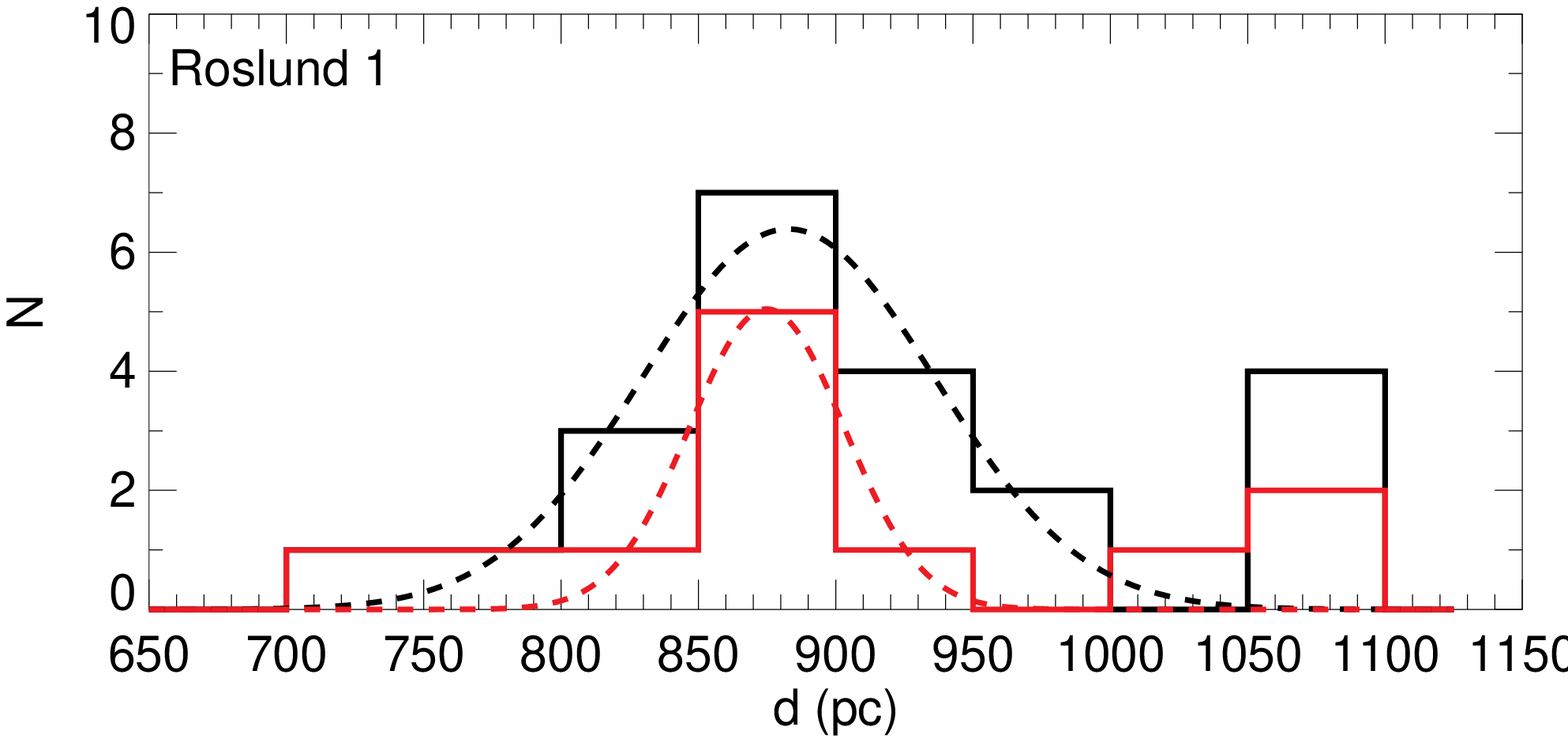}
\includegraphics[scale=.33, angle=0]{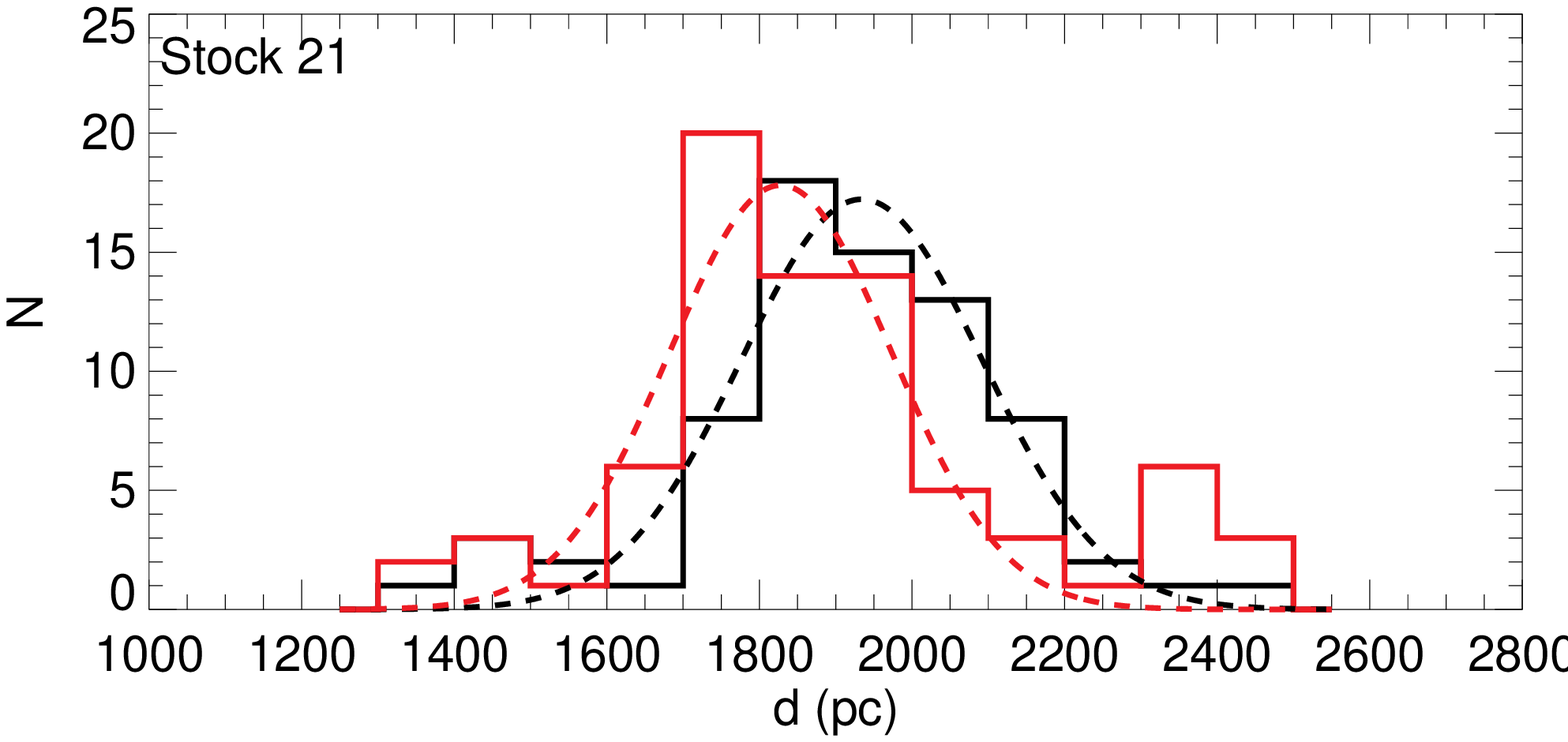}\\
\caption{The distance histograms of the cluster stars whose distances are calculated by linear \citep{Gaia18} and geometric parallax \citep{Bailer-Jones18} methods. The dashed lines show the Gaussian fits to the distance histograms.}
\end{figure*}

\section{Acknowledgments}
We thank the anonymous referees for their insightful and constructive suggestions, which significantly improved the paper. This study has been supported in part by the Scientific and Technological Research Council (T\"UB\.ITAK) 113F270. Part of this work was supported by the Research Fund of the University of Istanbul, Project Numbers: FDK-2016-22543, FBG-2017-23943, and BYP 48482. We thank to T\"UB\.ITAK National Observatory for a partial support in using T100 telescope with project number 15AT100-738. We also thank to the on-duty observers and members of the technical staff at the T\"UB\.ITAK National Observatory for their support before and during the observations. This study is a part of the PhD thesis of Talar Yontan. 

This work has made use of data from the European Space Agency (ESA) mission {\it Gaia} \\ 
(https://www.cosmos.esa.int/gaia), processed by the {\it Gaia} Data Processing and Analysis Consortium \\ 
(DPAC, https://www.cosmos.esa.int/web/gaia/dpac/consortium). Funding for the DPAC has been provided by national institutions, in particular the institutions participating in the {\it Gaia} Multilateral Agreement.

\end{document}